\documentclass[a4paper,fleqn]{cas-sc}
\ExplSyntaxOn
\cs_gset:Npn \__first_footerline:
  { \group_begin: \small \sffamily \__short_authors: \group_end: }
\ExplSyntaxOff

\usepackage{balance} 
\usepackage[numbers]{natbib}

\usepackage{float}

\usepackage[ruled,linesnumbered]{algorithm2e}
\usepackage[sort&compress,nameinlink,noabbrev]{cleveref}
\usepackage{paralist}
\usepackage{comment}
\usepackage{xspace}
\usepackage{xcolor}
\usepackage{nicematrix}
\usepackage{tabstackengine}
\usepackage{mathtools}
\usepackage{xintexpr}
\usepackage{subcaption}

\usepackage{amsthm}

\usepackage{thm-restate}

\usepackage[textsize=tiny,linecolor=green!70!black, backgroundcolor=green!10, bordercolor=black, disable]{todonotes}

\newcommand{\friendshipColor}{blue}
\newcommand{\enemyColor}{red}

\newcommand{\agent}{agent\xspace}
\newcommand{\agents}{agents\xspace}
\newcommand{\Agent}{Agent\xspace}
\newcommand{\Agents}{Agents\xspace}

\newcommand{\nbAgents}{\ensuremath{n}\xspace}

\newcommand{\agentSet}{\ensuremath{A}\xspace}

\newcommand{\agentSetInMaxCoalition}{\ensuremath{X}\xspace}
\newcommand{\partition}{\ensuremath{\Pi}\xspace}
\newcommand{\coalition}{\ensuremath{C}\xspace}
\newcommand{\blockingCoalition}{\ensuremath{B}\xspace}

\newcommand{\friendSuperscript}{{\color{\friendshipColor}g}}
\newcommand{\enemySuperscript}{{\color{\enemyColor}b}}

\newcommand{\friendshipGraph}{\ensuremath{G^{\friendSuperscript}}\xspace}
\newcommand{\enemyGraph}{\ensuremath{G^{\enemySuperscript}}\xspace}
\newcommand{\friendshipEdgeSet}{\ensuremath{E^{\friendSuperscript}}\xspace}
\newcommand{\enemyEdgeSet}{\ensuremath{E^{\enemySuperscript}}\xspace}

\newcommand{\coalitionOfi}[1]{\ensuremath{\partition(#1)}\xspace}

\DeclareMathOperator{\friendOf}{Fr}
\DeclareMathOperator{\enemiesOf}{En}
\DeclareMathOperator{\neutralsOf}{Ne}
\newcommand{\friendsOfi}[1]{\ensuremath{\friendOf(#1)}\xspace}
\newcommand{\enemiesOfi}[1]{\ensuremath{\enemiesOf(#1)}\xspace}
\newcommand{\neutralsOfi}[1]{\ensuremath{\neutralsOf(#1)}\xspace}

\newcommand{\coalA}{\ensuremath{A}}
\newcommand{\coalB}{\ensuremath{B}}

\newcommand{\labell}{ell}
\newcommand{\bipart}{$^\diamond$}

\newcommand{\decprob}[3]{
  \begin{center}    \begin{minipage}{0.94\linewidth}      \textsc{#1}\\[0.2ex]
      \textbf{Input:} #2\\[0.2ex]
      \textbf{Question:} #3
    \end{minipage}  \end{center}
}

\newcommand{\probname}[1]{\textsc{#1}}

\newcommand{\optprob}[3]{
  \begin{center}    \begin{minipage}{0.94\linewidth}      \textsc{#1}\\[0.2ex]
      \textbf{Input:} #2\\[0.2ex]
      \textbf{Task:} #3
    \end{minipage}  \end{center} 
}

\newcommand{\maxclique}{\probname{Max Clique}\xspace}
\newcommand{\maxIS}{\probname{Max Independent Set}\xspace}
\newcommand{\kclique}[1]{\probname{#1-Clique}\xspace}
\newcommand{\kIS}[1]{\probname{#1-Independent Set}\xspace}

\newcommand{\partitionIntokCliques}[1]{\probname{Partition Into #1 Cliques} \xspace}

\newcommand{\kColouring}[1]{\probname{#1-Coloring}\xspace}
\newcommand{\threeSAT}{\probname{3SAT}\xspace}

\newcommand{\Wxhard}[1]{W[#1]-hard\xspace}
\newcommand{\Wxhardshort}[1]{W[#1]h\xspace}

\newcommand{\CE}{CE\xspace}
\newcommand{\SCE}{SCE\xspace}
\newcommand{\CEBounded}{CE$_{\maxNumberOfCoalitions}$\xspace}
\newcommand{\SCEBounded}{SCE$_{\maxNumberOfCoalitions}$\xspace}
\newcommand{\CEBoundedCoal}{CE$_{\maxCoalitionSize}$\xspace}
\newcommand{\SCEBoundedCoal}{SCE$_{\maxCoalitionSize}$\xspace}
\newcommand{\SCV}{SCV\xspace}
\newcommand{\CF}{CF\xspace}
\newcommand{\CV}{CV\xspace}

\newcommand{\neut}[1]{#1\unskip$^{N}$}

\newcommand{\maxDegree}{\ensuremath{\Delta}}
\newcommand{\maxDegreeFriend}{\ensuremath{\maxDegree^{\friendSuperscript}}\xspace}
\newcommand{\maxDegreeEnemy}{\ensuremath{\maxDegree^{\enemySuperscript}}\xspace}

\newcommand{\maxDegreeFE}{\ensuremath{\maxDegree^{\friendSuperscript,\enemySuperscript}}\xspace}

\newcommand{\treewidth}{\ensuremath{\text{tw}}}
\newcommand{\treewidthFriend}{\ensuremath{\treewidth^{\friendSuperscript}}\xspace}
\newcommand{\treewidthEnemy}{\ensuremath{\treewidth^{\enemySuperscript}}\xspace}

\newcommand{\maxCoalitionSize}{\ensuremath{|\coalition|}\xspace}
\newcommand{\maxNumberOfCoalitions}{\ensuremath{|\partition|}\xspace}

\newcommand{\ri}{\ensuremath{{\lambda}}}

\newcommand{\tvarN}[1]{\ensuremath{x_{#1}}}
\newcommand{\fvarN}[1]{\ensuremath{\bar{x}_{#1}}}

\newcommand{\cladN}[2]{\ensuremath{c_{#1,#2}}}
\newcommand{\claeN}[2]{\ensuremath{d_{#1,#2}}}\newcommand{\clafN}[1]{\ensuremath{e_{#1}}}

\newcommand{\pickN}[1]{\ensuremath{p_{#1}}}
\newcommand{\qickN}[1]{\ensuremath{q_{#1}}}
\newcommand{\mainpickN}{\ensuremath{\hat{p}}}

\newcommand{\tvar}[2]{\ensuremath{\tvarN{#1}(#2)}}
\newcommand{\fvar}[2]{\ensuremath{\fvarN{#1}(#2)}}

\newcommand{\clad}[3]{\ensuremath{\cladN{#1}{#2}(#3)}}
\newcommand{\clae}[3]{\ensuremath{\claeN{#1}{#2}(#3)}}
\newcommand{\claf}[2]{\ensuremath{\clafN{#1}(#2)}}

\newcommand{\tvars}[1]{\ensuremath{X_{#1}}}
\newcommand{\fvars}[1]{\ensuremath{\bar{X}_{#1}}}

\newcommand{\clads}[2]{\ensuremath{C_{#1,#2}}}
\newcommand{\claes}[2]{\ensuremath{D_{#1,#2}}}
\newcommand{\clafs}[1]{\ensuremath{E_{#1}}}

\newcommand{\picks}[1]{\ensuremath{P_{#1}}}
\newcommand{\qicks}[1]{\ensuremath{Q_{#1}}}
\newcommand{\mainpicks}{\ensuremath{\hat{P}}}

\newcommand{\cla}[3]{\ifthenelse{\equal{#2}{1}}{\clad{#1}{1}{#3}}
{\ifthenelse{\equal{#2}{2}}{\clad{#1}{2}{#3}}
{\ifthenelse{\equal{#2}{3}}{\clad{#1}{3}{#3}}
{\ifthenelse{\equal{#2}{4}}{\clae{#1}{1}{#3}}
{\ifthenelse{\equal{#2}{5}}{\clae{#1}{2}{#3}}
{\ifthenelse{\equal{#2}{6}}{\claf{#1}{#3}}
{\textbf{\color{red}[Arg \ensuremath{#2} not supported]}}}}}}}}

\newcommand{\pick}[2]{\ensuremath{\pickN{#1}(#2)}}
\newcommand{\qick}[2]{\ensuremath{\qickN{#1}(#2)}}
\newcommand{\mainpick}[1]{\ensuremath{\mainpickN(#1)}}

\newcommand{\gi}{\ensuremath{{k}}}

\newcommand{\ii}{\ensuremath{{i}}}

\newcommand{\cia}{\ensuremath{{j}}}

\newcommand{\cib}{\ensuremath{{t}}}

\newcommand{\ria}{\ensuremath{{\theta}}}

\newcommand{\clausecount}{{\ensuremath{b}}}

\newcommand{\agentholder}{\ensuremath{\alpha}}

\newcommand{\varT}{\text{\color{blue!80!black}T}}
\newcommand{\varF}{\text{\color{red!60!black}F}}

\newcommand{\posVertexTikz}[1]{\tvar {#1}  0}
\newcommand{\negVertexTikz}[1]{\fvar {#1}  0}
\newcommand{\posVertexPrimeTikz}[1]{\tvar {#1} 1}
\newcommand{\negVertexPrimeTikz}[1]{\fvar {#1} 1}
\newcommand{\posVertexSecondTikz}[1]{\tvar {#1}  2}
\newcommand{\negVertexSecondTikz}[1]{\fvar {#1}  2}

\newcommand{\clauseNewVertexTikz}[1]{\cla \cia {#1} 0}
\newcommand{\clausePrimeNewVertexTikz}[1]{\cla \cia {#1} 1}
\newcommand{\clauseSecondNewVertexTikz}[1]{\cla \cia {#1} 2}

\newcommand{\trueVertexTikz}[1]{\pick {#1} 0}
\newcommand{\falseVertexTikz}[1]{\pick {#1} 2}
\newcommand{\neutralVertexTikz}[1]{\pick {#1} 1}
\newcommand{\trueVertexTikzQ}[1]{\qick {#1} 0}
\newcommand{\falseVertexTikzQ}[1]{\qick {#1} 2}
\newcommand{\neutralVertexTikzQ}[1]{\qick {#1} 1}

\crefname{section}{Section}{Sections}
\crefname{figure}{Figure}{Figures}
\crefname{lemma}{Lemma}{Lemmas}
\crefname{proposition}{Proposition}{Propositions}
\crefname{algorithm}{Algorithm}{Algorithms}
\crefname{obs}{Observation}{Observations}
\crefname{table}{Table}{Tables}
\crefname{theorem}{Theorem}{Theorems}
\crefname{claim}{Claim}{Claims}
\crefalias{AlgoLine}{line}

\newtheorem{proposition}{Proposition}
\newtheorem{obs}{Observation}
\newtheorem{lemma}{Lemma}
\newtheorem{corollary}{Corollary}

\newtheorem{claim}{Claim}
\newtheorem{definition}{Definition}

\newcommand{\myemph}[1]{{\color{green!25!black}\emph{#1}}}

\newcommand{\enemyav}{enemy aversion}

\newcommand{\ashg}{ASHG}

\newcommand{\intuition}[1]{\textit{#1}}

\newcommand{\shortProposition}{P}
\newcommand{\shortTheorem}{T}
\newcommand{\shortCorollary}{C}

\newcommand{\tableRef}[2]{
\ifthenelse{\equal{#1}{P}}{[\shortProposition #2]}
{\ifthenelse{\equal{#1}{T}}{[\shortTheorem #2]}
{\ifthenelse{\equal{#1}{C}}{[\shortCorollary #2]}
{\textbf{\color{red}[Arg \ensuremath{#2} not supported]}}}}

}

\newcommand{\cliqf}{\ensuremath{\hat{f}}}
\newcommand{\cliqparf}{\ensuremath{\hat{g}}}

\newcommand{\fcliq}{friendship clique\xspace}

\newcommand{\fcliqs}{friendship cliques\xspace}

\newcommand{\appsymb}{$\star$}
\newcommand{\knownressymb}{$\dagger$}

\newcommand{\sets}{\ensuremath{\mathcal{S}}}
\newcommand{\elements}{\ensuremath{[3n]}}
\newcommand{\sset}[1]{\ensuremath{S_{#1}}}
\newcommand{\ecov}{\ensuremath{\mathcal{K}}}

\tikzstyle{coal} = [fill = blue!15!white, draw=none]
\tikzstyle{coalline} = [draw = orange, line width = 1.5pt, opacity=0.5]
\tikzstyle{coalgreen} = [draw=none, pattern color = green!70!black, pattern=north west lines]

\newcommand{\asg}{\ensuremath{\sigma}}

\usetikzlibrary{calc,patterns}

\usepackage{etoolbox}

\newcommand{\appendixproofwithstatement}[3]{  \gappto{\appendixtext}{
    \subsection{Proof of \cref{#1}}\label{proof:#1}
    #2
    \begin{proof}
    #3\end{proof}
  }
}
\newcommand{\appendixproofwithstatementcontinued}[4]{\begin{proof}[Proof Sketch]
#3
\end{proof}
  \gappto{\appendixtext}{
    \subsection{Continuation of Proof of \cref{#1}}\label{proof:#1}
    #2
    \begin{proof}[Proof (Continued)]
    #4\end{proof}
  }
}

\newcommand{\appendixproofwithstatementsketch}[4]{\begin{proof}[Proof Sketch]
#3\end{proof}
  \gappto{\appendixtext}{
    \subsection{Proof of \cref{#1}}\label{proof:#1}
    #2
    \begin{proof}
    #4\end{proof}
  }
}

\newcommand{\appendixsection}[1]{  \gappto{\appendixtext}{
    \section{Additional material for Section~\ref{#1}}
    \label{appsec:#1}
  }
}

\newcommand{\vars}{\mathcal{X}}
\newcommand{\clas}{\mathcal{C}}

\newcommand{\labelTwo}{'}
\tikzstyle{friend}=[blue]
\tikzstyle{neutral}=[dashed,black]
\tikzstyle{enemy}=[dashed,red]

\newcommand{\sevenfig}{
    \foreach \i in {1,2,3}{
        
        \node at (\the\numexpr 40+2*\i*360/7:8) [vertex, fill=red!30!white] (\labell\i) {\footnotesize $\lvarn {\cib_{\i}} $};
    
    }

    \foreach \i in {0,...,6}{
        
        \node at (\the\numexpr 36+\i*360/7:3.5) [vertex] (\i) {\footnotesize $a^1_{\i}$};
        \node at (\the\numexpr 36+\i*360/7:5) [vertex] (\i\labelTwo) {\footnotesize $a^2_{\i}$};

        \draw[friend] (\i) -- (\i\labelTwo);
    }

    \foreach \i in {0,...,6}{
        
        \node at (\the\numexpr 72+\i*360/7:6.5) [vertex] (\the\numexpr 10+\i) {\footnotesize$b^1_{\i}$};
        \node at (\the\numexpr 72+\i*360/7:8.5) [vertex] (\the\numexpr 10+\i\labelTwo) {\footnotesize$b^2_{\i}$};

        \draw[friend] (\the\numexpr 10+\i) -- (\the\numexpr 10+\i\labelTwo);
    }
    
    \foreach \i in {0,...,5}{
                        \draw[friend] (\the\numexpr \i) -- (\the\numexpr \i+1);
        \draw[friend] (\the\numexpr \i) -- (\the\numexpr \i+1\labelTwo);
        \draw[friend] (\the\numexpr \i\labelTwo) -- (\the\numexpr \i+1);
                \draw[friend] (\the\numexpr \i\labelTwo) -- (\the\numexpr \i+1\labelTwo);
    }
            \draw[friend] (\the\numexpr 6) -- (\the\numexpr 0);
    \draw[friend] (\the\numexpr 6) -- (\the\numexpr 0\labelTwo);
    \draw[friend] (\the\numexpr 6\labelTwo) -- (\the\numexpr 0);
        \draw[friend] (\the\numexpr 6\labelTwo) -- (\the\numexpr 0\labelTwo);

    \foreach \i in {0,...,6}{
        \draw[friend] (\the\numexpr \i) -- (\the\numexpr 10+\i);
        \draw[friend] (\the\numexpr \i) -- (\the\numexpr 10+\i\labelTwo);
        \draw[friend] (\the\numexpr \i\labelTwo) -- (\the\numexpr 10+\i);
        \draw[friend] (\the\numexpr \i\labelTwo) -- (\the\numexpr 10+\i\labelTwo);
    }

    \foreach \i in {0,...,5}{
        \draw[friend] (\the\numexpr 10+\i) -- (\the\numexpr \i+1);
        \draw[friend] (\the\numexpr 10+\i\labelTwo) -- (\the\numexpr \i+1\labelTwo);
        
        \draw[neutral] (\the\numexpr 10+\i\labelTwo) -- (\the\numexpr \i+1);
    }
    
    \draw[friend] (\the\numexpr 16) -- (\the\numexpr 0);
    \draw[friend] (\the\numexpr 16\labelTwo) -- (\the\numexpr 0\labelTwo);
    
    \draw[neutral] (\the\numexpr 16\labelTwo) -- (\the\numexpr 0);

                        \draw[neutral] (\the\numexpr 10) -- (\the\numexpr 1\labelTwo);
    \draw[neutral] (\the\numexpr 11) -- (\the\numexpr 2\labelTwo);
    \draw[neutral] (\the\numexpr 12) -- (\the\numexpr 3\labelTwo);
    \draw[neutral] (\the\numexpr 13) -- (\the\numexpr 4\labelTwo);
    \draw[neutral] (\the\numexpr 14) -- (\the\numexpr 5\labelTwo);
    \draw[neutral] (\the\numexpr 15) -- (\the\numexpr 6\labelTwo);
    \draw[neutral] (\the\numexpr 16) -- (\the\numexpr 0\labelTwo);

    \foreach \i in {1,2,3}{
        \draw[friend] (\the \numexpr 2*\i) -- (\labell\i);
        \draw[friend] (\the \numexpr 2*\i\labelTwo) -- (\labell\i);
        \draw[friend] (\the \numexpr 9+2*\i) -- (\labell\i);
        \draw[friend] (\the \numexpr 9+2*\i\labelTwo) -- (\labell\i);
    }
        \foreach \i in {0,1,3,4,5,6} {
        \draw[neutral, opacity=0.3] (\i) -- (\labell1);
        \draw[neutral, opacity=0.3] (\i\labelTwo) -- (\labell1);
    }
        \foreach \i in {0,1,2,4,5,6} {
        \draw[neutral, opacity=0.3] (\i) -- (\labell2);
        \draw[neutral, opacity=0.3] (\i\labelTwo) -- (\labell2);
    }
        \foreach \i in {0,1,2,3,5,6} {
        \draw[neutral, opacity=0.3] (\i) -- (\labell3);
        \draw[neutral, opacity=0.3] (\i\labelTwo) -- (\labell3);
    }
            \foreach \i in {0,2,3,4,5,6} {
        \draw[neutral, opacity=0.3] (1\i) -- (\labell1);
        \draw[neutral, opacity=0.3] (1\i\labelTwo) -- (\labell1);
    }
        \foreach \i in {0,1,3,4,5,6} {
        \draw[neutral, opacity=0.3] (1\i) -- (\labell2);
        \draw[neutral, opacity=0.3] (1\i\labelTwo) -- (\labell2);
    }
        \foreach \i in {0,1,2,4,5,6} {
        \draw[neutral, opacity=0.3] (1\i) -- (\labell3);
        \draw[neutral, opacity=0.3] (1\i\labelTwo) -- (\labell3);
    }

}

\newcommand{\fivefig}{
    \foreach \i in {0,...,4}{
        \node at (\the\numexpr 36+\i*360/5:3.5) [vertex] (\i) {\footnotesize $a^1_{\i}$};
        \node at (\the\numexpr 36+\i*360/5:5) [vertex] (\i\labelTwo) {\footnotesize $a^2_{\i}$};
                
        \draw[friend] (\i) -- (\i\labelTwo);
    }

    \foreach \i in {0,...,4}{
        \node at (\the\numexpr 72+\i*360/5:6.5) [vertex] (\the\numexpr 10+\i) {\footnotesize $b^1_{\i}$};
        \node at (\the\numexpr 72+\i*360/5:8.5) [vertex] (\the\numexpr 10+\i\labelTwo) {\footnotesize $b^2_{\i}$};
        \draw[friend] (\the\numexpr 10+\i) -- (\the\numexpr 10+\i\labelTwo);
    }

    \foreach \i in {0,...,3}{
                        \draw[friend] (\the\numexpr \i) -- (\the\numexpr \i+1);
        \draw[friend] (\the\numexpr \i) -- (\the\numexpr \i+1\labelTwo);
        \draw[friend] (\the\numexpr \i\labelTwo) -- (\the\numexpr \i+1);
                \draw[friend] (\the\numexpr \i\labelTwo) -- (\the\numexpr \i+1\labelTwo);
    }
            \draw[friend] (\the\numexpr 4) -- (\the\numexpr 0);
    \draw[friend] (\the\numexpr 4) -- (\the\numexpr 0\labelTwo);
    \draw[friend] (\the\numexpr 4\labelTwo) -- (\the\numexpr 0);
        \draw[friend] (\the\numexpr 4\labelTwo) -- (\the\numexpr 0\labelTwo);

    \foreach \i in {0,...,4}{
        \draw[friend] (\the\numexpr \i) -- (\the\numexpr 10+\i);
        \draw[friend] (\the\numexpr \i) -- (\the\numexpr 10+\i\labelTwo);
        \draw[friend] (\the\numexpr \i\labelTwo) -- (\the\numexpr 10+\i);
        \draw[friend] (\the\numexpr \i\labelTwo) -- (\the\numexpr 10+\i\labelTwo);
    }

    \foreach \i in {0,...,3}{
        \draw[friend] (\the\numexpr 10+\i) -- (\the\numexpr \i+1);
        \draw[friend] (\the\numexpr 10+\i\labelTwo) -- (\the\numexpr \i+1\labelTwo);
        
        \draw[neutral] (\the\numexpr 10+\i\labelTwo) -- (\the\numexpr \i+1);
    }
    
    \draw[friend] (\the\numexpr 14) -- (\the\numexpr 0);
    \draw[friend] (\the\numexpr 14\labelTwo) -- (\the\numexpr 0\labelTwo);
    
    \draw[neutral] (\the\numexpr 14\labelTwo) -- (\the\numexpr 0);

                        \draw[neutral] (\the\numexpr 10) -- (\the\numexpr 1\labelTwo);
    \draw[neutral] (\the\numexpr 11) -- (\the\numexpr 2\labelTwo);
    \draw[neutral] (\the\numexpr 12) -- (\the\numexpr 3\labelTwo);
    \draw[neutral] (\the\numexpr 13) -- (\the\numexpr 4\labelTwo);
    \draw[neutral] (\the\numexpr 14) -- (\the\numexpr 0\labelTwo);

}

\newcommand{\colorPrime}{red}
\newcommand{\colorSecond}{blue}
\newcommand{\colorInsideTriangle}{black!50!white}

\newcommand{\pickcolor}{black}

\newcommand{\pickerLabel}[3]{

\ifthenelse{\equal{#1}{0}}{\renewcommand{\pickcolor}{black}}
{\ifthenelse{\equal{#1}{1}}{\renewcommand{\pickcolor}{\colorPrime}}
{\ifthenelse{\equal{#1}{2}}{\renewcommand{\pickcolor}{\colorSecond}}
{\renewcommand{\pickcolor}{green}}}}

\node[draw=\pickcolor, color=\pickcolor,#3= 1pt of #2, inner sep = 1, fill=white, fill opacity = 0.5, text opacity = 1] {{\footnotesize\Pick \gi #1 z}}; 
}

\newcommand{\labelPrime}{'}
\newcommand{\labelSecond}{''}

\newcommand{\labelFalse}{F}

\pgfdeclarelayer{background}
\pgfdeclarelayer{foreground}
\pgfsetlayers{background,main,foreground}

\tikzstyle{pn} = [draw=gray, circle, inner sep=1.5pt,fill=gray]
\tikzstyle{pndummy} = [draw=gray, circle, inner sep=1.5pt,fill=white]
\tikzstyle{lb} = [inner sep=-1pt, fill=white, fill opacity = 0.5, text opacity = 1]
\tikzstyle{pickeredge} = [opacity = 0.4]
\tikzstyle{varedge} = [opacity = 0.4]

\newcommand{\vertexRight}{0}
\newcommand{\vertexTopRight}{45}
\newcommand{\vertexTop}{90}
\newcommand{\vertexTopLeft}{135}
\newcommand{\vertexLeft}{180}
\newcommand{\vertexBottomLeft}{225}
\newcommand{\vertexBottom}{270}
\newcommand{\vertexBottomRight}{315}

\newcommand{\vertexNameTopLeft}{0}
\newcommand{\vertexNameTopRight}{1}
\newcommand{\vertexNameBottomLeft}{2}
\newcommand{\vertexNameBottomRight}{3}
\newcommand{\vertexNameBelow}{4}
\newcommand{\vertexNameAbove}{5}
\newcommand{\vertexNameLeft}{6}
\newcommand{\vertexNameRight}{7}

\newcommand{\myVertexNameSize}{\footnotesize}
\newcommand{\myVertexNameSizeHL}{\tiny}

\newcommand{\myVertexName}[3]{
	\ifthenelse{\equal{#3}{\vertexNameTopLeft}}{
	            \node [lb,above left= 1pt of #2] {{\myVertexNameSize #1}};
	        }{
	        	\ifthenelse{\equal{#3}{\vertexNameTopRight}}{
	            \node [lb,above right = 1pt of #2] {{\myVertexNameSize #1}};
	        }{
	        	\ifthenelse{\equal{#3}{\vertexNameBottomLeft}}{
	            \node [lb,below left = 1pt of #2] {{\myVertexNameSize #1}};
	        }{
	        	\ifthenelse{\equal{#3}{\vertexNameBottomRight}}{
	            \node [lb,below right=1pt of #2] {{\myVertexNameSize #1}};
	        }{
	        	\ifthenelse{\equal{#3}{\vertexNameBelow}}{
	            \node [lb,below = 1pt of #2] {{\myVertexNameSize #1}};
	        }{
	        	\ifthenelse{\equal{#3}{\vertexNameAbove}}{
	            \node [lb,above = 1pt of #2] {{\myVertexNameSize #1}};
	        }{
	        	\ifthenelse{\equal{#3}{\vertexNameLeft}}{
	            \node [lb,left = 1pt of #2] {{\myVertexNameSize #1}};
	        }{
	        	\ifthenelse{\equal{#3}{\vertexNameRight}}{
	            \node [lb,right = 1pt of #2] {{\myVertexNameSize #1}};
	        }{
	        	\node [lb,above left= -3pt of #2] {{\myVertexNameSize #1}};
	        }
	        }
	        }
            }
            }
	       }
		}
	}
}

\newcommand{\myVertexNameHL}[3]{
	\ifthenelse{\equal{#3}{\vertexNameTopLeft}}{
	            \node [lb,above left= -3pt of #2] {{\myVertexNameSizeHL #1}};
	        }{
	        	\ifthenelse{\equal{#3}{\vertexNameTopRight}}{
	            \node [lb,above right = 1pt of #2] {{\myVertexNameSizeHL #1}};
	        }{
	        	\ifthenelse{\equal{#3}{\vertexNameBottomLeft}}{
	            \node [lb,below left = 1pt of #2] {{\myVertexNameSizeHL #1}};
	        }{
	        	\ifthenelse{\equal{#3}{\vertexNameBottomRight}}{
	            \node [lb,below right=1pt of #2] {{\myVertexNameSizeHL #1}};
	        }{
	        	\ifthenelse{\equal{#3}{\vertexNameBelow}}{
	            \node [lb,below = 1pt of #2] {{\myVertexNameSizeHL #1}};
	        }{
	        	\ifthenelse{\equal{#3}{\vertexNameAbove}}{
	            \node [lb,above = 1pt of #2] {{\myVertexNameSizeHL #1}};
	        }{
	        	\node [lb,above left= -3pt of #2] {{\myVertexNameSizeHL #1}};
	        }
	        }
	        }
	       }
		}
	}
}

\newcommand{\offsetXSecondN}{1.25}
\newcommand{\offsetYSecondN}{1}
\newcommand{\offsetXPrimeN}{-1.25}
\newcommand{\offsetYPrimeN}{0.75}

\newcommand{\offsetXSecondS}{1}
\newcommand{\offsetYSecondS}{-0.5}
\newcommand{\offsetXPrimeS}{-0.75}
\newcommand{\offsetYPrimeS}{-1}

\newcommand{\myTriangleN}[9]{
  \begin{pgfonlayer}{foreground}
    \node[pn] at (#4,#5) [vertex] (#6) {}; 
  \end{pgfonlayer}

  \myVertexName{#1}{#6}{#7}

  \begin{pgfonlayer}{foreground}
    \node[pn, color=\colorPrime] at (#4+\offsetXPrimeN,#5+\offsetYPrimeN) [vertex] (#6\labelPrime) {};  
  \end{pgfonlayer}

  \myVertexName{#2}{#6\labelPrime}{#8}

  \begin{pgfonlayer}{foreground}
    \node[pn, color=\colorSecond] at (#4+\offsetXSecondN,#5+\offsetYSecondN) [vertex] (#6\labelSecond) {};
  \end{pgfonlayer}
  \myVertexName{#3}{#6\labelSecond}{#9}
  
  \begin{pgfonlayer}{background}
    \draw[\colorInsideTriangle, enemy] (#6) -- (#6\labelPrime); 
        \draw[\colorInsideTriangle, enemy] (#6) to (#6\labelSecond); 
    \draw[\colorInsideTriangle, enemy] (#6\labelPrime) -- (#6\labelSecond); 
  \end{pgfonlayer}
}

\newcommand{\myTriangleS}[9]{
  \begin{pgfonlayer}{foreground}
    \node[pn] at (#4,#5) [vertex] (#6) {}; 
  \end{pgfonlayer}

  \myVertexName{#1}{#6}{#7}

  \begin{pgfonlayer}{foreground}
    \node[pn, color=\colorPrime] at (#4+\offsetXPrimeS,#5+\offsetYPrimeS) [vertex] (#6\labelPrime) {};  
  \end{pgfonlayer}

  \myVertexName{#2}{#6\labelPrime}{#8}

  \begin{pgfonlayer}{foreground}
    \node[pn, color=\colorSecond] at (#4+\offsetXSecondS,#5+\offsetYSecondS) [vertex] (#6\labelSecond) {};
  \end{pgfonlayer}
  \myVertexName{#3}{#6\labelSecond}{#9}
  
  \begin{pgfonlayer}{background}
    \draw[\colorInsideTriangle, enemy] (#6) -- (#6\labelPrime); 
        \draw[\colorInsideTriangle, enemy] (#6) to (#6\labelSecond); 
    \draw[\colorInsideTriangle, enemy] (#6\labelPrime) -- (#6\labelSecond); 
  \end{pgfonlayer}
}

\newcommand{\myTrianglePickerS}[7]{
    \myTriangleN{$\trueVertexTikz{#1}$}{$\neutralVertexTikz{#1}$}{$\falseVertexTikz{#1}$}{#2}{#3}{#4}{#5}{#6}{#7}
}

\newcommand{\myTriangleQickerS}[7]{
    \myTriangleN{$\trueVertexTikzQ{#1}$}{$\neutralVertexTikzQ{#1}$}{$\falseVertexTikzQ{#1}$}{#2}{#3}{#4}{#5}{#6}{#7}
}

\newcommand{\myTrianglePosLiteralN}[7]{
    \myTriangleN{$\posVertexTikz{#1}$}{$\posVertexPrimeTikz{#1}$}{$\posVertexSecondTikz{#1}$}{#2}{#3}{#4}{#5}{#6}{#7}
}

\newcommand{\myTriangleNegLiteralN}[7]{
    \myTriangleN{$\negVertexTikz{#1}$}{$\negVertexPrimeTikz{#1}$}{$\negVertexSecondTikz{#1}$}{#2}{#3}{#4}{#5}{#6}{#7}
}

\newcommand{\myTriangleClauseNew}[7]{
    \myTriangleS{$\clauseNewVertexTikz{#1}$}{$\clausePrimeNewVertexTikz{#1}$}{$\clauseSecondNewVertexTikz{#1}$}{#2}{#3}{#4}{#5}{#6}{#7}
}

\newcommand{\myTriangleEdges}[5]{
    \draw[enemy] (#1) to [out=#3,in=#4,looseness=#5] (#2);
    \draw[\colorPrime, enemy] (#1\labelPrime) to [out=#3,in=#4,looseness=#5] (#2\labelPrime);
    \draw[\colorSecond, enemy] (#1\labelSecond) to [out=#3,in=#4,looseness=#5] (#2\labelSecond); 
}

\newcommand{\TvarHL}[3]{\ensuremath{\mathbf{x}^{#1,#3}_{#2}}}

\newcommand{\FvarHL}[3]{\ensuremath{\bar{\mathbf{x}}^{#1,#3}_{#2}}}

\newcommand{\ClaHL}[2]{\ensuremath{C^{#2}_{#1}}}
\newcommand{\rewHL}[2]{\ensuremath{r^{#1,#2}}}

\newcommand{\labelI}{'''}

\newcommand{\labelNminusone}{''''}
\newcommand{\labelN}{'''''}

\newcommand{\labelClause}{C}
\newcommand{\labelReward}{R}

\tikzstyle{gadget} = [circle, inner sep=3.5pt]

\tikzset{hfit/.style={draw,color=#1,rectangle, rounded corners, inner xsep=10pt, inner ysep=10pt, fill=#1!10},vfit/.style={rounded corners, fill=#1!5}}

\newcommand{\myVertexHL}[6]{
  \begin{pgfonlayer}{foreground}
    \node[pn,color=#5] at (#2,#3) [vertex] (#4) {}; 
  \end{pgfonlayer}

  \myVertexNameHL{#1}{#4}{#6}
}

\newcommand{\myVerticesVariableHL}[8]{
  \begin{pgfonlayer}{foreground}
    \node[pn,color=#6] at (#3,#4) [vertex] (#5) {};
  \end{pgfonlayer} 
  \myVertexNameHL{#1}{#5}{#7}

  \begin{pgfonlayer}{foreground}
    \node[pn,color=#6] at (#3+0.5,#4) [vertex] (#5\labelFalse) {}; 
  \end{pgfonlayer}

  \myVertexNameHL{#2}{#5\labelFalse}{#8}
  
  \draw[enemy,color=#6] (#5) -- (#5\labelFalse);
}

\newcommand{\myVerticesVariablesHL}[6]{

  \myVerticesVariableHL{\TvarHL{#1}{1}{#2}}{\FvarHL{#1}{1}{#2}}{#3}{#4}{#5\labelOne}{#6}{\vertexNameBottomLeft}{\vertexNameBelow}
  \myVerticesVariableHL{\TvarHL{#1}{2}{#2}}{\FvarHL{#1}{2}{#2}}{#3+2}{#4}{#5\labelTwo}{#6}{\vertexNameBelow}{\vertexNameBottomRight}
  
  \node at (#3+4.2,#4) {\dots};

  \myVerticesVariableHL{\TvarHL{#1}{i}{#2}}{\FvarHL{#1}{i}{#2}}{#3+6}{#4}{#5\labelI}{#6}{\vertexNameBottomLeft}{\vertexNameBottomRight}

  \node at (#3+8.2,#4) {\dots};

  \myVerticesVariableHL{\TvarHL{#1}{n-1}{#2}}{\FvarHL{#1}{n-1}{#2}}{#3+10}{#4}{#5\labelNminusone}{#6}{\vertexNameBottomLeft}{\vertexNameBelow}
  \myVerticesVariableHL{\TvarHL{#1}{n}{#2}}{\FvarHL{#1}{n}{#2}}{#3+12}{#4}{#5\labelN}{#6}{\vertexNameBelow}{\vertexNameBottomRight}
}

\newcommand{\myClauseGadgetHL}[6]{
  \begin{pgfonlayer}{foreground}
    \node[gadget,color=#5] at (#2,#3) [vertex] (#4) {}; 
  \end{pgfonlayer}
  \myVertexNameHL{#1}{#4}{#6}
}

\newcommand{\myClauseGadgetsHL}[5]{
  \myClauseGadgetHL{\ClaHL{1}{#1}}{#2+2.25}{#3}{#4\labelOne}{#5}{\vertexNameAbove}
  \myClauseGadgetHL{\ClaHL{2}{#1}}{#2+4.25}{#3}{#4\labelTwo}{#5}{\vertexNameAbove}
  \node at (#2+6.25,#3) {\dots};
  \myClauseGadgetHL{\ClaHL{m-1}{#1}}{#2+8.25}{#3}{#4\labelNminusone}{#5}{\vertexNameAbove}
  \myClauseGadgetHL{\ClaHL{m}{#1}}{#2+10.25}{#3}{#4\labelN}{#5}{\vertexNameAbove}
}

\newcommand{\myCNFRHL}[5]{
	\myVerticesVariablesHL{2}{#1}{#2}{#3}{#4}{#5}
	\myClauseGadgetsHL{2}{#2}{#3+1.5}{#4\labelClause}{#5}
	
	\myVertexHL{\rewHL{2}{#1}}{#2-1}{#3-2}{#4\labelReward}{#5}{\vertexNameBottomLeft}
	
	\ifthenelse{\equal{#1}{1}}{
		\draw[enemy] (#4\labelReward) -- (#4\labelOne\labelFalse);
	}{
		\ifthenelse{\equal{#1}{2}}{
			\draw[enemy] (#4\labelReward) -- (#4\labelTwo\labelFalse);
	}{
		\ifthenelse{\equal{#1}{i}}{
			\draw[enemy] (#4\labelReward) -- (#4\labelI\labelFalse);
	}{
		\ifthenelse{\equal{#1}{n-1}}{
			\draw[enemy] (#4\labelReward) -- (#4\labelNminusone\labelFalse);
	}{
		\ifthenelse{\equal{#1}{n}}{
			\draw[enemy] (#4\labelReward) -- (#4\labelN\labelFalse);
	}{
		\draw[enemy] (#4\labelReward) -- (#4\labelOne\labelFalse);
	}
	}
	}
	}
	}
	
	\begin{pgfonlayer}{background}
	  \node[fit=(#4\labelReward)(#4\labelN\labelFalse)(#4\labelClause\labelN), hfit=#5] {};
	\end{pgfonlayer}
	
}

\newcommand{\myCNFRHLbis}[5]{
	\myVerticesVariablesHL{z}{k}{#2}{#3}{#4}{#5}
	\myClauseGadgetsHL{z}{#2}{#3+1.5}{#4\labelClause}{#5}
	
	\myVertexHL{\rewHL{z}{k}}{#2-1}{#3-2}{#4\labelReward}{#5}{\vertexNameBottomRight}
	
	\node[align=left] at (#2-0.5,#3+2) {\tiny \color{#5} z=2,k=#1};
	
	\ifthenelse{\equal{#1}{1}}{
		\draw[enemy] (#4\labelReward) -- (#4\labelOne\labelFalse);
	}{
		\ifthenelse{\equal{#1}{2}}{
			\draw[enemy] (#4\labelReward) -- (#4\labelTwo\labelFalse);
	}{
		\ifthenelse{\equal{#1}{i}}{
			\draw[enemy] (#4\labelReward) -- (#4\labelI\labelFalse);
	}{
		\ifthenelse{\equal{#1}{n-1}}{
			\draw[enemy] (#4\labelReward) -- (#4\labelNminusone\labelFalse);
	}{
		\ifthenelse{\equal{#1}{n}}{
			\draw[enemy] (#4\labelReward) -- (#4\labelN\labelFalse);
	}{
		\draw[enemy] (#4\labelReward) -- (#4\labelOne\labelFalse);
	}
	}
	}
	}
	}
	
	\begin{pgfonlayer}{background}
	  \node[fit=(#4\labelReward)(#4\labelN\labelFalse)(#4\labelClause\labelN), hfit=#5] {};
	\end{pgfonlayer}
	
	\draw[color=black!40] (#4\labelOne) -- (#4\labelClause\labelOne);
	\draw[color=black!40] (#4\labelTwo\labelFalse) -- (#4\labelClause\labelOne);
	\draw[color=black!40] (#4\labelI) -- (#4\labelClause\labelOne);

	\draw[color=black!40] (#4\labelTwo\labelFalse) -- (#4\labelClause\labelTwo);
	\draw[color=black!40] (#2+4.2,#3) -- (#4\labelClause\labelTwo);
	\draw[color=black!40] (#4\labelI\labelFalse) -- (#4\labelClause\labelTwo);

	\draw[color=black!40] (#4\labelI\labelFalse) -- (#4\labelClause\labelNminusone);
	\draw[color=black!40] (#2+8.2,#3) -- (#4\labelClause\labelNminusone);
	\draw[color=black!40] (#4\labelN) -- (#4\labelClause\labelNminusone);
	
	\draw[color=black!40] (#2+8.2,#3) -- (#4\labelClause\labelN);
	\draw[color=black!40] (#4\labelNminusone) -- (#4\labelClause\labelN);
	\draw[color=black!40] (#4\labelN\labelFalse) -- (#4\labelClause\labelN);
	
}

\newcommand{\myCNFLHLbis}[5]{
	\myVerticesVariablesHL{z}{k}{#2}{#3}{#4}{#5}
	\myClauseGadgetsHL{z}{#2}{#3+1.5}{#4\labelClause}{#5}
	
	\myVertexHL{\rewHL{z}{k}}{#2+14.5}{#3-2}{#4\labelReward}{#5}{\vertexNameBottomLeft}
	
	\node[align=right] at (#2+14,#3+2) {\tiny \color{#5} z=1,k=#1};
	
	\ifthenelse{\equal{#1}{1}}{
		\draw[enemy] (#4\labelReward) -- (#4\labelOne\labelFalse);
	}{
		\ifthenelse{\equal{#1}{2}}{
			\draw[enemy] (#4\labelReward) -- (#4\labelTwo\labelFalse);
	}{
		\ifthenelse{\equal{#1}{i}}{
			\draw[enemy] (#4\labelReward) -- (#4\labelI\labelFalse);
	}{
		\ifthenelse{\equal{#1}{n-1}}{
			\draw[enemy] (#4\labelReward) -- (#4\labelNminusone\labelFalse);
	}{
		\ifthenelse{\equal{#1}{n}}{
			\draw[enemy] (#4\labelReward) -- (#4\labelN\labelFalse);
	}{
		\draw[enemy] (#4\labelReward) -- (#4\labelOne\labelFalse);
	}
	}
	}
	}
	}
	
	\begin{pgfonlayer}{background}
	  \node[fit=(#4\labelReward)(#4\labelOne)(#4\labelClause\labelOne), hfit=#5] {};
	\end{pgfonlayer}

	\draw[color=black!40] (#4\labelOne\labelFalse) -- (#4\labelClause\labelOne);
	\draw[color=black!40] (#4\labelTwo) -- (#4\labelClause\labelOne);
	\draw[color=black!40] (#2+4.2,#3) -- (#4\labelClause\labelOne);

	\draw[color=black!40] (#4\labelTwo\labelFalse) -- (#4\labelClause\labelTwo);
	\draw[color=black!40] (#2+4.2,#3) -- (#4\labelClause\labelTwo);
	\draw[color=black!40] (#4\labelI) -- (#4\labelClause\labelTwo);

	\draw[color=black!40] (#4\labelI\labelFalse) -- (#4\labelClause\labelNminusone);
	\draw[color=black!40] (#4\labelNminusone) -- (#4\labelClause\labelNminusone);
	\draw[color=black!40] (#4\labelN) -- (#4\labelClause\labelNminusone);
	
	\draw[color=black!40] (#2+8.2,#3) -- (#4\labelClause\labelN);
	\draw[color=black!40] (#4\labelNminusone\labelFalse) -- (#4\labelClause\labelN);
	\draw[color=black!40] (#4\labelN) -- (#4\labelClause\labelN);

}

\newcommand{\myCNFLHL}[5]{

	\myVerticesVariablesHL{1}{#1}{#2}{#3}{#4}{#5}
	\myClauseGadgetsHL{1}{#2}{#3+1.5}{#4\labelClause}{#5}
	
	\myVertexHL{\rewHL{1}{#1}}{#2+14.5}{#3-2}{#4\labelReward}{#5}{\vertexNameBottomLeft}
	
	\ifthenelse{\equal{#1}{1}}{
		\draw[enemy] (#4\labelReward) -- (#4\labelOne\labelFalse);
	}{
		\ifthenelse{\equal{#1}{2}}{
			\draw[enemy] (#4\labelReward) -- (#4\labelTwo\labelFalse);
	}{
		\ifthenelse{\equal{#1}{i}}{
			\draw[enemy] (#4\labelReward) -- (#4\labelI\labelFalse);
	}{
		\ifthenelse{\equal{#1}{n-1}}{
			\draw[enemy] (#4\labelReward) -- (#4\labelNminusone\labelFalse);
	}{
		\ifthenelse{\equal{#1}{n}}{
			\draw[enemy] (#4\labelReward) -- (#4\labelN\labelFalse);
	}{
		\draw[enemy] (#4\labelReward) -- (#4\labelOne\labelFalse);
	}
	}
	}
	}
	}
	
	\begin{pgfonlayer}{background}
	  \node[fit=(#4\labelReward)(#4\labelOne)(#4\labelClause\labelOne), hfit=#5] {};
	\end{pgfonlayer}
	
}

\begin{document}

\title[mode=title]{Parameterized Complexity of Hedonic Games with Enemy-Oriented Preferences}

\shorttitle{Parameterized Complexity of Hedonic Games with Enemy-Oriented Preferences}

\author[l1]{Martin Durand}

\ead{mdurand@ac.tuwien.ac.at}

\author[l1]{Laurin Erlacher}

\ead{laurin.erlacher@tuwien.ac.at}

\author[l1,l2]{ Johanne M\"uller Vistisen}
\ead{s191508@dtu.dk}

\author[l1]{Sofia Simola}

\ead{sofia.simola@tuwien.ac.at}

\affiliation[l1]{organization={TU Wien},            city={Vienna},
            country={Austria}}
\affiliation[l2]{organization={Technical University of Denmark},            city={Kongens Lyngby},
            country={Denmark}}
\shortauthors{Durand et al}

\begin{abstract}
Hedonic games model settings in which a set of agents have to be partitioned into groups which we call coalitions. In the enemy aversion model, each agent has friends and enemies, and an agent prefers to be in a coalition with as few enemies as possible and, subject to that, as many friends as possible. 
A partition should be stable, i.e., no subset of agents prefer to be together rather than being in their assigned coalition under the partition. We look at two stability concepts: core stability and strict core stability. This yields several algorithmic problems: determining whether a (strictly) core stable partition exists, finding such a partition, and checking whether a given partition is (strictly) core stable. Several of these problems have been shown to be NP-complete, or even beyond NP. 
This motivates the study of parameterized complexity. We conduct a thorough computational study using several parameters: treewidth, number of friends, number of enemies, partition size, and coalition size.
We give polynomial algorithms for restricted graph classes as well as FPT algorithms with respect to the number of friends an agent may have and the treewidth of the graph representing the friendship or enemy relations. 
We show W[1]-hardness or para-NP-hardness with respect to the other parameters.

We conclude this paper with results in the setting in which agents can have neutral relations with each other, including hardness-results for very restricted cases.
\end{abstract}


\maketitle 

\section{Introduction}\label{sec:intro}
\label{sec:introduction}

\todo{TODONOTES NOT DISABLED}

In this paper, we study \myemph{hedonic games with enemy aversion}.
Hedonic games describe coalition formation, where \agents' preferences only depend on the \agents in the coalition. This setting was introduced by Dr{\`e}ze and Greenberg~\cite{DG80hedonic}. 
Dimitrov et al~\cite{dimitrov2006simple} introduce \myemph{friend appreciation} and \myemph{\enemyav}: Each \agent divides other \agents into \myemph{friends} and \myemph{enemies}.
Ota et al~\cite{ohta2017core} add \myemph{neutrals} to the model.
Under friend appreciation, an \agent prefers a coalition containing more friends over one containing fewer friends. Subject to that, the \agent prefers the coalition containing fewer enemies.
Enemy aversion is the opposite: An \agent prefers fewer enemies, and subject to that, more friends. 
In both settings, if neutrals are included, then \agents are indifferent to them.

Under most stability concepts, enemy aversion leads to an \agent being in a coalition containing no enemies -- if an \agent has any enemies in her coalition, she would rather be on her own.

An example of hedonic games with \enemyav\ would be a setting where 
politicians form political coalitions: The more members a group has, the larger its weight. However, ideological conflicts can stop members from working together.
Similarly, an organization might wish to organize working groups from its members.
Larger working groups have more weight and can achieve more.
However, there are some conflicts of interest between members of the organization, and these members cannot be in the same group.
An example that includes neutrals could be students forming study groups.
A student cannot form a group with someone who is not free at the same time as her. 
Such a timetable conflict corresponds to an enemy relation.
She is friends with some other students and would like to be in the same group as them.
Other than her friends, she does not mind whether there are other students in the group: She considers them neutral.

When selecting partitions, we desire that the partition is in some sense \myemph{stable}, i.e., a group of \agents will not deviate and form a coalition together.
In this paper, we focus on \myemph{core stability} and \myemph{strict core stability}.
A partition is core stable, if no group of agents can leave their assigned coalitions and form a new coalition \blockingCoalition where every one of them prefers \blockingCoalition over their assigned coalitions.
Similarly, it is strictly core stable, if no group of \agents can leave their assigned coalitions and form a new coalition \blockingCoalition where no \agent prefers their assigned coalition over \blockingCoalition and at least one \agent prefers \blockingCoalition over their assigned coalition.

In this paper, we represent the friendship and enemy relations with graphs. Each \agent corresponds to a vertex and there is an edge between two vertices in the friendship graph (resp.\ enemy graph) if their corresponding \agents are friends (resp.\ enemies). Coalitions can then be interpreted as subsets of vertices. This allows us to exhibit a connection between this problem and graph problems.

Under \enemyav, Dimitrov et al~\cite{dimitrov2006simple} show that a core stable partition always exists  when there are no neutral relations, 
although finding one is NP-hard as it corresponds to repeatedly finding maximum cliques.
Finding a strictly core stable partition is harder: Rey et al~\cite{rey2016toward} show it is hard for complexity class DP, which is beyond NP.
Sung and Dimitrov~\cite{sung2007core} show that verifying whether a partition is core stable is NP-complete under the \enemyav, and Aziz et al~\cite{aziz2013computing} remark the result also holds for verifying strict core stability.
This computational hardness motivates us to look into parameterized complexity.

\paragraph{Our Contribution.}
In this paper, we study the parameterized complexity of hedonic games under \enemyav\ with and without neutrals.

In the setting without neutrals, we study the following problems:
\begin{compactitem}
\item \CF: Find a core stable partition. (Recall one always exists.)
\item \CV: Verify that a given partition is core stable.
\item \SCE: Determine whether a strictly core stable partition exists.
\item \SCV: Verify that a given partition is strictly core stable.
\end{compactitem}
We focus on the following parameters:
\begin{compactitem}
\item \maxDegreeFriend (resp.\ \maxDegreeEnemy): The maximum degree of the friendship graph (resp.\ the enemy graph). This corresponds to the maximum number of friends (resp.\ enemies) an agent may have. 
The parameter \maxDegreeFriend\ yields FPT-algorithms for all problems except strict core existence, whereas all four problems remain NP-hard for constant \maxDegreeEnemy.
\item \treewidthFriend (resp.\ \treewidthEnemy): The treewidth of the friendship graph (resp.\ the enemy graph). 
\item \maxCoalitionSize and \maxNumberOfCoalitions:
Maximum coalition size and maximum number of coalitions. 
For verification problems, these parameters correspond to the size of the largest coalition and the number of coalitions  in the given partition, respectively. For finding and existence problems they are additional constraints on the solution. 
The parameter~\maxCoalitionSize has a connection to the size of the maximum clique of the friendship graph that yields  XP-algorithms for all problems except \SCE.
The parameter \maxNumberOfCoalitions on the other hand makes all problems hard even when $\maxNumberOfCoalitions \geq 3$.
\end{compactitem}
We also show that the problems are easy when either the friendship or the enemy graph is bipartite or an interval graph.

In the setting with neutrals, we study the same four problems, except that we focus on core existence (\CE) instead of core finding, since a core stable partition is no longer guaranteed to exist.
We focus on the following parameters:
\begin{compactitem}
\item \maxDegreeFE: The number of non-neutral relations an \agent can have.
\item \maxCoalitionSize and \maxNumberOfCoalitions: These are defined analogously to the case without neutrals.
\end{compactitem}
All the problems remain hard for these parameters and even reasonable combinations of them.
We also show for all of our problems except \CE\ that they remain NP-hard even when the union of the friendship and enemy graphs is bipartite.
All of our hardness-results hold even when the relations are symmetric.
\CE on bipartite graphs remains an open question.

Our results are summarized in \cref{tab:results}. 

\begin{table*}[ht]
\caption{Overview of parameterized results. Rows correspond to instance restrictions, columns correspond to problems. For results that are conditional on parameter values, the condition is noted in parentheses. Known results are marked with a (\knownressymb) and a reference can be found in their cell. The symbol (\bipart) indicates that the result additionally holds even when $\friendshipGraph \cup \enemyGraph$ is bipartite.}
\begin{NiceTabular}{|l|cc|cc|cc|cc|}
\hline
    \rowcolor{gray!15} & \multicolumn{2}{c|}{\CF} & \multicolumn{2}{c|}{\CV} & \multicolumn{2}{c|}{\SCE} & \multicolumn{2}{c|}{\SCV} \\ 
    \hline
    bipartite \friendshipGraph / \enemyGraph & 
    P &\tableRef{P}{\ref{prop:SCE_alg_interval_bipartite}}& 
    P &\tableRef{P}{\ref{prop:SCE_alg_interval_bipartite}}& 
    P &\tableRef{P}{\ref{prop:SCE_alg_interval_bipartite}}& 
    P &\tableRef{P}{\ref{prop:SCE_alg_interval_bipartite}}\\
    
    \rowcolor{gray!15} interval  \friendshipGraph / \enemyGraph & 
    P &\tableRef{P}{\ref{prop:SCE_alg_interval_bipartite}}& 
    P &\tableRef{P}{\ref{prop:SCE_alg_interval_bipartite}}& 
    P &\tableRef{P}{\ref{prop:SCE_alg_interval_bipartite}}& 
    P &\tableRef{P}{\ref{prop:SCE_alg_interval_bipartite}}\\
    \hline
    \treewidthFriend / \treewidthEnemy & 
    FPT &\tableRef{P}{\ref{prop:TW_complexity}}& 
    FPT &\tableRef{P}{\ref{prop:TW_complexity}}& 
    FPT &\tableRef{P}{\ref{prop:TW_complexity}}& 
    FPT &\tableRef{P}{\ref{prop:TW_complexity}}\\ 

    \rowcolor{gray!15} \maxDegreeFriend & 
    FPT &\tableRef{C}{\ref{cor:CF_CV_SCV_DegreeFriend}}& 
    FPT &\tableRef{C}{\ref{cor:CF_CV_SCV_DegreeFriend}}& 
    \Block[]{}{NPh ($\geq 4$) \\ P ($\leq 3$)} & \Block[]{}{\tableRef{P}{\ref{prop:SCE_DegreeFriend_4}} \\ \tableRef{P}{\ref{prop:SCE_DegreeFriend_3}}}& 
    FPT &\tableRef{C}{\ref{cor:CF_CV_SCV_DegreeFriend}}\\

    \maxDegreeEnemy & 
    \Block[]{}{NPh ($\geq 3$) \\P ($ \leq 2$) } & \Block[]{}{\tableRef{P}{\ref{prop:CF_DegreeEnemy_3}}\\ \tableRef{P}{\ref{prop:CF_DegreeEnemy_2}}}& 
    \Block[]{}{NPc ($\geq 8$) \\P ($ \leq 2$) } & \Block[]{}{\tableRef{T}{\ref{prop:CV_SCV_DegreeEnemy}}\\ \tableRef{P}{\ref{prop:CF_DegreeEnemy_2}}} & 
    \Block[]{}{NPh ($\geq 6$)\\P ($ \leq 2$) } & \Block[]{}{ \tableRef{P}{\ref{prop:CE_Number_Coalitions_3+}}\\ \tableRef{C}{\ref{cor:SCE_DegreeEnemy_2}}}& 
    \Block[]{}{NPc ($\geq 16$) \\P ($ \leq 2$) } & \Block[]{}{\tableRef{T}{\ref{prop:CV_SCV_DegreeEnemy}}\\ \tableRef{P}{\ref{prop:CF_DegreeEnemy_2}}}\\

    \rowcolor{gray!15} \maxCoalitionSize & 
    \Block[]{}{\Wxhardshort{1}\\XP} & \Block[]{}{\tableRef{P}{\ref{prop:CF_Coalition_Size}}\\ \tableRef{P}{\ref{prop:CF_Coalition_Size_XP}}} & 
    \Block[]{}{\Wxhardshort{1} (\knownressymb)\\XP (\knownressymb)} & \Block[]{}{\cite{hanaka2024core}\\ \cite{hanaka2024core}} & 
    \Block[]{}{NPh ($\geq 3$) \\ P ($ \leq 2$)} & \Block[]{}{\tableRef{P}{\ref{prop:SCE_DegreeFriend_4}}\\ \tableRef{P}{\ref{prop:SCE_Coalition_Size_2}}}& 
    \Block[]{}{\Wxhardshort{1} (\knownressymb)\\XP (\knownressymb)} & \Block[]{}{\cite{hanaka2024core}\\ \cite{hanaka2024core}}\\

    \maxNumberOfCoalitions & 
    \Block[]{}{NPh ($\geq 3$) \\P ($\leq 2$)} & \Block[]{}{\tableRef{P}{\ref{prop:CE_Number_Coalitions_3+}}\\ \tableRef{C}{\ref{prop:CE_Number_Coalitions_2}}}& 
    \Block[]{}{NPc ($\geq 3$)\\P ($ \leq 2$) } & \Block[]{}{\tableRef{P}{\ref{prop:CV_SCV_Number_Coalitions_3+}}\\\tableRef{C}{\ref{prop:CE_Number_Coalitions_2}}}& 
    \Block[]{}{NPh ($\geq 3$)\\P ($ \leq 2$) } & \Block[]{}{\tableRef{P}{\ref{prop:CE_Number_Coalitions_3+}}\\ \tableRef{C}{\ref{prop:CE_Number_Coalitions_2}}}& 
    \Block[]{}{NPc ($\geq 3$)\\P ($\leq 2$)} & \Block[]{}{\tableRef{P}{\ref{prop:CV_SCV_Number_Coalitions_3+}}\\\tableRef{C}{\ref{prop:CE_Number_Coalitions_2}}}\\

    \bottomrule
    
    \toprule
    
    \rowcolor{gray!15} & \multicolumn{2}{c|}{\neut{\CE}} & \multicolumn{2}{c|}{\neut{\CV}} & \multicolumn{2}{c}{\neut{\SCE}} & \multicolumn{2}{c|}{\neut{\SCV}} \\ 
    \hline
    bipartite $\friendshipGraph \cup \enemyGraph$ & 
    ? & & 
    NPc &\tableRef{T}{\ref{thm:cvneutrhard}}& 
    NPh &\tableRef{T}{\ref{thm:scenbipartite}}& 
    NPc &\tableRef{T}{\ref{thm:scvneutrhard}}\\ \hline
    \rowcolor{gray!15}   $\maxDegreeFE + \maxCoalitionSize$ & 
    NPh & \tableRef{T}{\ref{cen:nph}}& 
    NPc (\bipart) &\tableRef{T}{\ref{thm:cvneutrhard}}& 
    NPh &\tableRef{P}{\ref{prop:scenpartsize}}& 
    NPc (\bipart) &\tableRef{T}{\ref{thm:scvneutrhard}}\\ 
     $\maxDegreeFE + \maxNumberOfCoalitions$ & 
    NPh & \tableRef{T}{\ref{cen:nph}}& 
    NPc (\bipart)  &\tableRef{T}{\ref{thm:cvneutrhard}}& 
    NPh &\tableRef{P}{\ref{prop:scencoalsize}}& 
    NPc (\bipart) &\tableRef{T}{\ref{thm:scvneutrhard}}\\ \bottomrule
    
\end{NiceTabular}

\label{tab:results}
\end{table*}

\paragraph{Related Work.}

Dimitrov et al~\cite{dimitrov2006simple} show that in the setting without neutrals and under friend appreciation, a (strictly) core stable partition is guaranteed to exist as the strongly connected components of the friendship relations form a (strictly) core stable partition. Chen et al~\cite{CheCsaRoySim2023Verif} show that under friend appreciation and without neutrals, verifying whether a partition is (strictly) core stable is NP-complete.

Ota et al~\cite{ohta2017core} study the setting with neutrals. They show that under friend appreciation, a core stable partition can still be found via strongly connected components, but determining the existence of a strictly core stable partition is $\Sigma^P_2$-complete and verification is NP-complete.
Under enemy-aversion and with neutrals, existence is $\Sigma^P_2$-complete and verification NP-complete for both core stability and strict core stability.

Chen et al~\cite{CheCsaRoySim2023Verif} study the parameterized complexity of hedonic games with friend appreciation, both with and without neutrals. 
They focus on the complexity of verifying whether a partition is (strictly) core stable and on determining the existence of Nash and individually stable partitions.
Their parameters include friendship degree, feedback arc set number, and the size and number of initial coalitions.

Hedonic games with \enemyav\ can be seen as a special case of hedonic games with additively separable preferences (\ashg). Their parameterized complexity has attracted prior study: For example Hanaka and Lampis~\cite{hanaka2022tree} and Hanaka et al~\cite{hanaka2024core} study the parameterized complexity of \ashg\ with respect to various graph theoretical parameters, such as treewidth and pathwidth. 
Peters~\cite{peters2016graphical} presents a parameterized algorithm with respect to treewidth + degree for a class of hedonic games that includes \ashg.

Brandt et al~\cite{brandt2024stability} study \ashg s, including friends appreciation and \enemyav, under stability concepts that rely on single agent deviations. These are for example Nash and individual stability.
They show that under \enemyav\ an individually stable partition always exists and can be found in polynomial time, whereas determining the existence of a Nash stable partition is NP-complete, even when we wish to find coalitions of at most size four. Their reduction can be modified to show that the hardness holds even when the degree of the friendship graph is bounded by a constant~\cite{comsocsurvey}.

\paragraph{Paper Structure.}
In \cref{sec:preliminaries} we present some preliminaries.
In \cref{sec:algorithmic_results} we present our main algorithms for the model without neutrals.
In \cref{sec:param_compl} we show parameterized hardness-results for the model without neutrals.
Finally in \cref{sec:neutrals} we show hardness-results for the case where neutrals are allowed.
For clarity, the proofs of results marked with (\appsymb) can be found in the appendix.

\section{Preliminaries}
\label{sec:preliminaries}
Let $[n]=\{1,2,\dots,n\}$. We call the \myemph{\agent set} $\agentSet=[\nbAgents]$. A \myemph{partition} of the \agent set $\partition=\{\coalition_1, \dots, \coalition_{|\partition|}\}$ is a set of disjoint subsets of \agents such that $\bigcup_{i=1}^{|\partition|} \coalition_i = \agentSet$. We call each subset in \partition a \myemph{coalition}. We denote by $\coalitionOfi{i}$ the coalition that contains \agent $i$ in partition \partition. Each \agent $i$ partitions other agents in three sets: Her friends \friendsOfi{i}, her enemies \enemiesOfi{i}, and neutrals \neutralsOfi{i}. We consider that the relations are \myemph{symmetric}, i.e., $j \in \friendsOfi{i} \text{ if and only if } i \in \friendsOfi{j}$, the same holds for enemies and neutrals. We use two simple graphs to represent relations: the \myemph{friendship graph}~\friendshipGraph and the \myemph{enemy graph} \enemyGraph. The friendship graph is defined as follows: $\friendshipGraph=(V=\{v_1,\dots,v_{\nbAgents}\},\friendshipEdgeSet)$ where $\forall \{i,j\} \subseteq \agentSet, \{v_i,v_j\} \in \friendshipEdgeSet \text{ if and only if } i \in \friendsOfi{j}$. Symmetrically, the enemy graph is defined as follows: $\enemyGraph=(V=\{v_1,\dots,v_{\nbAgents}\},\enemyEdgeSet)$ where $\forall \{i,j\} \subseteq \agentSet, \{v_i,v_j\} \in \enemyEdgeSet$ if and only if  $i \in \enemiesOfi{j}$. As the relations are symmetric, the graphs are undirected. If the set of neutrals is empty for all \agents, then \enemyGraph is the complement of \friendshipGraph.

We define the preference relation, $\succ_i$, for every \agent $i$. We use a definition equivalent to the one of Woeginger~\cite{woeginger2013core}.
Given two subsets $S_1,S_2 \subseteq \agentSet$, with $i\in S_1\cap S_2$, \agent $i$ \myemph{prefers} $S_1$ over $S_2$, denoted $S_1\succ_i S_2$, if and only if: 
\begin{align*}
    |\enemiesOfi{i} \cap S_1| &< |\enemiesOfi{i} \cap S_2| 
   \qquad \text{ or}\\
    |\enemiesOfi{i}\cap S_1| = |\enemiesOfi{i} \cap S_2|
    &\text{ and }~|\friendsOfi{i}\cap S_1| > |\friendsOfi{i}\cap S_2|.    
\end{align*}
Given two subsets $S_1,S_2 \subseteq \agentSet$, with $i\in S_1\cap S_2$, \agent $i$ \myemph{weakly prefers} $S_1$ over $S_2$, denoted by $S_1 \succeq_i S_2$, if $i$ does not prefer $S_2$ over $S_1$, i.e., $\lnot (S_2 \succ_i S_1) \implies S_1 \succeq_i S_2$.
This implies that if $S_1 \succeq_i S_2$, then either $i$ has fewer enemies in~$S_1$ \emph{or} $i$ has the same number of enemies in both, and at least as many friends in $S_1$ as in $S_2$.

We consider two notions of blocking coalitions with their corresponding stability notions: \myemph{core stability} and \myemph{strict core stability}.
\begin{definition}[Blocking coalition]
    Given some \agent set \agentSet and partition \partition, a nonempty set of \agents $\coalition \subseteq \agentSet$ \emph{blocks} \partition if all \agents in \agentSet prefer \coalition over their coalition under \partition. We call \coalition a \emph{blocking coalition}.
\end{definition}
\begin{definition}[Weakly blocking coalition]
    A set of \agents \coalition \emph{weakly blocks} a partition \partition if all agents in \coalition weakly prefer \coalition over their coalition under \partition and at least one \agent $i$ in \coalition prefers \coalition over $\coalitionOfi{i}$. Then we say \coalition is a \emph{weakly blocking coalition}. 
\end{definition}

\begin{definition}[Core stability]
	A partition \partition of \agentSet is \textit{core stable} if there is no blocking coalition $\coalition \subseteq \agentSet$.
\end{definition}

\begin{definition}[Strict core stability]\label{def:SCE}
    A partition \partition of \agentSet is \textit{strictly core stable} if there is no weakly blocking coalition $\coalition \subseteq \agentSet$.
\end{definition}

We observe here that it is natural to assume the relations to be symmetric.
If one agent considers another an enemy, they can never be in the same coalition in a (strictly) core stable partition, as the one who considers the other an enemy would form a blocking coalition on her own.
Thus we can assume that the enemy relations are symmetric.
In the case without neutrals, this also implies that friendship relations are symmetric.
In the case with neutrals we only present hardness-results, so they also extend to the case where friendship relations are not necessarily symmetric. 

If all the agents in a coalition are mutual friends, we say that the coalition is a \myemph{\fcliq}. Observe that, when there are no neutrals, if a partition \partition is core stable or strictly core stable, then every coalition in it must be a \fcliq; otherwise an agent who has an enemy forms a strictly blocking coalition alone.

The stability concepts above yield several algorithmic problems: determining whether a stable partition exists, finding such a partition, and verifying that a given partition is indeed core stable. Here we give the definitions for the problems without neutrals: 

\decprob{\CE (\SCE)}{An enemy oriented hedonic game without neutrals $I=(\agentSet,\friendshipGraph,\enemyGraph)$.}{Is there a (strictly) core stable partition of \agentSet?}

\optprob{\CF}{An enemy oriented hedonic game without neutrals $I=(\agentSet,\friendshipGraph,\enemyGraph)$.}{Find a core stable partition of \agentSet, if one exists.}

\decprob{\CV (\SCV)}{An enemy oriented hedonic game without neutrals $I=(\agentSet,\friendshipGraph,\enemyGraph)$ and a partition \partition of \agentSet.}{Is there a coalition that (weakly) blocks \partition?}

 Even though the answer to \CE is always ``yes" as there always exists a core stable partition, this does not hold if we bound the number of coalitions in the partition we look for. We therefore introduce two problems related to the maximum number of coalitions parameter \maxNumberOfCoalitions, which we abbreviate with \CEBounded and \SCEBounded.

\decprob{\CEBounded (\SCEBounded)}{An enemy oriented hedonic game without neutrals $I=(\agentSet,\friendshipGraph,\enemyGraph)$, an integer $k$.}{Is there a (strictly) core stable partition \partition of \agentSet such that $|\partition| \leq k$?}

We also introduce the problems \CEBoundedCoal\ and \SCEBoundedCoal:
\decprob{\CEBoundedCoal  (\SCEBoundedCoal)}{An enemy oriented hedonic game without neutrals $I=(\agentSet,\friendshipGraph,\enemyGraph)$, an integer $k$.}{Is there a (strictly) core stable partition \partition of \agentSet such that $\max_{\coalition \in \partition}|\coalition| \leq k$?}
Observe that in the case without neutrals, the largest coalition in every (strictly) core stable partition is always of the same size as the largest clique of the friendship graph.
Thus the problem is mainly interesting in relation to the parameterized complexity with respect to $k$.
When neutrals are allowed, not every (strictly) core stable partition has the same maximum coalition size, and the decision problem becomes interesting in its own right.

In \cref{sec:algorithmic_results,sec:results_param}, we consider that there are no neutral relations. In \cref{sec:neutrals}, we consider that neutral relations are possible; given a problem \probname{Problem}, we use \neut{\probname{Problem}} to denote the problem \probname{Problem} in which we allow neutral relations. 

We assume that the reader is familiar with parameterized complexity, including treewidth. Otherwise, one can refer to the books of Downey and Fellows~\cite{downey2012parameterized} and Cygan et al~\cite{cygan2020parameterized}. Intuitively, treewidth is a parameter indicating how close to a tree a graph is. It is commonly used in hedonic games papers \cite{peters2016graphical,hanaka2022tree,hanaka2024core}.

Before moving to the main sections, we give a brief example.

\paragraph{Example.}
Let $\agentSet = [3]$, and let  $\friendshipGraph = (V = \{v_1, v_2, v_3\}, \friendshipEdgeSet = \{\{v_1, v_2\}, \{v_2, v_3\}\})$. In other words, the agent~$1$ is friends with $2$, $2$ friends with both $1$ and $3$, and $3$ is friends with $1$. The agents $1$ and $3$ are enemies.
We assume first that there are no neutrals. Thus the enemy graph is the complement of $\friendshipGraph$, i.e., $\enemyGraph = (V, \enemyEdgeSet = \{\{v_1, v_3\}\})$.

Since $1$ and $3$ are enemies, they cannot be in the same coalition in a (strictly) core stable partition.
Thus only one of them can be with the agent $2$.
Let $\partition \coloneqq\{\{1,2\}, \{3\}\}$. The agents $1$ and $2$ obtain one friend, and the agent $3$ obtains no friends. No agent obtains enemies.
The coalition $\coalB = \{2,3\}$ weakly blocks $\partition$, because $2$ does not prefer~$\partition(2)$ over~$\coalB$, and $3$ prefers~\coalB\ over~$\partition(3)$.
In fact, this instance admits no strictly core stable partition.
Partition \partition\ is however core stable: Because~$2$ does not prefer $\coalB$ over $\partition(2)$, the coalition~\coalB\ does not block \partition. 
In fact, the agent $1$ cannot be in a blocking coalition, because she cannot obtain more friends or fewer enemies.
The agent $2$ can only prefer a coalition containing both $1$ and $3$, but they are enemies.
The agent $3$ would prefer \coalB, but we showed it does not block.

Let us now allow for neutrals, and let $\friendshipGraph = (V = \{v_1, v_2, v_3\}, \friendshipEdgeSet = \{\{v_1, v_2\}\})$ and $\enemyGraph = (V, \enemyEdgeSet = \{\{v_1, v_3\}\})$. Since neither of the graphs has edge $\{v_2, v_3\}$, the agents $2$ and $3$ have a neutral relation.
Now $\{\{1,2\}, \{3\}\}$ is strictly core stable.

\section{Algorithms without neutrals}
\label{sec:algorithmic_results}
\appendixsection{sec:algorithmic_results}

\subsection{Generic algorithms}
We start by presenting three generic algorithms solving \CF, \CV, and \SCE without neutrals. These algorithms exhibit a connection between these problems and classical problems in graph theory that we define now.
A \myemph{clique} is a set of pairwise connected vertices, i.e., there is an edge between each pair of vertices in a clique. An \myemph{independent set} is a set of pairwise disconnected vertices.

\decprob{\kclique{$k$}}{A graph $G=(V,E)$, an integer $k$}{Is there a clique of size $k$ in $G$?}

\decprob{\kIS{$k$}}{A graph $G=(V,E)$, an integer $k$}{Is there an independent set of size $k$ in $G$?}

\decprob{Partition Into $k$ Cliques}{A graph $G=(V,E)$, an integer $k$}{Can $V$ be partitioned into $k$ disjoint cliques?}

\decprob{\kColouring{$k$}}{A graph $G=(V,E)$, an integer $k$}{Can $V$ be partitioned into $k$ disjoint independent sets?}

We only present the algorithms for the friendship graph but it is possible to adapt them for the enemy graph by replacing ``clique" by ``independent set".

\begin{algorithm}[h!]
    \SetAlgoVlined
    \SetKwInput{Input}{Input}
    \SetKwBlock{Block}{}{}
    \SetKw{KwReturn}{return}
    \SetKw{KwNo}{no}
    \SetKw{KwYes}{yes}
    
    \SetKw{KwElse}{else}
        \SetKw{KwReturn}{return}
    \SetKwFunction{FReset}{Reset}
    \SetKwFunction{FMain}{Main}
    \SetKwProg{Fn}{}{:}{}

    \Input{An \agent set $\agentSet$ and a friendship graph $\friendshipGraph$}
    $\partition \leftarrow \emptyset$
    
    \For{$k$ in $\{\nbAgents, \nbAgents-1,\dots,1\}$}{
        
        \While{$\friendshipGraph$ contains a clique of size $k$}{
            $\agentSetInMaxCoalition \leftarrow$ clique of size $k$ in $G$
            
            $\partition \leftarrow \partition \cup \agentSetInMaxCoalition; \friendshipGraph \leftarrow \friendshipGraph[ V(\friendshipGraph) \setminus \agentSetInMaxCoalition]$
        }
    }
    \KwReturn $\partition$
  \caption{General Algorithm for \CF}\label{alg:CF}
\end{algorithm}

\cref{alg:CF} returns a core stable partition \cite{woeginger2013core}.

\begin{algorithm}[h!]
    \SetAlgoVlined
    \SetKwInput{Input}{Input}
    \SetKwBlock{Block}{}{}
    \SetKw{KwReturn}{return}
    \SetKw{KwNo}{no}
    \SetKw{KwYes}{yes}
    
    \SetKw{KwElse}{else}
        \SetKw{KwReturn}{return}
    \SetKwFunction{FReset}{Reset}
    \SetKwFunction{FMain}{Main}
    \SetKwProg{Fn}{}{:}{}

    \Input{An \agent set $\agentSet$ and a friendship graph $\friendshipGraph$, a partition $\partition=\{\coalition_1, \dots, \coalition_{|\partition|}\}$ of \agentSet, we assume $|C_i| \geq |C_j|$ if $i<j$}
        \For{$k$ in $\{1,\dots,|\partition|\}$}{
        \If{$\{v_i \mid i \in \coalition_k\}$ is not a clique in \friendshipGraph}{ \KwReturn \KwYes  \label{alg:CVlcliquecheck}}
        $s \leftarrow |\coalition_k|$
        
        \If{$\friendshipGraph$ contains a clique of size $s+1$}{
            \KwReturn \KwYes\label{alg:CVcheck}
        }
        $\friendshipGraph \leftarrow G[V(\friendshipGraph) \setminus \coalition_k]$
    }
    \KwReturn \KwNo
  \caption{General Algorithm for \CV}\label{alg:CV}
\end{algorithm}

\begin{restatable}[\appsymb]{proposition}{propCValg}
    \cref{alg:CV} returns ``yes" if and only if there exists a coalition blocking the input partition \partition.
    \label{prop:CV_alg}
\end{restatable}

\appendixproofwithstatement{prop:CV_alg}{\propCValg*}{

(Only if) Let us assume that there exists a coalition \blockingCoalition which blocks partition \partition. 
If there is an \agent~$i \in \blockingCoalition$ that has enemies in \coalitionOfi{i}, then the set of vertices corresponding to \coalitionOfi{i} is not a clique in the friendship graph, and the algorithm returns ``yes'' on line~\ref{alg:CVlcliquecheck}.

Now let us assume that no agent in $\blockingCoalition$ has enemies under \partition.
Thus $\blockingCoalition$ must be a friendship clique, as otherwise it would not block \partition.
Since each \agent~$i$ in \blockingCoalition prefers \blockingCoalition over \coalitionOfi{i}, it means that $|\coalitionOfi{i}| < |\blockingCoalition|$. Let us call $\coalition_j$ the first coalition in \partition such that $\blockingCoalition \cap \coalition_j \neq \emptyset$.
In \cref{alg:CV}, when $k=j$ all \agents in \blockingCoalition are still in $\friendshipGraph$. \cref{alg:CV} then clearly returns ``yes" as a clique of size strictly greater than $\coalition_j$ exists, e.g.,~\blockingCoalition. 

    (If) Let us now assume that there exists no coalition blocking \partition. Let us assume towards a contradiction that \cref{alg:CV} returns ``yes". 
    If this is returned on line~\ref{alg:CVlcliquecheck}, then there is a coalition where all the \agents are not mutual friends. Then an \agent~$i$ who has an enemy in the coalition prefers $\{i\}$ over her coalition under \partition, a contradiction.
    For \cref{alg:CV} to return ``yes" on ~\cref{alg:CVcheck}, a friendship clique of size $s+1$ must exist in $G$ at a given iteration~$k$. Every agent in such a coalition would obtain $s$ friends and no enemies, whereas they obtain at most $s - 1$ friends in \partition and would therefore be blocking \partition, a contradiction.
}

We now move to the algorithm for \SCE.

\begin{algorithm}[h!]
    \SetAlgoVlined
    \SetKwInput{Input}{Input}
    \SetKwBlock{Block}{}{}
    \SetKw{KwReturn}{return}
    \SetKw{KwNo}{no}
    \SetKw{KwYes}{yes}
    
    \SetKw{KwElse}{else}
        \SetKw{KwReturn}{return}
    \SetKwFunction{FReset}{Reset}
    \SetKwFunction{FMain}{Main}
    \SetKwProg{Fn}{}{:}{}

    \Input{An \agent set $\agentSet$ and a friendship graph \friendshipGraph.}
            $V' \gets V(\friendshipGraph)$ \label{alg:SCEXstart}\\
    \For{$k$ in $\{\nbAgents, \nbAgents-1,\dots,1\}$}{
        $\agentSetInMaxCoalition_k \leftarrow \emptyset$
        
                \For{$v \in V$}{
            \If{$v$ is in a clique of size $k$\label{algline:SCE_AgentSetMaxCoal} in $\friendshipGraph[V']$}{
                $\agentSetInMaxCoalition_k \leftarrow \agentSetInMaxCoalition_k \cup \{v\}$ \label{alg:SCEcliquesize}
            }
        }
        $V' \gets V' \setminus X_k$\\
    } \label{alg:SCEXend}
        \For{$k$ in $\{\nbAgents, \nbAgents-1,\dots,1\}$\label{alg:SCEcheckstart}}{
                \If{$\friendshipGraph[\agentSetInMaxCoalition_k]$ \textbf{cannot} be partitioned in cliques of size $k$\label{algline:SCE_Packing}}{ \KwReturn \KwNo  }
            }
    \KwReturn \KwYes\label{alg:SCEcheckend}\
  \caption{General Algorithm for \SCE}\label{alg:SCE}
\end{algorithm}

\begin{lemma}
Let $\agentSetInMaxCoalition_1, \dots, \agentSetInMaxCoalition_n$ be the sets constructed on lines~\ref{alg:SCEXstart}--\ref{alg:SCEXend} of \cref{alg:SCE}.
    A partition $\partition$ is strictly core stable if and only if its every coalition is a \fcliq and for every $j \in [n]$, every \agent in~$\agentSetInMaxCoalition_j$ is in a coalition of size $j$ with other \agents of $\agentSetInMaxCoalition_j$.
        \label{lem:SCE_Partition}
\end{lemma}

\begin{proof}
    (Only if) Suppose, for the sake of contradiction, that there exists a strictly core stable partition $\partition$, a value $i \in [n]$, and an \agent $p \in \agentSetInMaxCoalition_i$ which is not in a coalition of size $i$. We consider, without loss of generality, that there exists no $j > i$ such that an \agent from $\agentSetInMaxCoalition_j$ is not in a coalition of size $j$.

    Assume first, towards a contradiction, that $p$ is in a coalition of size $j' > j$. This coalition must be a \fcliq, as otherwise~\partition\ would not be stable.
    However, on~\cref{alg:SCEcliquesize}, the \agent $p$ was not placed in $\agentSetInMaxCoalition_{j'}$. Thus there must be an \agent $q \in \coalitionOfi{p}$ whose vertex had already been removed from $V'$ on some iteration $j'' > j'$.
    By our assumption, agent $q$ must be in a coalition of size $j''$, and this coalition is a \fcliq. This contradicts that $q \in \coalitionOfi{p}$, since $|\coalitionOfi{p}| = j' \neq j''$.

    We now assume, towards a contradiction, that $p$ is in a coalition of size $j'<j$. Since $p$ is part of $\agentSetInMaxCoalition_j$, it means that she is part of a \fcliq $\blockingCoalition_j$ of size~$j$ with some other \agents of $\agentSetInMaxCoalition_j$.
    As no \agent of $\agentSetInMaxCoalition_j$ is in a coalition with any agent of $\agentSetInMaxCoalition_\ell$ for any $\ell>j$, all agents in $\blockingCoalition_j$ are in coalitions of size at most $j$ in $\partition$.
    This means that~$\blockingCoalition_j$ is weakly blocking as $p$ would prefer $\blockingCoalition_j$ over \coalitionOfi{p} and all other \agents in $\blockingCoalition_j$ would weakly prefer $\blockingCoalition_j$ over their coalition under~\partition.

    (If) Assume, for the sake of contradiction, that there is a partition~$\partition$ of the \agents such that for all~$j \in [n]$, each agent in $\agentSetInMaxCoalition_j$ is in a coalition of size~$j$ which is a \fcliq and a subset of $\agentSetInMaxCoalition_j$, and there is coalition~$\blockingCoalition$ that weakly blocks~\partition.
    Since no agent has enemies under \partition, the coalition~\blockingCoalition must be a \fcliq too.
    Clearly, if $\blockingCoalition \subseteq \agentSetInMaxCoalition_j$ for some~$j$, then $\blockingCoalition$ cannot block: By construction the largest clique induced by agents in $\agentSetInMaxCoalition_j$ is of size~$j$.
    Let $i'= \max\{j \mid \blockingCoalition \cap \agentSetInMaxCoalition_j \neq \emptyset\}$, i.e., $i'$ in the largest index of a set $\agentSetInMaxCoalition_j$ set such that \blockingCoalition contains an \agent from $\agentSetInMaxCoalition_j$.     We have shown that $|\blockingCoalition| \leq i'$. 
    In iteration~$i'$, every \agent in $\blockingCoalition$ is still in~$V'$ by our choice of~$i'$. Thus if $|\blockingCoalition| = i'$, then $\blockingCoalition \subseteq \agentSetInMaxCoalition_{i'}$ as~$\blockingCoalition$ forms a clique of size $i'$. This is a contradiction by earlier reasoning.
    If $|\blockingCoalition| < i'$, then the \agents in $\agentSetInMaxCoalition_{i'} \cap \blockingCoalition$ prefer their coalitions under \partition over \blockingCoalition, a contradiction.
       \end{proof}

\begin{proposition}
    \cref{alg:SCE} returns ``yes" if and only if the input instance admits a strictly core stable partition.
    \label{prop:correctness_alg_SCE}
\end{proposition}

\begin{proof}
Observe that \cref{alg:SCE} returns ``yes'' if and only if for every $i \in [n]$, the set $\agentSetInMaxCoalition_i$ can be partitioned into cliques of size~$i$. By \cref{lem:SCE_Partition}, a partition is strictly core stable if and only if for every $i \in [n]$, every agent in $\agentSetInMaxCoalition_i$ is in a coalition of size~$i$, and these coalitions are \fcliqs. This concludes the proof.
\end{proof}

The following observation follows directly from \cref{lem:SCE_Partition}:
\begin{obs}
\cref{alg:SCE} can be adapted to solve \SCV as follows:
Replace the lines~\ref{alg:SCEcheckstart}--\ref{alg:SCEcheckend} with checking that for every $i \in [n]$, every agent in $\agentSetInMaxCoalition_i$ is in a coalition of size $i$ and this coalition is a \fcliq. 
    \label{obs:alg_SCE_to_SCV}
\end{obs}

We now look at the complexity of \cref{alg:CF,alg:CV,alg:SCE}.

\begin{lemma}
\CF, \CV, and \SCV\ can be solved in time $n^{O(1)} \cliqf(n)$, where the function~$\cliqf(n)$ denotes the complexity of solving \kclique{$k$} or \kIS{$k$} for an arbitrary $k \in [n]$.

     \label{lem:alg_CF_complexity}
\end{lemma}

\begin{proof}
\cref{alg:CF} solves the \kclique{$k$} problem at most $2n$ times as, for each call, we either remove at least one vertex if the answer is yes, or decrease the value of $k$ by one if the answer is no, and all other operations run in polynomial time. \cref{alg:CV} solves the \kclique{$k$} problem at most $n$ times, and all other operations in polynomial time.
The modified algorithm for \SCV described in \cref{obs:alg_SCE_to_SCV} solves the \kclique{$k$} problem at most~$n^2$ times, and the modified part can be performed in polynomial time.

Since \friendshipGraph and \enemyGraph are complements, instead of solving \kclique{$k$} on \friendshipGraph, we can also solve \kIS{$k$} on \enemyGraph. 
\end{proof}

We now exhibit a connection between \SCE and \kColouring{$k$}.
We claim that when solving \cref{alg:SCE} with \enemyGraph, \cref{algline:SCE_Packing} corresponds to solving the \kColouring{$\frac{|\agentSetInMaxCoalition_k|}{k}$} problem on the subgraph induced by~$\agentSetInMaxCoalition_k$. 
Since the maximum size of an independent set in this subgraph is $k$, each color class contains at most $k$ vertices.
In a valid $\frac{|\agentSetInMaxCoalition_k|}{k}$-coloring we have at most $\frac{|\agentSetInMaxCoalition_k|}{k}$ color classes, therefore each of these color classes is of size precisely $k$. 

Note that this also holds for \friendshipGraph. We are looking for a partition into cliques of size $k$, because of the structure of the graph any partition into $\frac{|\agentSetInMaxCoalition_k|}{k}$ cliques would only contain cliques of size $k$. Therefore solving the \partitionIntokCliques{$\frac{|\agentSetInMaxCoalition_k|}{k}$} problem in the subgraph induced by $\agentSetInMaxCoalition_k$ allows us to compute \cref{algline:SCE_Packing}.

\begin{lemma}
\SCE can be solved in time $n^{O(1)}( \cliqparf(n) + \cliqf(n))$, where~$\cliqparf(n)$ is the complexity of solving \partitionIntokCliques{$\frac{|\agentSetInMaxCoalition_k|}{k}$} or \kColouring{$\frac{|\agentSetInMaxCoalition_k|}{k}$} for an arbitrary $k \in [n]$, and where $\cliqf(n)$ is the complexity of solving \kclique{$k$} or \kIS{$k$} for an arbitrary $k \in [n]$.

          \label{lem:alg_SCE_complexity}
\end{lemma}

\begin{proof}
\cref{alg:SCE} solves the \kclique{$k$} problem at most $n^2$ times and the \partitionIntokCliques{$\frac{|\agentSetInMaxCoalition_k|}{k}$} at most $n$ times. All other operations run in polynomial time.
      
Since \friendshipGraph and \enemyGraph are complements, instead of solving \kclique{$k$} and \partitionIntokCliques{$\frac{|\agentSetInMaxCoalition_k|}{k}$} on \friendshipGraph, we can also solve \kIS{$k$} and \kColouring{$\frac{|\agentSetInMaxCoalition_k|}{k}$} on \enemyGraph. 
\end{proof}

\cref{lem:alg_CF_complexity,lem:alg_SCE_complexity} imply that if the \kclique{$k$} and \partitionIntokCliques{$\frac{|\agentSetInMaxCoalition_k|}{k}$} (resp. \kIS{$k$} and \kColouring{$\frac{|\agentSetInMaxCoalition_k|}{k}$}) problems are tractable on a specific graph class, then \CF, \CV, \SCE, and \SCV are tractable as well if the friendship graph (resp. enemy graph) belongs to this class. 
\cref{lem:alg_SCE_complexity} also implies that \SCE is contained in the complexity class $\Delta^P_2$.

\subsection{Application of the algorithms}

We start with two restricted graph classes: interval graphs and bipartite graphs. An interval graph is built from intervals on the real line, e.g., time periods in a day. Each vertex corresponds to an interval and there is an edge between two vertices if the corresponding intervals overlap. Interval graphs can represent timetable conflicts and are therefore relevant in some realistic settings.
As all the problems mentioned above can be solved in polynomial time on interval and bipartite graphs, we obtain the following result.

\begin{proposition}
\CF, \SCE, \CV, and \SCV can all be solved in polynomial time if \friendshipGraph (resp. \enemyGraph) is bipartite or an interval graph.
\label{prop:SCE_alg_interval_bipartite}
\end{proposition}

\begin{proof}
On bipartite and interval graphs, the \maxclique and \maxIS problems are polynomial time solvable \cite{mohring1985algorithmic}. Thus by \cref{lem:alg_CF_complexity}, \CF, \CV, and \SCV can be solved in polynomial time. 

To show the result for \SCE,
it remains to show that \partitionIntokCliques{$\frac{|\agentSetInMaxCoalition_k|}{k}$} and \kColouring{$\frac{|\agentSetInMaxCoalition_k|}{k}$} can be solved in polynomial time.
If \friendshipGraph is bipartite, no vertex is in a clique of size strictly larger than two, each singleton is put into its own coalition. Thus checking whether there is a \partitionIntokCliques{$\frac{|\agentSetInMaxCoalition_2|}{2}$} amounts to checking that every vertex with degree at least one is matched with another vertex, i.e., it is equivalent to finding a perfect matching of vertices of degree one or more, which can be done in polynomial time~\cite{kleinberg2014algorithm}.
If \enemyGraph is bipartite, this amounts to determining if the maximum independent set is unique if it is of size strictly greater than $\nbAgents/2$, which can be done in polynomial time. Otherwise, we need to check whether the graph admits a perfect matching; in that case, there are two disjoint maximum independent sets of size $\nbAgents/2$.

If \friendshipGraph is an interval graph, then we can solve \partitionIntokCliques{$\frac{|\agentSetInMaxCoalition_k|}{k}$} by ordering the intervals of the \agents in the set $\agentSetInMaxCoalition_k$ by non-decreasing end point. Observe that the agent with the smallest end point in the interval representation is only part of a single maximum clique. We pick the clique corresponding to the largest overlap for the first interval, removing this clique and repeating the process. If \enemyGraph is an interval graph the \kColouring{$\frac{|\agentSetInMaxCoalition_k|}{k}$} problem can be solved in polynomial time~\cite{cormen2022introduction}. We also note that an interval graph can be identified in polynomial time~\cite{booth1976testing}.
\end{proof}

As we know that \kclique{$k$}, \kIS{$k$}, \partitionIntokCliques{$\frac{|\agentSetInMaxCoalition_k|}{k}$}, and \kColouring{$\frac{|\agentSetInMaxCoalition_k|}{k}$} are in FPT with respect to treewidth~\cite{courcelle1993monadic}, we get the following result.

\begin{restatable}[\appsymb]{proposition}{propComplexityTreewidth}
\CF, \CV, \SCE, and \SCV are in FPT with respect to \treewidthFriend (resp. \treewidthEnemy).
\label{prop:TW_complexity}
\end{restatable}

\appendixproofwithstatement{prop:TW_complexity}{\propComplexityTreewidth*}{
The \maxclique and \maxIS problems can be solved in FPT time with respect to the treewidth of the input graph $\treewidth(G)$ \cite{courcelle1993monadic}. Therefore \cref{alg:CF,alg:CV} and the adaptation of \cref{alg:SCE} to \SCV can be run in FPT-time with respect to treewidth by \cref{lem:alg_CF_complexity}.

We give more details about \cref{algline:SCE_Packing} of \cref{alg:SCE}. For friendship graph, we solve the \partitionIntokCliques{$\frac{|\agentSetInMaxCoalition_k|}{k}$} problem, which is in FPT with respect to treewidth \cite{courcelle1993monadic}, on the graph $\friendshipGraph[\agentSetInMaxCoalition_k]$ induced by the subset $\agentSetInMaxCoalition_k$. Since this is a subgraph of $\friendshipGraph$, its treewidth is smaller or equal to $\treewidthFriend$. 

For the enemy graph, we have to solve \kColouring{$\frac{|\agentSetInMaxCoalition_k|}{k}$}. 
We show that $|\agentSetInMaxCoalition_k|/k \leq \treewidthEnemy+1$. We know that the maximum size of an independent set in $\enemyGraph[\agentSetInMaxCoalition_k]$ is exactly $k$. Since $\enemyGraph[\agentSetInMaxCoalition_k]$ has treewidth at most $\treewidthEnemy$, and as any graph with treewidth $\treewidth$ can be colored with $\treewidth+1$ colors \cite{dallard2021treewidth}, we know that the maximum independent set is of size at least $|\agentSetInMaxCoalition_k|/(\treewidthEnemy+1)$. This implies $k \geq |\agentSetInMaxCoalition_k|/(\treewidthEnemy+1)$ and $\treewidthEnemy+1 \geq |\agentSetInMaxCoalition_k|/k$. As it is possible to determine if a graph of treewidth $\treewidth$ can be colored with $c$ colors in time $O^*(c^{\treewidth})$, we can solve the problem in FPT-time by \cref{lem:alg_SCE_complexity}.
}

\begin{proposition}\label{prop:SCE_DegreeFriend_3}
    \SCE\ can be solved in polynomial time when $\maxDegreeFriend\!\leq~3$. 
\end{proposition}
\begin{proof}
We start the execution of \cref{alg:SCE} with $k=4$ as no \fcliq\ of size $5$ or greater can exist. Detecting whether a given vertex is in a clique of size 4, i.e., $K_4$, can be done in polynomial time. As any $K_4$ is necessarily disconnected of the rest of the graph since the maximum degree is 3, this means that the \cref{algline:SCE_Packing} can be skipped since no~$K_4$ overlap. The next step is $k=3$. In that case the \cref{algline:SCE_Packing} can be computed in polynomial time thanks to the triangle packing algorithm of Rooij et al~\cite{Rooij2011PartitionIT} on graphs of maximum degree 3.
For $k=2$, any vertex that has degree one or more has to be matched, which is equivalent to looking for a perfect matching of the non-isolated vertices, which can be done in polynomial time \cite{kleinberg2014algorithm}. The final iteration with $k=1$ leaves all the isolated vertices in coalitions on their own.
\end{proof}

Since \kclique{$k$} is in FPT with respect to the maximum degree of the graph we get the following result by \cref{lem:alg_CF_complexity}.
\begin{corollary}
    \CF, \CV, and \SCV are in FPT with respect to \maxDegreeFriend.
    \label{cor:CF_CV_SCV_DegreeFriend}    
\end{corollary}

Since \kclique{k} is in XP with respect to the solution size, we can show the following result.

\begin{restatable}[\appsymb]{proposition}{propCFCoalitionSizeXP}
\CEBoundedCoal\ is in XP with respect to~\maxCoalitionSize.
        \label{prop:CF_Coalition_Size_XP}
\end{restatable}
\appendixproofwithstatement{prop:CF_Coalition_Size_XP}{\propCFCoalitionSizeXP*}{
We can check in XP time with respect to $|C|$ whether the friendship graph contains a clique of size $|C|+1$. If this is the case, then the answer to the problem is ``no", as the coalition corresponding to such a clique would be blocking a partition containing coalitions of size at most $|C|$. Otherwise, the answer is ``yes", and we can solve the problem in XP-time by \cref{alg:CF}.
}

If $\maxCoalitionSize \leq 2$, then \SCE can be solved by checking that there is no $K_3$ and finding a perfect matching of vertices with degree at least one in the friendship graph.

\begin{restatable}[\appsymb]{proposition}{propSCECoalitionSize}
    \SCEBoundedCoal\ can be solved in polynomial time if $\maxCoalitionSize \leq 2$.
    \label{prop:SCE_Coalition_Size_2}
\end{restatable}
\appendixproofwithstatement{prop:SCE_Coalition_Size_2}{\propSCECoalitionSize*}{
We can verify in time $n^3$ that there is no $K_3$.
    If the maximum coalition size is two, then \SCE amounts to finding a perfect matching of vertices with degree at least one in the friendship graph, which can be done in polynomial time.
}

As the \kIS{$k$} problem is polynomial time solvable on graphs of maximum degree 2, we get the following result.

\begin{restatable}[\appsymb]{proposition}{propCFCVSCVEnemyDegree}
    \CF, \CV, and \SCV\ can be solved in polynomial time if $\maxDegreeEnemy \leq 2$.
    \label{prop:CF_DegreeEnemy_2}
\end{restatable}

\appendixproofwithstatement{prop:CF_DegreeEnemy_2}{\propCFCVSCVEnemyDegree*}{
    Graphs of maximum degree 2 consist of chains, cycles, and singletons. A maximum independent set can be found in polynomial time in such a graph by: (1) Picking all singleton vertices; (2) for each chain, picking one of the extremities and then every other vertex on the chain; (3) picking every other vertex on each cycle. Removing this independent set does not increase the degree of the graph, we can then repeat the procedure until the graph is empty.
    Thus by \cref{lem:alg_CF_complexity} the problem is in P.
}

We show that when $\maxDegreeEnemy=2$, it is possible to characterize enemy graphs for which a strictly core stable partition exists.

\begin{restatable}[\appsymb]{proposition}{propSCEexistenceDegreeTwo}
    If $\maxDegreeEnemy=2$, there exists a strictly core stable partition if and only if the connected components of~\enemyGraph are either: (1) Only cycles of size 3; (2) only paths with an odd number of vertices; (3) only paths and cycles with an even number of vertices. 
    \label{prop:SCE_DegreeEnemy_2_characterization}
\end{restatable}

\begin{proof}[Proof Sketch]
A maximum independent set can be found by grouping the maximum independent sets from the connected components, which are only paths and cycles. 
A strictly core stable partition exists if and only if either the enemy graph can be partitioned into disjoint maximum independent sets (cases (1) and (3)) or if there is one unique maximum independent set and one other disjoint independent set (case (2)).
\end{proof}

\appendixproofwithstatement{prop:SCE_DegreeEnemy_2_characterization}{\propSCEexistenceDegreeTwo*}{

   A graph of maximum degree 2 consists of singletons, paths, and cycles. A maximum independent set of the graph can be obtained by taking the union of the maximum independent sets of each connected component. In this proof, we write path or cycle of size $k$ to denote the number of vertices in the path or cycle.

   We start by showing that any strictly core stable partition would contain at most three coalitions. Indeed let us consider a strictly core stable partition $\partition=\{\coalition_1, \coalition_2, \dots, \coalition_{|\partition|}\}$, with $|C_i| \geq C_j \text{ if } i>j$ and $|\partition| \geq 4$. As the maximum degree of the enemy graph is two, each \agent can have at most two enemies. We consider an agent $i$ in~$\coalition_4$, such that~$i$ is enemies with at most two other agents, and therefore it has no enemies in at least one coalition in $\{\coalition_1,\coalition_2,\coalition_3\}$. Therefore, the coalition~\blockingCoalition consisting of \agent $i$ and one coalition $\coalition_j$, with $j \in [3]$, in which $i$ has no enemy, weakly blocks \partition.

   Throughout this proof, we will consider only partitions consisting of at most three coalitions $\coalition_1,\coalition_2,\coalition_3$ with $|\coalition_1| \geq |\coalition_2| \geq |\coalition_3|$.\\
   
   Firstly, we show that if the graph contains an odd cycle $S$ of size~$5$ or more, then there can be no strictly core stable partition. We call $s$ the number of vertices in~$S$. Coalition $\coalition_1$ necessarily contains a set of agents whose corresponding vertices form a maximum independent set in $S$, otherwise the partition would be weakly blocked by a coalition containing the same \agents corresponding to vertices in the other connected components and a larger independent set in~$S$. The maximum independent set in $S$ necessarily contains $\lfloor s/2 \rfloor$ vertices of $S$. Removing this independent set from $S$ leaves only singletons and a path of two vertices. 
   
   Indeed, if a path of at least three vertices is left, then one vertex which is not at the extremity of the path can be added to the independent set. 
   If several paths of length two are left, then the independent set is not maximum. 
   Indeed let us consider $P_1$ and $P_2$, which are two paths of length $2$ left after the removal of the vertices of $\coalition_1$. As they are part of the same cycle $S$, they are linked by a path $P'=(x_1, \dots,x_p)$, where $p$ is the number of vertices in the path, $x_1$ in a neighbor of a vertex $v_1$ of $P_1$ and $x_p$ is a neighbor of a vertex $v_2$ in $P_2$. 
   The independent set contains $\lceil p/2 \rceil$ vertices of $P'$ and none from $P_1$ and $P_2$. Let us consider the path $P^*=(v_1,x_1,x_2, \dots, x_p, v_2)$. Clearly a maximum independent set on~$P^*$ contains $\lceil (p+2)/2 \rceil > \lceil p/2 \rceil$ vertices therefore, the previous independent set was not maximum. There are then only singletons and a single path of length $2$ left of $S$ after the removal of vertices of~$\coalition_1$.   
   
   We call the vertices on the two-vertex-path~$v$ and~$w$; note that the other neighbors of~$v$ and~$w$ are in the set of vertices corresponding to~$\coalition_1$. Coalition $\coalition_2$ also contains $\lfloor s/2 \rfloor$ \agents corresponding to other vertices from the cycle: The newly formed singletons and one vertex from the path; assume without loss of generality that it contains the agent corresponding to $w$. This leaves only one vertex $v$ of the cycle. 
   Coalition $\coalition_3$ is of size strictly smaller than the two other coalitions as it contains strictly less \agents corresponding to vertices from~$S$ and at most as many \agents corresponding to vertices from any other connected component (path or cycle). 
   We call $\blockingCoalition$ the coalition consisting of $(\coalition_2 \cup \{a_v\}) \setminus \{a_w\}$, where $a_v$ and $a_w$ are the agents corresponding to $v$ and $w$ respectively. Coalition \blockingCoalition weakly blocks~\partition as~$a_v$ prefers \blockingCoalition over $\coalition_3$, no agent in \blockingCoalition prefers $\coalition_2$ over \blockingCoalition and \blockingCoalition does not contain any pair of enemies since the other enemies of $a_v$ and~$a_w$ are in $\coalition_1$.\\

   If the graph contains an odd cycle of length $3$, then there exists a strictly core stable partition if and only if the graph only consists of cycles of length $3$. Indeed let us assume, towards a contradiction, that there exists a strictly core stable partition of an \agent set when the enemy graph contains a cycle of length $3$ and at least one path, of odd or even number of vertices, or one cycle of even size. The \agents corresponding to the vertices of the paths or even cycles are split in $\coalition_1$ and $\coalition_2$, both coalitions containing one \agent corresponding to a vertex in each cycle of length $3$. The \agent corresponding to the last vertex of the cycle of length $3$ is necessarily in $\coalition_3$ whose size is strictly smaller than $\coalition_1$ as it contains no \agent corresponding to vertices from the paths. Therefore the coalition obtained from $\coalition_1$ in which the \agent corresponding to the last vertex of the cycle of length $3$ replaces the \agent corresponding to the first vertex of the same cycle is weakly blocking. Let us now assume that the graph contains only cycles of size $3$, then a strictly core stable partition can be obtained by forming three coalitions each containing exactly one \agent corresponding to a vertex from each cycle. In such a partition, each vertex is in an independent set of maximum size, and therefore no \agent can prefer another coalition over her coalition under \partition.\\

   Finally, we show that a graph that does not contain any cycles of odd length contains a core stable partition if and only if it contains either only paths of odd length or only paths and cycles of even number of vertices. 
   
   We first prove that if the enemy graph contains both a path with an odd number of vertices and either a path or a cycle with an even number of vertices, then there exists no strictly stable partition. The \agents corresponding to the vertices of the paths with an odd number of vertices are split into two coalitions~$\coalition_1$ and~$\coalition_2$. Coalition $\coalition_1$ contains the \agents corresponding to the two extremities and every other vertex and $\coalition_2$ contains the rest. The \agents corresponding to the vertices of the even cycles and paths are split into the same two coalitions by taking every other vertex, the number of these \agents is the same in both coalitions. We clearly have $|\coalition_1|>|\coalition_2|$. 
   The coalition $\blockingCoalition$ consisting of the \agents corresponding to the vertices of the odd paths from $\coalition_1$ and the \agents corresponding to the vertices of the even paths and cycles of $\coalition_2$ form a weakly blocking coalition as all agents in $\blockingCoalition \cap \coalition_2$ prefer~\blockingCoalition over $\coalition_2$ and agents in $\blockingCoalition \cap \coalition_1$ weakly prefer~\blockingCoalition over $\coalition_1$. 
   
   We now show that an \agent set with an enemy graph containing only odd paths has a strictly core stable partition. We form $\coalition_1$ by taking the \agents corresponding to extremities and every other vertex of each path. Coalition $\coalition_2$ contains the rest of the \agents. Clearly the vertices corresponding to $\coalition_1$ form a maximum independent set of the graph, therefore no vertex in it can prefer another coalition over it. The maximum size of an independent set containing any vertex corresponding to an \agent of $\coalition_2$ is necessarily strictly smaller than the size of $\coalition_1$ as for each path of size $s$, the size of the independent set containing the extremities and every other vertex is the unique maximum independent set of size $\lceil s/2 \rceil$, while the other vertices are only part of an independent set of size $\lfloor s/2 \rfloor$. Therefore any \agent of $\coalition_1$ would prefer $\coalition_1$ to any coalition containing at least one \agent of $\coalition_2$.
   
   When the graph contains only paths and cycles of even length then a maximum independent set can be obtained by taking every other vertex in each component. An independent set of the exact same size can be obtained by taking the remaining vertices, each vertex is in a maximum size independent set. Therefore a partition~\partition consisting of two coalitions obtained that way cannot be weakly blocked as no \agent could prefer a coalition over her coalition under~\partition.
}

As such structures can be found in polynomial time with a breadth-first search algorithm, we get the following corollary.

\begin{corollary}
    \SCE can be solved in polynomial time if $\maxDegreeEnemy=2$.
    \label{cor:SCE_DegreeEnemy_2}
\end{corollary}

We observe  that if $\maxNumberOfCoalitions \leq 2$, then a (strictly) core stable partition exists only if the enemy graph is bipartite. Thus by \cref{prop:SCE_alg_interval_bipartite}, we obtain the following results. 

\begin{restatable}[\appsymb]{corollary}{propCENumberCoalition}
    \CEBounded can be solved in polynomial time when $\maxNumberOfCoalitions \leq 2$. This also holds for \SCEBounded, \CV, and \SCV.
    \label{prop:CE_Number_Coalitions_2}
\end{restatable}

\appendixproofwithstatement{prop:CE_Number_Coalitions_2}{\propCENumberCoalition*}{
Note that a (strictly) core stable partition of size $2$ would necessarily correspond to a pair of disjoint independent sets of the enemy graph. Therefore, we can check the enemy graph for bipartiteness, if it is not bipartite, then we return ``no". Otherwise, we use \cref{alg:CF} (or \cref{alg:SCE} or \cref{alg:CV}) as described in \cref{prop:SCE_alg_interval_bipartite}.
}

We move on to the parameterized complexity study of the algorithmic problems.

\section{Hardness-results without neutrals}\label{sec:param_compl} 
\appendixsection{sec:param_compl}
\label{sec:results_param}

We look at the maximum degree of the enemy graph~\maxDegreeEnemy. We show a dichotomy as \CF is polynomial time solvable when~\maxDegreeEnemy is at most~2 and NP-hard when \maxDegreeEnemy is at least~3.
\begin{proposition}
     \CF is NP-hard when $\maxDegreeEnemy=3$.
    \label{prop:CF_DegreeEnemy_3}
\end{proposition}

\begin{proof}
A core stable partition necessarily contains a maximum size \fcliq, which corresponds to a maximum size independent set in the enemy graph \enemyGraph. Therefore a straightforward reduction from \kIS{$k$} on cubic graphs~\cite{garey1976simplified} to \CF with an enemy graph isomorphic to the input graph of \kIS{$k$} shows that \CF is NP-hard, even if the maximum degree of the enemy graph is 3.
\end{proof}

We now look at the maximum coalition size parameter~\maxCoalitionSize and show W[1]-hardness of \CEBoundedCoal, using an argument similar to \cref{prop:CF_DegreeEnemy_3}.

\begin{restatable}[\appsymb]{proposition}{propCFCoalitionSize}
    \CEBoundedCoal\  is \Wxhard{1} with respect to \maxCoalitionSize.
    \label{prop:CF_Coalition_Size}
\end{restatable}
\appendixproofwithstatement{prop:CF_Coalition_Size}{\propCFCoalitionSize*}{
We reduce from a variant of \kclique{$k$} called \probname{Multicolor $k$-Clique}. This problem is NP-complete, and \Wxhard{1} with respect to $k$~\cite{fellows2009parameterized}.
In this variant, the largest clique of the input graph may be of size at most $k$.

We use the same argument as in the proof of \cref{prop:CF_DegreeEnemy_3}. A core stable partition contains a coalition which corresponds to a maximum clique of the friendship graph.
Thus finding a core stable partition allows us to answer the question of \probname{Multicolor $k$-Clique}.
}

We show that \CEBounded is NP-hard even when both $\maxNumberOfCoalitions$ and \maxDegreeEnemy are constant.  
\begin{restatable}[\appsymb]{proposition}{CENumberCoalitionThree}
    \CEBounded and \SCEBounded are NP-hard even when $\maxNumberOfCoalitions=3$ and $\maxDegreeEnemy=6$. \SCE\ is NP-hard even when $\maxDegreeEnemy = 6$.
    \label{prop:CE_Number_Coalitions_3+}
\end{restatable}

\appendixproofwithstatementsketch{prop:CE_Number_Coalitions_3+}{\CENumberCoalitionThree*}{
We reduce from \probname{3-Coloring} on a graph of degree 4~\cite{garey1976simplified}.
Let $n'$ be the number of vertices in the original graph.
We interpret the original graph as an enemy graph and give each vertex two dummies who are mutual enemies and enemies with the original vertex.
These dummies are used to enforce that the original graph admits a 3-coloring if and only if the enemy graph of the reduced instance admits three pairwise disjoint independent sets of size $n'$.
These correspond to a (strictly) core stable partition.}{
    By reduction from \probname{3-Coloring} of a graph of degree 4 (which is NP-hard \cite{garey1976simplified}) to \SCE on enemy graphs of degree 6.
    \decprob{3-Coloring}{A graph $G=(V,E)$}{Is it possible to partition the graph into 3 color classes $C_1,C_2,C_3$ such that no pair of vertices belonging to the same color class are linked by an edge?}
    
    Given the original graph $G=(V,E)$ we create a reduced instance as follows. For each vertex $v_i$ of $V$, we create three \agents: one ``main \agent"~$i$ and two ``dummy \agents" $d_i$ and $d'_i$, in total $3|V|$ \agents. Each main \agent $i$ is enemies with the two dummy \agents $d_i$ and $d'_i$. The dummies $d_i$ and $d'_i$ are also enemies with each other. Two main \agents $i$ and $j$ are enemies if and only if $\{v_i,v_j\} \in E$. All other pairs of \agents are friends with each other. We observe that the enemy graph \enemyGraph has degree at most 6 since each dummy \agent is enemies with $2$ other \agents and main \agents are enemies with at most $6$ \agents, 4 corresponding to the edges of $G$ and $2$ with the dummy \agents. Note that the enemy graph of the reduced instance is similar to $G$ except that each vertex now forms a triangle with two additional vertices, corresponding to the dummy vertices. We first note that in the reduced instance the maximum size of a \fcliq is $|V|$ as there are $|V|$ disjoint triangles, any coalition of size strictly greater than $|V|$ would necessarily contain at least two \agents from a single triangle, and these \agents would then prefer to leave this coalition. We now observe that all \agents are part of a \fcliq of size $|V|$: Starting with any \agent, pick a dummy \agent from each triangle she is not part of; this gives a \fcliq of size $|V|$.     
    We claim that there is a core stable partition consisting of 3 coalitions if and only if $G$ admits a 3-coloring.
    
    (If) First, we prove that if $G$ is 3-colorable, then the reduced instance has a core stable partition with 3 coalitions.
    We extend the coloring $C_1,C_2,C_3$ to the dummy vertices such that the vertices in each triangle corresponding to \agents $i, d_i, d'_i$ are assigned different colors, and we call the new color classes $C'_1,C'_2,C'_3$. We now show that the partition $\partition = \{C'_1, C'_2, C'_3\}$ is a core stable partition. Observe that each coalition contains exactly a third of the \agents and that, due to the properties of a coloring, no adjacent vertices have the same color, which means no enemy \agents are together. This means that no \agent could join a larger coalition and no \agent is in a coalition with an enemy; thus the partition is core stable.

    (Only If) We now assume that there exists a core stable partition $\partition = \{C_1, C_2, C_3\}$ of the reduced instance with at most 3 coalitions. Additionally, since this partition is core stable no pair of enemies are part of the same coalition. We now build a coloring of $G$ such that vertex $v_i$ has color $j$ if $\coalitionOfi{i}=C_j$. This coloring uses at most 3 colors, as there are at most three coalitions and no neighbors have the same color as two enemies are not in the same coalition in \partition.

    We extend this reduction by noting that in the reduced instance any core stable partition \partition with at most 3 coalitions would also be strictly core stable as no agent is part of a \fcliq of size larger than $|V|$, meaning that no \agent can prefer a coalition over her coalition under \partition. This means that \SCE is NP-hard even if the enemy degree is $6$ and the partition size is $3$.
}

\begin{restatable}[\appsymb]{proposition}{SCEDegreeFriendFour}
    \SCE is NP-hard even if $\maxDegreeFriend \geq 4$.
    \label{prop:SCE_DegreeFriend_4}
\end{restatable}

\begin{proof}[Proof Sketch]
We reduce from \probname{Triangle Packing}, which is NP-complete even if the maximum degree of the input graph $G$ is~3 and each vertex has one of two specific neighborhoods~\cite{Rooij2011PartitionIT}. In both neighborhoods, the vertex is part of at least one triangle and no $K_4$. By building an instance with a friendship graph isomorphic to $G$, we obtain that there is a strictly core stable partition if and only if it is possible to partition $G$ into triangles. 
\end{proof}

\appendixproofwithstatement{prop:SCE_DegreeFriend_4}{\SCEDegreeFriendFour*}{

We reduce from the following problem:
\decprob{Triangle Packing}{A graph $G=(V,E)$}{Is there a partition $P=\{T_1, \dots, T_{|V|/3}\} $ of $V$ such that $\forall i \in [|V/3|], T_i$ is a 3-clique?}

\probname{Triangle Packing} is NP-complete \cite{Rooij2011PartitionIT}, even if each vertex of the input graph has one of two specific neighborhoods, described in \cref{fig:2b_3a}.

\begin{figure}
        \centering
        \begin{tikzpicture}[black, node distance={15mm}, thick, main/.style = {draw, circle}] 
        \node[main] (4) {$v$}; 
        \node[main] (1) [above left of=4] {}; 
        \node[main] (2) [above right of=4] {}; 
        \node[main] (3) [below left of=4] {}; 
        \node[main] (5) [below right of=4] {}; 
        \draw (1) -- (4);
        \draw (2) -- (4); 
        \draw (3) -- (4); 
        \draw (5) -- (4); 
        \draw (2) -- (5);
        \draw (3) to  (1);
        \end{tikzpicture} 
                \begin{tikzpicture}[black, node distance={15mm}, thick, main/.style = {draw, circle}] 
        \node[main] (4) {$v$}; 
        \node[main] (1) [above left of=4] {}; 
        \node[main] (2) [above right of=4] {}; 
        \node[main] (3) [below left of=4] {}; 
        \node[main] (5) [below right of=4] {}; 
        \draw (1) -- (4);
        \draw (2) -- (4); 
        \draw (3) -- (4); 
        \draw (5) -- (4); 
        \draw (2) to [out=130,in=140,looseness=1.95] (3);
        \draw (3) -- (5);
        \draw (3) to  (1);
        \end{tikzpicture} 
            \caption{The two possible neighborhoods.}
    \label{fig:2b_3a}
\end{figure}

Every vertex has one of two possible neighborhoods. It is visible from \cref{fig:2b_3a} that all vertices have degree 4, but also that no vertex is part of a 4- or 5-clique. It is also straightforward that every vertex is going part of at least two triangles. 

The reduced instance is very straightforward: For each vertex $v_i$ of $G$, we create an \agent $i$, two \agents $i$ and $j$ are friends if and only if $\{v_i,v_j\} \in E$. This construction means that the input graph $G$ and the friendship graph \friendshipGraph of the reduced instance are isomorphic.

We now want to show that the instance of \probname{Triangle Packing} problem is a yes-instance if and only if the reduced instance of \SCE is a yes-instance.

(Only If) assume that we have a yes-instance of the \probname{Triangle Packing} problem, meaning that there is a partitioning of the vertices into triangles $T_1,\dots T_{|V|/3}$. The partition $\partition=\{\coalition_1,\dots,\coalition_{|V/3|}\}$ consisting of a coalition for each triangle is strictly core stable as all \agents are in coalitions of size $3$ and no \agent can be in a coalition of size 4 or more since there are no 4 cliques in the friendship graph.

(If) Assume that the reduced instance of \SCE is a yes-instance. Thus there exists a partitioning into coalitions $\Pi=\{\coalition_1,\dots \coalition_{|\pi|}\}$ that is strictly core stable. Since no clique of size greater than 3 exists in the friendship graph, all coalitions must have size at most 3. Now assume for the sake of contradiction, that some coalition $\coalition \in \Pi$ is such that $|\coalition|<3$.
Since each vertex of the friendship graph is part of at least one 3-clique, the members of $\coalition$ prefer to be in a coalition of size 3. As no \agent can be in a coalition of size strictly larger than 3, no agent would strictly prefer their coalition under \partition over any coalition of size 3. Therefore there necessarily exists a coalition weakly blocking \partition. Since~\partition is strictly core stable, no coalition with less than 3 members exists, each coalition corresponds to a 3-clique in the friendship graph and all these coalitions are disjoint, giving us a partition into triangles of $G$.

}

We finally focus on verification problems.
We  show that \CV and \SCV are NP-hard even for constant \maxDegreeEnemy and \maxNumberOfCoalitions.

\begin{restatable}[\appsymb]{theorem}{propCVSCVEnemyDegree}
 \CV and \SCV are NP-hard even if the $\maxDegreeEnemy=8$ (resp.\ $\maxDegreeEnemy=16$) and $\maxNumberOfCoalitions=6$.
    \label{prop:CV_SCV_DegreeEnemy}
\end{restatable}

\appendixproofwithstatementsketch{prop:CV_SCV_DegreeEnemy}{\propCVSCVEnemyDegree*}{
We reduce from \threeSAT, where every literal appears at most twice~\cite{BKS-2bal3sat-2003}.
Through some clever gadgets, we can construct an instance with bounded $\maxDegreeEnemy$ that admits three \fcliqs\ of size $\hat{M}$ regardless of whether the SAT-formula is satisfiable.
We use these as the initial partition (and leave three agents alone).
This instance only admits an \fcliq\ of size $\hat{M} + 1$ if and only if the SAT-formula is satisfiable.
This \fcliq\ then blocks the initial partition.}{

We reduce from \threeSAT, where every literal appears at most twice. This problem is NP-complete~\cite{BKS-2bal3sat-2003}. 

Let $I = (\vars = \{x_1, \dots, x_n\}, \clas = \{C_1, \dots, C_m\})$ be an instance of \threeSAT.

We will construct an instance of \CV\ that admits three coalitions of size $3n + 7m$ regardless of the satisfiability of the instance, which will be used as an initial partition, together with coalitions for so-called consistency agents.
There is a coalition of size $3n + 7m + 1$ if and only if the formula is satisfiable. This will then block the initial partition.
The construction is also illustrated in \cref{fig:CV_SCV_DegreeEnemy}.
\begin{figure} 
\centering
    \begin{tikzpicture}[black, scale=0.5,node distance={15mm}, thick, vertex/.style = {draw, circle},enemy/.style={}]

            \myTrianglePosLiteralN{\ii_1}{-1}{0}{1}{\vertexNameTopLeft}{\vertexNameTopLeft}{\vertexNameTopLeft}
            \myTriangleNegLiteralN{\ii_1}{-3}{-2}{2}{\vertexNameTopLeft}{\vertexNameTopLeft}{\vertexNameTopLeft}

            \myTrianglePosLiteralN{\ii_2}{7}{0}{11}{\vertexNameTopLeft}{\vertexNameTopLeft}{\vertexNameTopLeft}
            \myTriangleNegLiteralN{\ii_2}{5}{-2}{12}{\vertexNameTopLeft}{\vertexNameTopLeft}{\vertexNameTopLeft}

            \myTrianglePosLiteralN{\ii_3}{14}{0}{21}{\vertexNameTopLeft}{\vertexNameTopLeft}{\vertexNameTopLeft}
            \myTriangleNegLiteralN{\ii_3}{12}{-2}{22}{\vertexNameTopLeft}{\vertexNameTopLeft}{\vertexNameTopLeft}

            \myTriangleClauseNew{1}{-3}{4}{101}{\vertexNameTopLeft}{\vertexNameTopLeft}{\vertexNameTopLeft}
            \myTriangleClauseNew{2}{5}{4}{102}{\vertexNameTopLeft}{\vertexNameTopLeft}{\vertexNameTopRight}

            \myTriangleClauseNew{3}{12}{4}{112}{\vertexNameTopLeft}{\vertexNameTopLeft}{\vertexNameTopLeft}

            \myTriangleClauseNew{4}{1}{7}{103}{\vertexNameTopLeft}{\vertexNameTopLeft}{\vertexNameTopLeft}
            \myTriangleClauseNew{5}{16}{7}{111}{\vertexNameTopLeft}{\vertexNameTopLeft}{\vertexNameTopRight}

            \myTriangleClauseNew{6}{20}{4}{113}  {\vertexNameTopLeft}{\vertexNameTopLeft}{\vertexNameTopLeft}       

            \myTrianglePickerS{\ii_1}{-1}{-4}{0}{\vertexNameTopLeft}{\vertexNameTopLeft}{\vertexNameBottomLeft}
            \myTrianglePickerS{\ii_2}{7}{-4}{01}{\vertexNameTopLeft}{\vertexNameTopLeft}{\vertexNameBottomLeft}
            \myTrianglePickerS{\ii_3}{14}{-4}{02}{\vertexNameTopLeft}{\vertexNameTopLeft}{\vertexNameBottomLeft}

            \myTriangleQickerS{\cia}{20}{0}{Q}{\vertexNameTopLeft}{\vertexNameTopLeft}{\vertexNameBottomLeft}

\begin{pgfonlayer}{background}

            \myTriangleEdges{1}{2}{\vertexBottom}{\vertexTop}{0}
            \myTriangleEdges{11}{12}{\vertexBottomLeft}{\vertexTopRight}{0}
            \myTriangleEdges{21}{22}{\vertexBottom}{\vertexTop}{0}

            \myTriangleEdges{101}{102}{\vertexRight}{\vertexLeft}{0}
            \myTriangleEdges{101}{103}{\vertexTopRight}{\vertexBottomLeft}{0}
            \myTriangleEdges{102}{103}{\vertexTopLeft}{\vertexBottomRight}{0}

            \myTriangleEdges{103}{111}{\vertexRight}{\vertexLeft}{0}
            \myTriangleEdges{111}{113}{\vertexBottomRight}{\vertexTopLeft}{0}
            \myTriangleEdges{112}{111}{\vertexTopRight}{\vertexBottomLeft}{0}
            \myTriangleEdges{112}{113}{\vertexRight}{\vertexLeft}{0}

            \myTriangleEdges{1}{101}{\vertexTop}{\vertexBottom}{0}
            \myTriangleEdges{12}{102}{\vertexTop}{\vertexBottom}{0}
            \myTriangleEdges{21}{112}{\vertexTop}{\vertexBottom}{0}

            \draw[black, enemy, pickeredge] (0) to [out=\vertexTop,in=\vertexRight,looseness=0] (1\labelSecond);
            \draw[black, enemy, pickeredge] (0) to [out=\vertexTop,in=\vertexRight,looseness=0] (2\labelSecond);
            \draw[black, enemy, pickeredge] (01) to [out=\vertexTopRight,in=\vertexBottom,looseness=0] (11\labelSecond);
            \draw[black, enemy, pickeredge] (01) to [out=\vertexTopRight,in=\vertexBottomRight,looseness=0] (12\labelSecond);
            \draw[black, enemy, pickeredge] (02) to [out=\vertexTopRight,in=\vertexBottomRight,looseness=0] (21\labelSecond);
            \draw[black, enemy, pickeredge] (02) to [out=\vertexTopRight,in=\vertexBottomRight,looseness=0] (22\labelSecond);

            \draw[\colorPrime, enemy, pickeredge] (0\labelPrime) to [out=\vertexLeft,in=\vertexLeft,looseness=0] (1);
            \draw[\colorPrime, enemy, pickeredge] (0\labelPrime) to [out=\vertexLeft,in=\vertexLeft,looseness=0] (2);
            \draw[\colorPrime, enemy, pickeredge] (01\labelPrime) to [out=\vertexTop,in=\vertexLeft,looseness=0] (11);
            \draw[\colorPrime, enemy, pickeredge] (01\labelPrime) to [out=\vertexTop,in=\vertexLeft,looseness=0] (12);
            \draw[\colorPrime, enemy, pickeredge] (02\labelPrime) to [out=\vertexTopRight,in=\vertexBottom,looseness=0] (21);
            \draw[\colorPrime, enemy, pickeredge] (02\labelPrime) to [out=\vertexTopRight,in=\vertexBottomLeft,looseness=0] (22);

            \draw[\colorSecond, enemy, pickeredge] (0\labelSecond) to [out=\vertexTopLeft,in=\vertexRight,looseness=0] (1\labelPrime);
            \draw[\colorSecond, enemy, pickeredge] (0\labelSecond) to [out=\vertexTopLeft,in=\vertexRight,looseness=0] (2\labelPrime);
            \draw[\colorSecond, enemy, pickeredge] (01\labelSecond) to [out=100,in=\vertexLeft,looseness=0] (11\labelPrime);
            \draw[\colorSecond, enemy, pickeredge] (01\labelSecond) to [out=\vertexTop,in=\vertexBottomLeft,looseness=0] (12\labelPrime);
            \draw[\colorSecond, enemy, pickeredge] (02\labelSecond) to [out=\vertexTopRight,in=\vertexBottomLeft,looseness=0] (21\labelPrime);
            \draw[\colorSecond, enemy, pickeredge] (02\labelSecond) to [out=\vertexTopRight,in=\vertexLeft,looseness=0] (22\labelPrime);

			\draw[enemy, pickeredge] (Q\labelSecond) to [out=\vertexBottomRight,in=\vertexBottom,looseness=0] (113);
            \draw[\colorPrime,enemy, pickeredge] (Q) to [out=330,in=\vertexBottom,looseness=0] (113\labelPrime);
            \draw[\colorSecond,enemy, pickeredge] (Q\labelPrime) to [out=\vertexBottomRight,in=\vertexBottom,looseness=0] (113\labelSecond);

\end{pgfonlayer}

\end{tikzpicture} 

    \caption{A figure for the proof of \cref{prop:CV_SCV_DegreeEnemy}: The enemy-relations of the gadget for the clause $C_{\cia} = x_{\ii_1} \vee \neg x_{\ii_2} \vee x_{\ii_3}$. }
    \label{fig:CV_SCV_DegreeEnemy}
\end{figure}

We construct the following agents for every $\ri \in \{0,1,2\}$:
\begin{compactitem}
\item For every $x_\ii \in \vars$, create the literal agents $\tvar \ii \ri, \fvar \ii \ri$, and the variable enforcer agent \pick \ii \ri. 
\item For every $C_\cia \in \clas$, create the clause agents $\clad \cia 1 \ri,$ $\clad \cia 2 \ri,$ $\clad \cia 3 \ri,$ $\clae \cia 1 \ri,$ $\clae \cia 2 \ri,$ $\claf \cia \ri$, and the clause enforcer agent \qick \cia \ri.
\item Finally we create the consistency agent \mainpick\ri.
\end{compactitem}

For ease of presentation, we name the following agent sets:
\begin{compactitem}
\item For every $x_\ii \in \vars$, let $\tvars \ii \coloneqq \{\tvar \ii \ri \mid \ri \in \{0,1,2\}\}$, $\fvars \ii \coloneqq \{\fvar \ii \ri \mid \ri \in \{0,1,2\}\}$, and $\picks \ii \coloneqq \{\pick \ii \ri \mid \ri \in \{0,1,2\}\}$.
\item For every $C_\cia \in \clas$, $\cib \in [3], \cib' \in [2]$, $\clads \cia \cib \coloneqq \{\clad \cia \cib \ri \mid \ri \in \{0,1,2\}\}$, $\claes \cia {\cib'} \coloneqq \{\clae \cia {\cib'} \ri \mid \ri \in \{0,1,2\}\}$, $\clafs \cia \coloneqq \{ \claf \cia \ri \mid \ri \in \{0,1,2\}\}$, and $\qicks \cia \coloneqq \{\qick \cia \ri \mid \ri \in \{0,1,2\}\}$.
\item Finally, $\mainpicks \coloneqq \{\mainpick \ri \mid \ri \in \{0,1,2\}\}$.
\end{compactitem}

We will construct the enemy relations so that if a coalition contains a variable enforcer agent $\pick 1 \ri$ for some $\ri \in \{0,1,2\}$, then the only agent it can contain from \picks{2} is \pick{2}{\ri}, and so on until~\pick{n}{\ri}.
Similarly, if a coalition contains clause enforcer agent $\qick 1 {\ria}$ for some $\ria \in \{0,1,2\}$, then the only agent it can contain from \qicks{2} is~\qick{2}{\ria}, and so on until~\pick{n}{\ria}.

We will also construct the enemy relations so that if a coalition contains \pick 1 \ri, \dots, \pick n \ri, then it can, for each $x_\ii \in \vars$, only contain either $\tvar \ii \ri, \fvar \ii {\ri + 1}$ or $\tvar \ii {\ri + 1},  \fvar \ii \ri$, where the former corresponds to putting the variable to true and latter to false. If there exists $\ri \in \{0,1,2\}$ such that a coalition $\coalB$ contains $\pick \ii \ri$ for all $\ii \in [n]$, then we say that the \myemph{variable enforcers are consistent} in~$\coalB$. 
If there exists $\ri \in \{0,1,2\}$ such that coalition $\coalB$ contains $\qick \cia \ri$ for all $\cia \in [m]$, then we say that the \myemph{clause enforcers are consistent} in $\coalB$. 
In this case \coalB\ cannot contain \claf \cia {\ri + 1} for any $C_\cia \in \clas$.

If there is a $\ri \in \{0,1,2\}$ such that a coalition \coalB\ contains  $\pick \ii \ri$ for all $\ii \in [n]$ and $\qick \cia \ri$ for all $\cia \in [m]$, then we say that the clause and variable enforcers are \myemph{consistent with each other}.
We will later use this to enforce that a \coalB\ with $|\coalB| = 3n + 7m + 1$ corresponds to a satisfying assignment.
Moreover, if clause and variable enforcers are consistent, but not consistent with each other, we can construct coalitions of size $3n + 7m$ regardless of whether the SAT-instance is satisfiable. We will use these coalitions to construct the initial partition.

In the description of the enemy relation, the text in italic gives an intuition of why the relations are built that way.
The agents are all friends with each other, except for the following enemy relations:
For every $\ri \in \{0,1,2\}$ (addition involving $\ri$ is modulo 3):
\begin{compactitem}
\item For every $\agentholder \in \{\tvars i, \fvars i, \picks i \mid i \in [n]\} \cup \{\clads \cia 1, \clads \cia 2, \clads \cia 3, \claes \cia 1, \claes \cia 2,$ $\clafs \cia, \qicks \cia \mid \cia \in [m]\} \cup \{\mainpicks\}$, the agents in \agentholder\ are mutual enemies. 
\intuition{Thus we can have only one agent from each of these sets in any coalition.}
\item For every $x_\ii \in \vars$, the agents \tvar \ii \ri\ and \fvar \ii \ri\ are enemies.
\item For every clause $C_\cia = \{{\ell}_{\cib_1}, {\ell}_{\cib_2}, \ell_{\cib_3}\} \in \clas$, we construct the following enemy relations:
\begin{compactitem}
\item For every $\clausecount \in [3]$,
if $\ell_{\cib_\clausecount}$ is a true (resp.\ false) literal,
then $\tvar {\cib_\clausecount} \ri$ (resp.\ $\fvar {k_\clausecount} \ri$) 
is enemies with \clad \cia \clausecount \ri.
\item The agents \cla \cia 1 \ri, \cla \cia 2 \ri, and \cla \cia 4 \ri\ are mutually enemies.
\item The agents \cla \cia 3 \ri, \cla \cia 5 \ri, and \cla \cia 6 \ri\ are mutually enemies.
\item The agent \cla \cia 4 \ri\ is enemies with \cla \cia 5 \ri.
\end{compactitem}
\intuition{These are used to enforce that the clauses are satisfied.}
\item For every $x_\ii \in \vars$, the agent \pick \ii \ri\ is enemies with \tvar \ii {\ri + 2} and \fvar \ii {\ri + 2}. Also, if $\ii \neq n$, then \pick \ii \ri\ is enemies with \pick {\ii + 1} {\ri + 1} and \pick {\ii + 1} {\ri + 2}.
\intuition{This way, the variable enforcers enforce that the variables can be set to either true or false. Moreover, the variable enforcers are consistent with each other.}
\item For every $C_\cia \in \clas$, the agent \qick \cia \ri\ is enemies with \claf \cia {\ri + 1}. Also, if $\cia \neq m$, then \qick \cia \ri\ is enemies with \qick {\cia + 1} {\ri + 1} and \qick {\cia + 1} {\ri + 2}.
\intuition{This way, the clause enforcers enforce, together with clause agents, that the clause is satisfied. Moreover, the clause enforcers are consistent with each other.}
\item The agent \mainpick \ri\ is enemies with \pick 1 {\ri + 1}, \pick 1 {\ri + 2}, \qick 1 {\ri + 1}, and \qick 1 {\ri + 2}.
\intuition{This way, if an agent from \mainpicks\ is in a coalition, then the variable and clause enforcers in this coalition must be consistent with each other.}
\end{compactitem}

Observe that the amount of enemies each agent may have is as follows, for each $\ri \in \{0,1,2\}$:
\begin{compactitem}
\item For every $x_\ii \in \vars$, the agent~\tvar \ii \ri\ (resp.\ \fvar \ii \ri) is enemies with the two other agents from~\tvars \ii\ (resp.\ \fvars \ii), the agent~$\fvar \ii \ri$ (resp.\ \tvar \ii \ri), one agent from~\picks \ii,  and at most two agents corresponding to clauses that contain $x_\ii$ (resp.\ $\bar{x}_\ii$), in total \textbf{six} agents.
\item For every $C_\cia \in \clas$ we have the following: For every $\cib \in [3]$, the agent~$\clad \cia \cib \ri$ is enemies with the two other agents from~\clads \cia \cib, two further clause agents, and one literal agent, in total \textbf{five} agents. For every $\cib' \in [2]$, the agent~$\clae \cia {\cib'} \ri$ is enemies with the two other agents from~\claes \cia {\cib'}, and three further clause agents, in total \textbf{five} agents. The agent~\claf \cia \ri\ is enemies with the two other agents from~\clafs \cia, the agents \cla \cia 3 \ri, \cla \cia 5 \ri, and one agent from~\qicks \cia, in total \textbf{five} agents.
\item For every $x_\ii \in \vars$, the agent \pick \ii \ri\ is enemies with the two other agents in \picks \ii, the agents \pick {\ii + 1} {\ri + 1} and \pick {\ii + 1} {\ri + 2} (unless $\ii = n$), the agents \pick {\ii - 1} {\ri + 1} and \pick {\ii - 1} {\ri + 2} (unless $\ii = 1$, in which case the enemies are \mainpick {\ri + 1} and \mainpick {\ri + 2}), and two literal agents corresponding to $x_\ii$, at most \textbf{eight} agents.
\item Similarly, for every $C_\cia \in \clas$, the agent \qick \ii \ri\ is enemies with at most four other enforcer agents, the two other agents from \qicks \ii, and one agent from \clafs \cia, in total \textbf{seven} agents.
\item The agent \mainpick \ri\ is enemies with two agents from \picks 1, two agents from \qicks 1, and the two other agents from \mainpicks, in total \textbf{six} agents.
\end{compactitem}
Thus $\maxDegreeEnemy = 8$.

For every $\ri \in \{0,1,2\}$, let us construct the following coalition~$\coalA_\ri$:
\begin{compactitem}
\item For every $\ii \in [n]$, add the agents $\tvar \ii \ri$, \fvar \ii {\ri + 2}, and \pick \ii {\ri + 2} to \coalA.
\item For every clause $C_\cia = \{\ell_{\cib_1}, \ell_{\cib_2}, \ell_{\cib_3}\} \in \clas$:
\begin{compactitem}
\item If $\ell_{\cib_1}$ is a true literal, add \cla \cia 1 {\ri + 2},  \cla \cia 4 \ri, \cla \cia 5 {\ri + 2}, and \cla \cia 6 \ri\  to \coalA.
\item Otherwise, add \cla \cia 1 \ri, \cla \cia 4 {\ri + 2}, \cla \cia 5 \ri,  and \cla \cia 6 {\ri + 2} to \coalA.
\item In both cases, add \cla \cia 2 {\ri + 1}, \cla \cia 3 {\ri + 1}, and \qick \cia {\ri} to \coalA.
\end{compactitem}
\end{compactitem}
Note that variable enforcers are consistent and clause enforcers are consistent, but they are not consistent with each other.
Observe that $|\coalA_\ri| = 3n + 7m$ and that all agents in $\coalA_\ri$ are friends with each other.
Recall that for every $\ria \in \{0,1,2\}, C_\cia \in \clas$, the clause agents in $\{\cla \cia 1 \ria, \cla \cia 2 \ria, \cla \cia 4 \ria\}$, $\{\cla \cia 4 \ria, \cla \cia 5 \ria\}$, $\{\cla \cia 3 \ria, \cla \cia 5 \ria, \cla \cia 6 \ria\}$ form enemy cliques, respectively. Observe that we only pick one agent from each of these cliques.
Also, we never have a literal and clause agent who are enemies: If $\ell_\ii$ is the first literal of a clause $C_\cia \in \clas$, then we choose the clause agents so that there is no conflict. Otherwise, we add either $\ell_\ii(\ri)$ or $\ell_\ii(\ri + 2)$ and $\clad \cia \cib {\ri + 1}$, where $\cib \in \{2,3\}$.
These agents are not enemies with each other.

The variable and clause agents are not consistent with each other, and thus these coalitions do not need to correspond to a satisfying assignment.

Let the initial partition $\partition \coloneqq \{\coalA_0, \coalA_1, \coalA_2, \{\mainpick 0\}, \{\mainpick{1}\}, \{\mainpick 2\}\}$.

We claim that there exists a coalition blocking \partition if and only if $I$ is satisfiable.\\
    (If) Let us suppose $I$ is satisfiable with an assignment $\asg \in \{\varF, \varT\}^n$.

    Then we construct a blocking coalition \coalB\ as follows:
\begin{compactitem}
\item For every $x_\ii \in \vars$, if $\asg[\ii] = \varT$, then add \tvar \ii 0 and \fvar \ii {1} in \coalB. Otherwise, add \tvar \ii {1} and \fvar \ii {0} in \coalB.
\item For every $C_{\cia} = \{{\ell}_{\cib_1}, {\ell}_{\cib_2}, \ell_{\cib_3}\}$, we proceed as follows:
\begin{compactitem}
\item if $\ell_{\cib_1}$ is true according to $\asg$, then $
\cla \cia 1 1,
\cla \cia 2 2,
\cla \cia 3 2,$ $
\cla \cia 4 0,
\cla \cia 5 1,
\cla \cia 6 0 \in \coalB$,
\item else if $\ell_{\cib_2}$ is true according to $\asg$, then
$
\cla \cia 1 2,
\cla \cia 2 1,
\cla \cia 3 2,$ $
\cla \cia 4 0,
\cla \cia 5 1,
\cla \cia 6 0 \in~\coalB$,
\item else if $\ell_{\cib_3}$ is true according to $\asg$, then
$
\cla \cia 1 0,
\cla \cia 2 2,
\cla \cia 3 1,$ $
\cla \cia 4 1,
\cla \cia 5 2,
\cla \cia 6 0 \in~\coalB.
$
\end{compactitem}
\item For every $x_\ii \in \vars$, add $\pick \ii 0$ in $\coalB$.
\item For every $C_\cia \in \clas$, add $\qick \cia 0$ in $\coalB$.
\item Add \mainpick 0 in \coalB.
\end{compactitem}

See that the variable and clause enforcers are consistent with each other.
Observe that $|\coalB| = 2n + 6m + n + m + 1 = 3n + 7m + 1 > |\coalA_\ri|$ for every $\ri \in \{0,1,2\}$. 
Moreover, all the agents in $\coalB$ are mutual friends:
\begin{compactitem}
\item For every $\agentholder \in \{\tvarN i, \fvarN i, \pickN i \mid i \in [n]\} \cup \{\cladN \cia 1, \cladN \cia 2, \cladN \cia 3, \claeN \cia 1, \claeN \cia 2,$ $\clafN \cia, \qickN \cia \mid \cia \in [m]\} \cup \{\mainpickN\}$, $\ri \in \{0,1,2\}$ such that $\agentholder(\ri) \in \coalB$, the agent $\agentholder(\ri)$ is enemies with $\agentholder(\ri + 1)$ and $\agentholder(\ri + 2)$ (addition is taken modulo 3). Observe that neither $\agentholder(\ri + 1)$ nor $\agentholder(\ri + 2)$ is in \coalB.
\item The agent \mainpick 0 is enemies with \mainpick 1, \mainpick 2, \pick 1 1, \pick 1 2, \qick 1 1, and \qick 1 2. None of them is in \coalB.
\item For every $x_\ii \in \vars$, the agent \pick \ii 0 is enemies with the remaining agents \tvar \ii 1, \fvar \ii 1, \pick {\ii - 1} 1 and \pick {\ii - 1} 2 (unless $\ii = 1$), and \pick {\ii + 1} 1 and \pick {\ii + 1} 2 (unless $\ii = n$). None of them is in \coalB. 
\item For every $x_\ii \in \vars$:
\begin{compactitem}
\item If $\asg[\ii] = \varT$, then
the agent \tvar \ii 0 is in \coalB\ and is enemies with \fvar \ii 0. However, the agent $\fvar \ii 0$ is not in $\coalB$.
Similarly, the agent \fvar \ii 1 is in \coalB\ and is enemies with \tvar \ii 1. However, the agent $\tvar \ii 1$ is not in~$\coalB$.
\item Otherwise the agent~\fvar \ii 0 is in \coalB\ and is enemies with~\tvar \ii 0. However, the agent~$\tvar \ii 0$ is not in $\coalB$.
Similarly, the agent~\tvar \ii 1 is in \coalB\ and is enemies with~\fvar \ii 1. However, the agent~$\fvar \ii 1$ is not in~$\coalB$.
\end{compactitem}
For $\ell_\ii \in \{x_\ii, \bar{x}_\ii\}$, for every $C_\cia \in \clas, \clausecount \in [3]$ such that $\ell_\ii$ is the~$\clausecount^{\text{th}}$ literal of $C_\cia$:
\begin{compactitem}
\item If $\ell_\ii$ is true according to \asg, then $\ell_\ii(0) \in \coalB$ and $\ell_\ii(0)$ is enemies with $\clad \cia \clausecount 0$. However, by construction $\clad \cia \clausecount 0 \notin \coalB$ if $\ell_\ii$ is true according to \asg.
\item Otherwise, $\ell_\ii(1) \in \coalB$ and $\ell_\ii(1)$ is enemies with $\clad \cia \clausecount 1$.  However, by construction $\clad \cia \clausecount 1 \notin \coalB$ if $\ell_\ii$ is false according to \asg.
\end{compactitem}
\item For every $C_\cia \in \clas$, the included clause agents are not enemies with each other:\begin{compactitem}
\item For every $\ri \in \{0,1,2\}$, the agents \cla {\cia} 1 \ri, \cla {\cia} 2 \ri, and \cla {\cia} 4 \ri\ are enemies with each other. Regardless of the case in the construction, we never have more than one of these agents in \coalB.
\item The same holds for the agents  \cla {\cia} 4 \ri, \cla {\cia} 5 \ri, and~\cla {\cia} 6 \ri.
\item For every $\ri \in \{0,1,2\}$, the agents \cla {\cia} 4 \ri\ and~\cla {\cia} 5 \ri\ are enemies with each other. Regardless of the case in the construction, we never have more than one of these agents in \coalB.
\end{compactitem}
\end{compactitem}
Thus every agent strictly prefers \coalB\ over her coalition under \partition, and therefore \coalB\ is a blocking coalition. 

 (Only if) Let us suppose there is a coalition \coalB\ that blocks \partition.

We make the following observation of the structure of \coalB:
\begin{obs}\label{obs:verifcoalreqs}
The coalition $\coalB$ must contain
\begin{compactitem}
\item one agent from $\mainpicks$,
\item for each $x_\ii \in \vars$, $\agentholder \in \{\tvars \ii, \fvars \ii, \picks \ii\}$, one agent from \agentholder, in total~$3n$ agents.
\item for every $C_\cia \in \clas$, $\agentholder \in \{\clads \cia 1, \clads \cia 2, \clads \cia 3, \claes \cia 1, \claes \cia 2, \clafs \cia,$ $\qicks \cia\}$, one agent from \agentholder, in total $7m$ agents.
\end{compactitem}
Overall \coalB\ contains $3n + 7m + 1$ agents.
\end{obs}
\begin{proof}\renewcommand{\qedsymbol}{$\diamond$}
Since \mainpick 0, \mainpick 1, and \mainpick 2\ are enemies with each other, a blocking coalition must contain an agent $x \in \coalA_0 \cup \coalA_1 \cup \coalA_2$. Since~$x$ has $3n + 7m - 1$ friends under \partition, a blocking coalition must contain $3n + 7m$ friends, thus containing at least $3n + 7m + 1$ agents overall.

Observe that in each of the above enumerated triples, all of the agents are mutual enemies. Thus we cannot obtain more than one agent from each of these sets.
Therefore the maximum number of agents \coalB\ may contain is $3n + 7m +1$. As this is also the minimum number of agents \coalB\ must contain, it follows that $|\coalB| = 3n + 7m + 1$. Moreover, it must contain one agent of each of the above enumerated sets.
\end{proof}

By \cref{obs:verifcoalreqs}, there is $\ri \in \{0,1,2\}$ such that $\mainpick \ri \in \coalB$.

We make the following further observation that states that variable and clause enforcers must be consistent with each other:
\begin{obs}\label{obs:verifcoalconst}
For every $x_\ii \in \vars$, $\pick \ii \ri \in \coalB$, and for every $C_\cia \in \clas$, $\qick \cia \ri \in \coalB$.
\end{obs}
\begin{proof}\renewcommand{\qedsymbol}{$\diamond$}
By \cref{obs:verifcoalreqs}, the coalition \coalB\ must contain at least one agent from $\picks 1$.
Observe that \mainpick \ri\ is enemies with \pick 1 {\ri + 1} and \pick 1 {\ri + 2} (addition modulo 3). Thus it must be that $\pick 1 \ri \in \coalB$. 
Similarly, the agent $\pick 1 \ri$ is enemies with \pick 2 {\ri + 1} and \pick 2 {\ri + 2} (addition modulo 3). Thus it must be that $\pick 2 \ri \in \coalB$.
By repeating this argument, we obtain that $\pick \ii \ri \in \coalB$ for every $x_\ii \in \vars$.
The argument for $\qick \cia \ri \in \coalB$ for every $C_\cia \in \clas$ is analogous.
\end{proof}

This leads us to the following property, which we will use to show that \coalB\ corresponds to a satisfying assignment:
\begin{obs}\label{obs:coalverifasg}
For every $x_\ii \in \vars$, either $\tvar \ii \ri, \fvar \ii {\ri + 1} \in \coalB$ or $\tvar \ii {\ri + 1}, \fvar \ii {\ri} \in \coalB$.
For every $C_\cia \in \clas$, $\claf \cia {\ri + 1} \notin \coalB$.
\end{obs}
\begin{proof}\renewcommand{\qedsymbol}{$\diamond$}
We start by showing the statement for every $x_\ii \in \vars$.
By \cref{obs:verifcoalconst}, we have that \pick \ii \ri\ is in \coalB.
Since \pick \ii \ri\ is enemies with \tvar \ii {\ri + 2} and \fvar \ii {\ri + 2}, neither of them can be in \coalB.
By \cref{obs:verifcoalreqs}, either $\tvar \ii \ri$ or \tvar \ii {\ri + 1} must be in \coalB.
Similarly, either $\fvar \ii \ri$ or \fvar \ii {\ri + 1} must be in \coalB.
If \tvar \ii \ri\ is in \coalB, then \fvar \ii \ri\ cannot be in \coalB, because \tvar \ii \ri\ and~\fvar \ii \ri\ are enemies. Thus \tvar \ii \ri\ and \fvar \ii {\ri + 1} are in \coalB.
Similarly, if \tvar \ii {\ri + 1} is in \coalB, then \fvar \ii \ri\ must also be in \coalB.

For every $C_\cia \in \clas$, the agent \qick \cia \ri\ is enemies with \claf \cia {\ri + 1}. This concludes the statement.
\end{proof}

Let us construct an assignment $\asg$ as follows: For every $x_\ii \in \vars$,
\begin{align*}
\asg[\ii] = \varT \text{ if and only if } \tvar \ii \ri \in \coalB.
\end{align*}

We show that \asg\ is a satisfying assignment.
Assume, towards a contradiction, that \asg\ is not a satisfying assignment.
Then there is some clause $C_\cia = \{\ell_{\cib_1}, \ell_{\cib_2}, \ell_{\cib_3}\}\in \clas$ such that every literal in $C_\cia$ is false under $\asg$. 
Then by definition of~\asg\ and \cref{obs:coalverifasg}, 
for every $\clausecount \in [3]$, $\ell_{\cib_\clausecount}(\ri + 1) \in \coalB$.
Moreover, by \cref{obs:verifcoalreqs}, the coalition~$\coalB$ must contain one agent from \clads \cia \clausecount. Since \clad \cia \clausecount {\ri + 1} is enemies with $\ell_{\cib_\clausecount}(\ri + 1)$, either $\clad \cia \clausecount \ri \in \coalB$ or $\clad \cia \clausecount {\ri + 2} \in \coalB$.

Since \clad \cia 1 \ri\ and \clad \cia 2 \ri\ are enemies, we can have at most one of them in \coalB. Similarly, \clad \cia 1 {\ri + 2} and \clad \cia 2 {\ri + 2} are enemies, and we can have at most one of them in \coalB.
Thus, either $\clad \cia 1 \ri, \clad \cia 2 {\ri + 2} \in \coalB$ or $\clad \cia 1 {\ri + 2}, \clad \cia 2 \ri \in \coalB$.
In either case $\clae \cia 1 \ri$ and \clae \cia 1 {\ri + 2} have an enemy in \coalB. Thus by \cref{obs:verifcoalreqs}, we have that $\clae \cia 1 {\ri + 1} \in \coalB$.

By \cref{obs:verifcoalreqs}, the coalition~$\coalB$ must contain one agent from~\claes \cia 2. Since \clae \cia 1 {\ri + 1} is enemies with $\clae \cia 2 {\ri + 1}$, either $\clae \cia 2 \ri \in \coalB$ or $\clae \cia 2 {\ri + 2} \in \coalB$.
We have already shown that either $\clad \cia 3 \ri \in \coalB$ or $\clad \cia 3 {\ri + 2} \in \coalB$.
Since \cla \cia 3 \ri\ and \cla \cia 5 \ri\ are enemies, we can have at most one of them in \coalB. 
Similarly, \cla \cia 3 {\ri + 2} and \cla \cia 5 {\ri + 2} are enemies, and we can have at most one of them in \coalB.
Thus, either $\cla \cia 3 \ri, \cla \cia 5 {\ri + 2} \in \coalB$ or $\cla \cia 3 {\ri + 2}, \cla \cia 5 \ri \in \coalB$.
In either case~$\cla \cia 6 \ri$ and~\cla \cia 6 {\ri + 2} have an enemy in \coalB.
Thus by \cref{obs:verifcoalreqs}, we have that $\cla \cia 6 {\ri + 1} \in \coalB$.
However, by \cref{obs:coalverifasg}, $\cla \cia 6  {\ri + 1} \notin \coalB$, a contradiction. 

This concludes the proof for \CV.

For \SCV, we construct the same instance but duplicate all the agents except the agents in \mainpicks.
The duplicates are friends with the original.
We construct their enemy relations as follows: For every pair of original agents $a, b$ which are enemies and their respective duplicates $a', b'$, we add the enemy relations $\{a, b'\}, \{a', b\}$, and $\{a',b'\}$. For every agent $c$  that is enemies with an agent $\hat{p}$ from \mainpicks, the copy of $c$ is enemies with $\hat{p}$ as well.
We construct the initial partition as previously, always including both the original agent and the possible duplicate.
Thus the coalitions $\coalA_\ri, \ri \in \{0,1,2\}$ are if size $2(3n + 7m)$.
This construction doubles the enemy-degree, and thus $\maxDegreeEnemy = 16$.

A weakly blocking coalition can be of size $2(3n + 7m)$ or $2(3n + 7m) + 1$.
Observe that if there is a weakly blocking coalition that does not contain both the original and the duplicate for some agent, then we can construct a larger weakly blocking coalition which contains both, since they both must be friends of all the agents in the blocking coalition.

Thus if an inclusionwise maximal weakly blocking coalition is of size $2(3n + 7m)$, then it cannot contain agents from~\mainpicks, as otherwise the cardinality of the coalition would be odd. Then none of these agents would prefer the coalition over their coalition under~\partition, a contradiction.
Therefore an inclusionwise maximal blocking coalition must be of size $2(3n + 7m) + 1$ and include an agent from~\mainpicks.
The rest of the proof proceeds as previously.

}

\begin{restatable}[\appsymb]{proposition}{propCVSCVNumberCoalitionThree}
    \CV and \SCV  are NP-hard even if $\maxNumberOfCoalitions=3$.
    \label{prop:CV_SCV_Number_Coalitions_3+}
\end{restatable}

\begin{proof}[Proof Sketch]
We reduce from \kIS{$k$} on cubic graphs. A 3-coloring of a cubic graph can be found in polynomial time Lov{\'a}sz~\cite{lovasz1975three}. We complete each of the color classes with dummies until each reaches a size of $k-1$. Enemy relations are built so that no friendship clique of size $k$ can contain dummies. A blocking coalition in the reduced instance corresponds to an independent set of size $k$ in the original graph.
\end{proof}

\appendixproofwithstatement{prop:CV_SCV_Number_Coalitions_3+}{\propCVSCVNumberCoalitionThree*}{
    We reduce from the \probname{Independent Set} problem, which is known to be NP-hard, even on graphs of maximum degree 3. We call $G=(V,E)$ the input graph of the \probname{Independent Set} problem. We also observe that the problem remains hard if the value of the solution size $k$ is strictly greater than $3|V|/4$. Indeed, it is possible to reduce from \probname{Independent Set} to \probname{Independent Set} and add singleton vertices until $k$ reaches the desired value.

    We perform the reduction from \probname{Independent Set} in two steps. Firstly, we get a 3-coloring of $G$ by running the algorithm of ~\cite{lovasz1975three}. This gives us three color classes of $G$. We then proceed to a ``balance" operation. Since the maximum degree of a vertex is 3, the maximum number of vertices connected to a set of size strictly less than $|V|/4$ vertices is strictly less than $3|V|/4$. If a color class has at least $3|V|/4$ vertices, then at least one of its vertices is not connected to any vertex of one of the other two color classes, as at least one of the two other color classes has strictly less than $|V|/4$ vertices. We therefore reduce the size (if needed) of the maximum sized color class until it is strictly less than $3|V|/4$ and obtain a new valid 3-coloring of the graph, we call the obtained color classes $C_1,C_2$ and~$C_3$.

    The second step consists of generating the reduced instance for \CV or \SCV. For each vertex $v_i$ of $G$, we create a ``main" \agent $i$, two main \agents $i$ and $j$ are enemies if and only if $\{v_i,v_j\} \in E$. For each color class $C_j$, we create $k-1-|C_j|$ dummy \agents $d_j^{1}$ to $d_j^{k-1-|C_j|}$. For all $\{j,j'\} \subset \{1,2,3\}$, each dummy \agent in $\{d_j^1, \dots, d_j^{k-1-|C_j|}\}$ is enemies with each dummy \agent in $\{d_{j'}^1, \dots d_{j'}^{k-1-|C_{j'}|}\}$ as well as every \agent $i$ such that $v_i \notin C_j$. We then create the starting partition \partition as follows: We create three coalitions $\coalition'_1$, $\coalition'_2$ and $\coalition'_3$. \Agent $i$ is in coalition $\coalition'_j$ if and only if the vertex $v_i$ is in the color class $C_j$. Additionally, for all $j \in \{1,2,3\}$ and for all $i \in [k-1-|C_j|]$, dummy \agent $d_j^i$ is in coalition $\coalition'_j$.

    We now show that there exists a coalition (weakly) blocking \partition if and only if there is an independent set of size $k$ in the graph $G$. We first note that all coalitions in \partition are of size $k-1$ and therefore all \agents are in coalitions of size $k-1$. Furthermore, any coalition of size strictly larger than $k-1$ that would not contain any pair of enemies can not contain any dummy \agent, as dummy \agents are friends with exactly $k-2$ other \agents.

    (Only if) Let us first assume that there exists a coalition $\blockingCoalition=\{b(1), \dots, b(|\blockingCoalition|)\}$ blocking \partition. As argued above, this coalition contains only main \agents. For this coalition to be (weakly) blocking it needs to be of size at least $k$, otherwise no \agent would prefer it over her coalition under \partition and to contain no pair of \agents that are enemies. Since two \agents~$i$ and~$j$ are enemies if and only if $(v_i,v_j) \in E$, this implies that the set $\{v_{b(1)}, \dots, v(b(|\blockingCoalition|)\}$ is an independent set, furthermore, since \blockingCoalition is at least of size $k$, this is an independent set of size at least $k$.

    (If) Let us now assume that there exists an independent set $S=\{v_{s(1)}, \dots, v_{s(k)}\}$ of size $k$, then by construction, the coalition $\blockingCoalition=\{s(1),\dots,s(k)\}$ (weakly) blocks \partition as all \agents in it prefer \blockingCoalition over their coalition under \partition.
}

\paragraph{Parameter combinations.} We conclude this section with a remark on parameter combinations. Firstly, one can observe that the number of coalitions and coalition size parameters combined yield a direct FPT algorithm as it is possible to upper bound the number of \agents we can put in coalitions by $\maxNumberOfCoalitions \cdot \maxCoalitionSize$, and therefore the number of possible partitions by $f(\maxNumberOfCoalitions \cdot \maxCoalitionSize)$. Now, observe that~\maxDegreeFriend upper bounds the coalition size and that \maxDegreeEnemy upper bounds the number of coalitions.
Indeed, if \maxDegreeEnemy is $k$ and there are more than $k+1$ coalitions, then one agent of the smallest coalition has no enemy in at least one other coalition and therefore prefers to join another coalition which is at least as large and would therefore be larger after she joins.
This observation implies that the combination of the parameters $\maxDegreeEnemy$ and~$\maxNumberOfCoalitions$ or $\maxDegreeFriend$ and $\maxCoalitionSize$ is redundant. On the other hand, combining \maxDegreeEnemy and \maxCoalitionSize or \maxDegreeFriend and \maxNumberOfCoalitions yield FPT algorithms as it is possible to upper bound the number of agents with a combination of these parameters.
As \treewidthFriend and \treewidthEnemy already yield FPT algorithms on their own, we do not investigate parameter combinations.

\section{Friends, Enemies, and Neutrals}
\label{sec:neutrals}
\appendixsection{sec:neutrals}

From this point, we consider that \agents can have friendship, enemy, and neutral relations with other agents. Two agents are neutrals if they do not consider each other as friends or enemies. \Agents still want to minimize the number of enemies in their coalition and to maximize the number of friends, they are indifferent to neutrals. 

We obtain that the studied problems are NP-hard even in very restricted cases. All of our results hold when the relations are symmetric.

We start by showing that the existence result for core stable partitions \cite{dimitrov2006simple} in the friends and enemies case does not hold in the presence of neutrals when the relations are symmetric.
When the relations are not necessarily symmetric, Ota et al~\cite{ohta2017core} have already shown that a core stable partition may not exist.

\begin{restatable}[\appsymb]{theorem}{thmfensymno}
\label{thm:fensymno}
An enemy oriented hedonic game with neutrals and symmetric relations $(\agentSet, \friendshipGraph, \enemyGraph)$ may not admit a core stable partition.
\end{restatable}

\begin{figure}
    \centering
    \begin{tikzpicture}[black, scale=0.4,vertex/.style = {draw, circle, inner sep = 0.5mm}]
        
    \fivefig;

\end{tikzpicture}
    \caption{An instance with symmetric friendship (blue solid edges), enemy (absence of edge), and neutral (black and dashed edges) relations, admitting no core stable partition.}
    \label{fig:no_instance_neutral}
\end{figure}

\appendixproofwithstatementcontinued{thm:fensymno}{\thmfensymno*}{
We illustrate the instance in \cref{fig:no_instance_neutral}.
Let $I$ be an instance with 20 agents, constructed as follows for every $i \in \{0,1,2,3,4\}$: the main pair $a^1_i, a^2_i$, and the connectors $b^1_i, b^2_i$.

Next, we construct the friendship relations. Recall that the relations are symmetric. In what follows, let addition be modulo 5. For every $i \in \{0,1,2,3,4\}$: \begin{compactitem}
\item $a^1_i$ and $a^2_i$ are friends; similarly $b^1_i$ and $b^2_i$ are friends,
\item both $a^1_i$ and $a^2_i$ are friends with both $a^1_{i + 1}$ and $a^2_{i + 1}$,
\item both $a^1_i$ and $a^2_i$ are friends with both $b^1_{i}$ and $b^2_{i}$, and
\item the agent $a^1_i$ is friends with $b^1_{i - 1}$, similarly the agent $a^2_i$ is friends with $b^2_{i - 1}$
\end{compactitem}

The enemy relations are constructed as follows, for every $i \in \{0,1,2,3,4\}$:
\begin{compactitem}
\item The agents $b^1_i$ and $b^2_i$ are enemies with everyone except each other,  $a^1_i$, $a^2_i$,  $a^1_{i + 1}$ and $a^2_{i + 1}$, and
\item additionally the agents $a^1_i$ and $a^2_i$ are both enemies with $a^1_{i + 2}$, $a^2_{i + 2}$, $a^1_{i - 2}$ and $a^2_{i - 2}$.
\end{compactitem}

For every $i \in \{0,\dots,4\}$, let $T_i \coloneqq \{a^1_i, a^2_i, b^1_i, b^2_i,a^1_{i+ 1}, a^2_{i + 1}\}$, where addition is taken modulo 5.
In other words, the set~$T_i$ contains two consecutive main pairs and the connectors between them.
Intuitively, for~$a^1_i$ and~$a^2_i$, their best coalition is~$T_i$. However, this coalition contains~$a^1_{i + 1}$ and~$a^2_{i + 1}$, who would rather be in~$T_{i + 1}$.
Since there are five main pairs, no matter how we divide them into consecutive pairs, there will be one main pair which is left alone.
This will block with another main pair that is not in its most preferred coalition.}{

Assume, towards a contradiction, that $I$ admits a core stable partition $\partition$. 

We start by showing that every coalition in \partition\ must be a subset of some $T_i$.
\begin{obs}\label{obs:ticont}
Every coalition in \partition must be a subset of $T_i$ for some $i \in \{0,1,2,3,4\}$. 
\end{obs}
\begin{proof}\renewcommand{\qedsymbol}{$\diamond$}
First observe that for every $z \in [2], i \in \{0,\dots,4\}$, the agent $b^z_i$ is enemies with all the agents except those in~$T_i$. Thus the statement holds for the coalition containing $b^z_i$.

Now assume towards a contradiction that \partition contains a coalition that has $a^z_i, a^w_j$ such that $z, w \in [2], i, j \in \{0,\dots,4\}$ and there is no~$T_{\ell}$ containing both of them. Then $ i \neq j + 1$ and $i + 1 \neq j$. By construction~$a^z_i$ and~$a^w_j$ are enemies, a contradiction to \partition being stable.
\end{proof}

Next, we show that for every two consecutive $T_{i-1}, T_{i}$, exactly one of them (possibly with minor modifications) must be in \partition.
\begin{claim}\label{clm:trianglecoals}
For every $i \in \{0,1,2,3,4\}$, exactly one of the following coalitions is in $\partition$:
\begin{compactenum}
\item $T_i$, $T_i \setminus \{b^1_i\}$, $T_i \setminus \{b^2_i\}$, $T_i \setminus \{a^1_{i + 1}\}$, $T_i \setminus \{a^2_{i + 1}\}$,\label{forwardcoal}
\item $T_{i - 1}$, $T_{i - 1} \setminus \{b^1_{i - 1}\}$, $T_{i - 1} \setminus \{b^2_{i - 1}\}$, $T_{i - 1} \setminus \{a^1_{i}\}$ or $T_{i - 1} \setminus \{a^2_{i}\}$.\label{backwardcoal}
\end{compactenum}
\end{claim}
\begin{proof}\renewcommand{\qedsymbol}{$\diamond$}
It is obvious that \partition cannot contain more than one coalition from Option~\eqref{forwardcoal} or more than one coalition from Option~\eqref{backwardcoal}.
If~\partition contains any of the sets in Option~\eqref{forwardcoal}, then $\Pi(a^1_i) = \Pi(a^2_i) \subseteq T_i$.
Since each of the coalitions in Option~\eqref{backwardcoal} contains $a^1_i$ or $a^2_i$ and none of them is a subset of $T_i$, none of them can be contained in a partition containing a coalition from Option~\eqref{forwardcoal}.
Thus at most one of the listed coalitions can be in \partition.

Assume, towards a contradiction, that none of the listed coalitions is in \partition.
Since $T_{i - 1}$ does not block $\partition$, one of the following must hold:
\begin{description}
\item[Case 1:] $a^1_{i - 1}$ or $a^2_{i - 1}$ weakly prefers her coalition under \partition over~$T_{i - 1}$,
\item[Case 2:] $b^1_{i}$ or $b^2_{i}$ weakly prefers her coalition under \partition over~$T_{i - 1}$, or
\item[Case 3:]$a^1_{i}$ or $a^2_{i}$ weakly prefers her coalition under \partition over~$T_{i - 1}$.
\end{description}
Let us show the Case 1 for~$a^1_{i - 1}$.
She must obtain five friends and no enemies under \partition, as she obtains five friends and no enemies in~$T_{i-1}$.
By \cref{obs:ticont}, the coalition $\partition(a^1_{i - 1})$  is a subset of either~$T_{i-2}$ or $T_{i-1}$. As~$a^1_{i - 1}$  has only four friends in~$T_{i-2}$, the coalition~$\partition(a^1_{i - 1})$  cannot be a subset of $T_{i-2}$.
However, since all the five other agents in $T_{i-1}$ are friends of $a^1_{i - 1}$, it must hold that $\partition(a^1_{i - 1}) = T_{i-1}$, a contradiction.
The proof is analogous for the agent~$a^2_{i - 1}$.

Let us show Case 2 for~$b^1_{i - 1}$.
She obtains all of her friends and no enemies under \partition, as she obtains all of her friends and no enemies in~$T_{i-1}$.
Thus $\{a^1_{i - 1}, a^2_{i - 1}, b^1_{i - 1}, b^2_{i - 1}, a^1_i\} \subseteq \partition(b^1_{i - 1})$. By \cref{obs:ticont}, it holds that $\partition(b^1_{i - 1}) \subseteq T_{i - 1}$. Thus $\partition(b^1_{i-1}) = T_i$ or $\partition(b^1_{i-1}) = T_i \setminus \{a^2_i\}$. However, this contradicts the assumption that neither of them is in~\partition. 
The proof is analogous for~$b^2_{i - 1}$.

Let us show Case 3 for~$a^1_{i}$.
She  must obtain four friends and no enemies under \partition, as she obtains four friends and no enemies in~$T_{i-1}$.
By \cref{obs:ticont}, the coalition~$\partition(a^1_{i})$ is a subset of either~$T_{i-1}$ or~$T_{i}$. 

First assume $\partition(a^1_{i})$ is a subset of $T_{i-1}$. Since $a^1_i$ has four friends in~$T_{i-1}$, she must obtain all of them in her coalition.
Thus $\{a^1_{i - 1}, a^2_{i - 1},$ $b^1_{i - 1},$ $a^1_i, a^2_i\} \subseteq \partition(a^1_{i})$.  Thus $\partition(a^1_{i}) = T_i$ or $\partition(a^1_{i}) = T_i \setminus \{b^2_i\}$. However, this contradicts the assumption that neither of them is in~\partition. 

Now assume $\partition(a^1_{i})$ is a subset of $T_{i}$.
Since $a^1_i$ must obtain at least four friends and she has five friends in $T_i$, at most one of them may not be included. These five friends are precisely all the other agents in $T_i$.
However, if the friend who is not included is $b^1_{i}, b^2_i, a^1_{i+1}$, or $a^2_{i+1}$, then we contradict the assumption that no coalition of Option~\eqref{forwardcoal} is in \partition. This assumption is also contradicted if $a^1_i$ obtains all of her friends. Thus $\partition(a^1_i) = T_i \setminus \{a^2_i\}$. However, then $a^2_i$ has at most four friends in her current coalition, and would obtain five in $T_i$. Also, every other agent in $T_i$ is a friend of $a^2_i$ so they would also strictly prefer $T_i$ over their coalition. Thus \partition is not stable, a contradiction.
The proof for~$a^2_{i}$ is analogous.

This concludes the proof of the claim.
\end{proof}

By \cref{clm:trianglecoals} exactly one of the following must hold:
\begin{description}
\item[Case 1:] $T_1$, $T_1 \setminus \{b^1_1\}$, $T_1 \setminus \{b^2_1\}$, $T_1 \setminus \{a^1_{2}\}$ or $T_1 \setminus \{a^2_{2}\}$ is in \partition.

In this case, we cannot have by \cref{clm:trianglecoals} that $T_2$, $T_2 \setminus \{b^1_2\}$, $T_2 \setminus \{b^2_2\}$, $T_2 \setminus \{a^1_{3}\}$ or $T_2 \setminus \{a^2_{3}\}$ is in \partition.
Again by \cref{clm:trianglecoals}, we have that $T_3$, $T_3 \setminus \{b^1_3\}$, $T_3 \setminus \{b^2_3\}$, $T_3 \setminus \{a^1_{4}\}$ or $T_3 \setminus \{a^2_{4}\}$ is in \partition.
Again we cannot have by \cref{clm:trianglecoals} that $T_4$, $T_4 \setminus \{b^1_4\}$, $T_4 \setminus \{b^2_4\}$, $T_4 \setminus \{a^1_{0}\}$ or $T_4 \setminus \{a^2_{0}\}$ is in \partition.
Again by \cref{clm:trianglecoals}, we have that $T_0$, $T_0 \setminus \{b^1_0\}$, $T_0 \setminus \{b^2_0\}$, $T_0 \setminus \{a^1_{1}\}$ or $T_0 \setminus \{a^2_{1}\}$ is in \partition.
Now by by \cref{clm:trianglecoals} $T_1$, $T_1 \setminus \{b^1_1\}$, $T_1 \setminus \{b^2_2\}$, $T_1 \setminus \{a^1_{2}\}$ and $T_1 \setminus \{a^2_{2}\}$ cannot be in \partition, a contradiction.
\item[Case 2:] $T_{0}$, $T_{0} \setminus \{b^1_{0}\}$, $T_{0} \setminus \{b^2_{0}\}$, $T_{0} \setminus \{a^1_{1}\}$ or $T_{0} \setminus \{a^2_{1}\}$ is in \partition.
This leads to a contradiction through arguments analogous to the previous case.
\end{description}
This concludes the proof.}

By using this construction, we reach the following hardness-result.

\begin{restatable}[\appsymb]{theorem}{cennph}
\neut{CE} is NP-hard even when the relations are symmetric, and the degree is bounded by a constant. The results hold even for \neut{\CEBounded}\ and \neut{\CEBoundedCoal} when \maxNumberOfCoalitions\ and \maxCoalitionSize\ are respectively constant.\label{cen:nph}\end{restatable}

\newcommand{\tvarn}[1]{\ensuremath{y_{#1}}}
\newcommand{\fvarn}[1]{\ensuremath{\bar{y}_{#1}}}
\newcommand{\lvarn}[1]{\ensuremath{\ell_{#1}}}
\newcommand{\nlvarn}[1]{\ensuremath{\bar{\ell}_{#1}}}
\newcommand{\Ix}[1]{\ensuremath{I^{#1}_{x}}}
\newcommand{\Ic}[1]{\ensuremath{I^{#1}_{C}}}
\newcommand{\rew}[2]{\ensuremath{r^{#2}_{#1}}}
\newcommand{\Rew}[1]{\ensuremath{R_{#1}}}
\newcommand{\Ixx}[2]{\ensuremath{\Ix{#1}(#2)}}
\newcommand{\Icx}[2]{\ensuremath{\Ic{#1}(#2)}}

\appendixproofwithstatementsketch{cen:nph}{\cennph*}{
We reduce from \threeSAT, where every literal appears exactly twice~\cite{BKS-2bal3sat-2003}. 
We construct a copy of the previous no-instance for each variable and each clause.
Additionally, for every variable, we create an agent for both the true and the false literal, and some other agents for technical purposes.
They all need either the true or false literal agents to ``save" them from blocking the whole instance.
However, we also create clause blockers, which require that at least one of the corresponding literal variables ``saves'' them.
A literal agent cannot save both the clause blocker and the variable blocker: If it saves the variable, it can no longer save the clause.
Thus a literal saving the variable corresponds to setting the corresponding literal to false.}{

We reduce from \threeSAT, where every literal appears at most twice. This problem is NP-complete~\cite{BKS-2bal3sat-2003}. 
Let $I = (\vars = \{x_1, \dots,$ $x_n\}, \clas = \{C_1, \dots, C_m\})$ be an instance of \threeSAT.

We construct an instance $I' = (\agentSet, \friendshipGraph, \enemyGraph)$ with symmetric relations.

We construct agents as follows:
\begin{compactitem}
\item For every variable $x_\ii \in \vars$, construct a true literal agent \tvarn \ii\ and a false literal agent \fvarn \ii. Create additional four rewards agents \rew \ii 1, \dots, \rew \ii 4. Moreover, construct a copy of the instance from \cref{thm:fensymno}, which we call the variable blocker \Ix \ii.
\item For every $C_\cia \in \clas$, construct a copy the instance from \cref{thm:fensymno}, \emph{except} construct the agents and their relations for every $i \in \{0,\dots,6\}$ instead of every $i \in \{0,\dots,4\}$. We call this the clause blocker \Ic \cia.
\end{compactitem}

The idea of the proof is that each of the variable blockers corresponds to the previous no-instance for \neut{\CE}. 
They all need either the true or false literal agents to ``save" them from blocking the whole instance.
However, we also create the clause blockers, which require that at least one of the corresponding literal variables ``saves'' them.
A literal agent cannot save both the clause blocker and the variable blocker: If it saves the variable, it can no longer save the clause.
Thus a literal saving the variable corresponds to setting the corresponding literal to false.

In a slight abuse of notation, for every $x_\ii \in \vars, C_\cia \in \clas$, let $\Ix \ii$ and $\Ic \cia$ also refer to the set of agents contained in them.
Moreover, for every $p \in \Ix \ii$, let $\Ixx \ii p$ denote $p$ when $\Ix \ii$ is not otherwise clear.
Similarly, for every $q \in \Ic \cia$, let $\Icx \cia q$ denote $q$ when $\Ic \cia$ is not otherwise clear.
Let $\Rew \ii \coloneqq \{\rew \ii 1, \dots, \rew \ii 4\}$.
Given a literal $\ell_i$, let $\lvarn{\ii} \coloneqq \tvarn \ii$ if $\ell_\ii = x_\ii$ and $\lvarn{\ii} \coloneqq \fvarn \ii$ otherwise.

Next, we describe the friendship and the enemy relations:
\begin{compactitem}
\item For every $x_\ii \in \vars, C_\cia \in \clas$, the sets \Ix \ii\ and \Ic \cia\ have the friendship and enemy relations within the gadget defined as in \cref{thm:fensymno}, except that for \Ic \cia\ the relations are constructed for every $i \in \{0, \dots, 6\}$.
\item
For every $x_\ii \in \vars$:
\begin{compactitem}
\item The agent \tvarn \ii\ is enemies with \fvarn \ii.

\item The agents $\Ixx \ii {b^1_0}$, $\Ixx \ii {b^2_0}$, $\Ixx \ii {a^1_1}$, and $\Ixx \ii {a^2_1}$ are friends with \tvarn \ii\ and \fvarn \ii. 

\item The agents in \Rew \ii\ are friends with each other and the agents \tvarn \ii\ and \fvarn \ii. 
\end{compactitem}
\item For every clause $C_\cia = \{\ell_{\cib_1}, \ell_{\cib_2}, \ell_{\cib_3}\}$, for every $k \in [3]$,
\begin{compactitem}
\item the agents in $\lvarn{\cib_k}$ are friends with the agents  $\Icx \cia {b^1_{2k}}$, $\Icx \cia {b^2_{2k}}$, $\Icx \cia {a^1_{2k + 1}}$, and~$\Icx \cia {a^2_{2k + 1}}$,
\item the agents  $\Icx \cia {b^1_{2k}}$, $\Icx \cia {b^2_{2k}}$, $\Icx \cia {a^1_{2k + 1}}$ and~$\Icx \cia {a^2_{2k + 1}}$ are all enemies with all of the following agents:
$\Ixx {\cib_k}{b^1_0}$, $\Ixx{\cib_k}{b^2_0}$, $\Ixx {\cib_k}{a^1_1}$, $\Ixx{\cib_k}{a^2_1}$, and \Rew {\cib_k}.
\end{compactitem}

\end{compactitem}
All the unmentioned relations are neutral relations.

Observe that the maximum total number of friends and enemies any agent may have is $26 + 1 + 8= 35$. This is for every $C_\cia \in \clas, k \in [3]$, where the agents $\Icx \cia {a^1_{2k - 1}}, \Icx \cia {a^2_{2k - 1}}, \Icx \cia {b^1_{2k}}, \Icx \cia {b^2_{2k}}$ each have 26 non-neutral relations to other agents in \Ic \cia, one variable agent friend, and eight enemies among \Rew \ii\ and \Ix \ii\ for some $x_\ii \in \vars$.

We show that $I$ admits a satisfying assignment if and only if $I'$ admits a core stable partition.

\begin{figure}
    \centering
    \begin{tikzpicture}[black, scale=0.4,vertex/.style = {draw, circle, inner sep=0.5mm}]
    \input{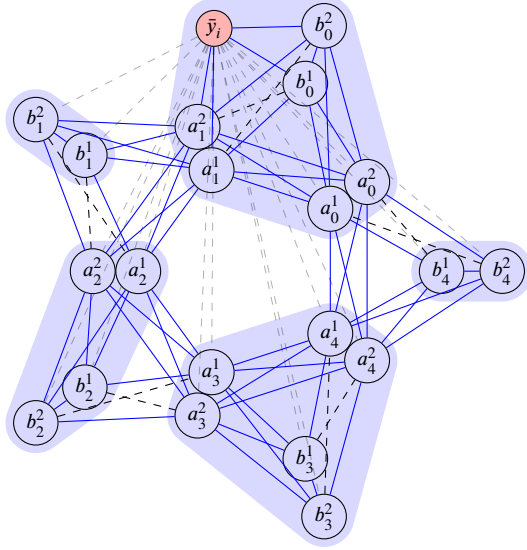}
            \node at (-1,8)  [vertex, fill=red!30!white] (fvar) {\footnotesize \fvarn \ii};

    \fivefig;

    \draw[friend] (fvar) -- (1);
    \draw[friend] (fvar) -- (1\labelTwo);
    \draw[friend] (fvar) -- (10);
    \draw[friend] (fvar) -- (10\labelTwo);

    \foreach \i in {0,2,3,4,11,12,13,14}{
        \draw[neutral, opacity=0.3] (fvar) -- (\i);
        \draw[neutral, opacity=0.3] (fvar) -- (\i\labelTwo);
    }
    \begin{pgfonlayer}{background}
            \draw[coal] \hedgeii{14}{14\labelTwo}{10mm};
    \draw[coal] \hedgeii{11}{11\labelTwo}{10mm};
    \draw[coal] \hedgem{2\labelTwo}{2}{12,12\labelTwo}{10mm};
    \draw[coal] \hedgem{3\labelTwo}{3}{4,4\labelTwo,13\labelTwo}{10mm};
    \draw[coal] \hedgem{0\labelTwo}{0}{1,1\labelTwo,fvar,10\labelTwo}{10mm};
    \end{pgfonlayer}

\end{tikzpicture}
    \caption{Partition from Step~\eqref{ceconstr:varblock} of the proof of \cref{cen:nph}. Blue solid edges denote friendship relations and dashed black edges neutral relations. The lack of an edge denotes an enemy relation.}
    \label{fig:ceconstrix}
\end{figure}

\begin{figure*}
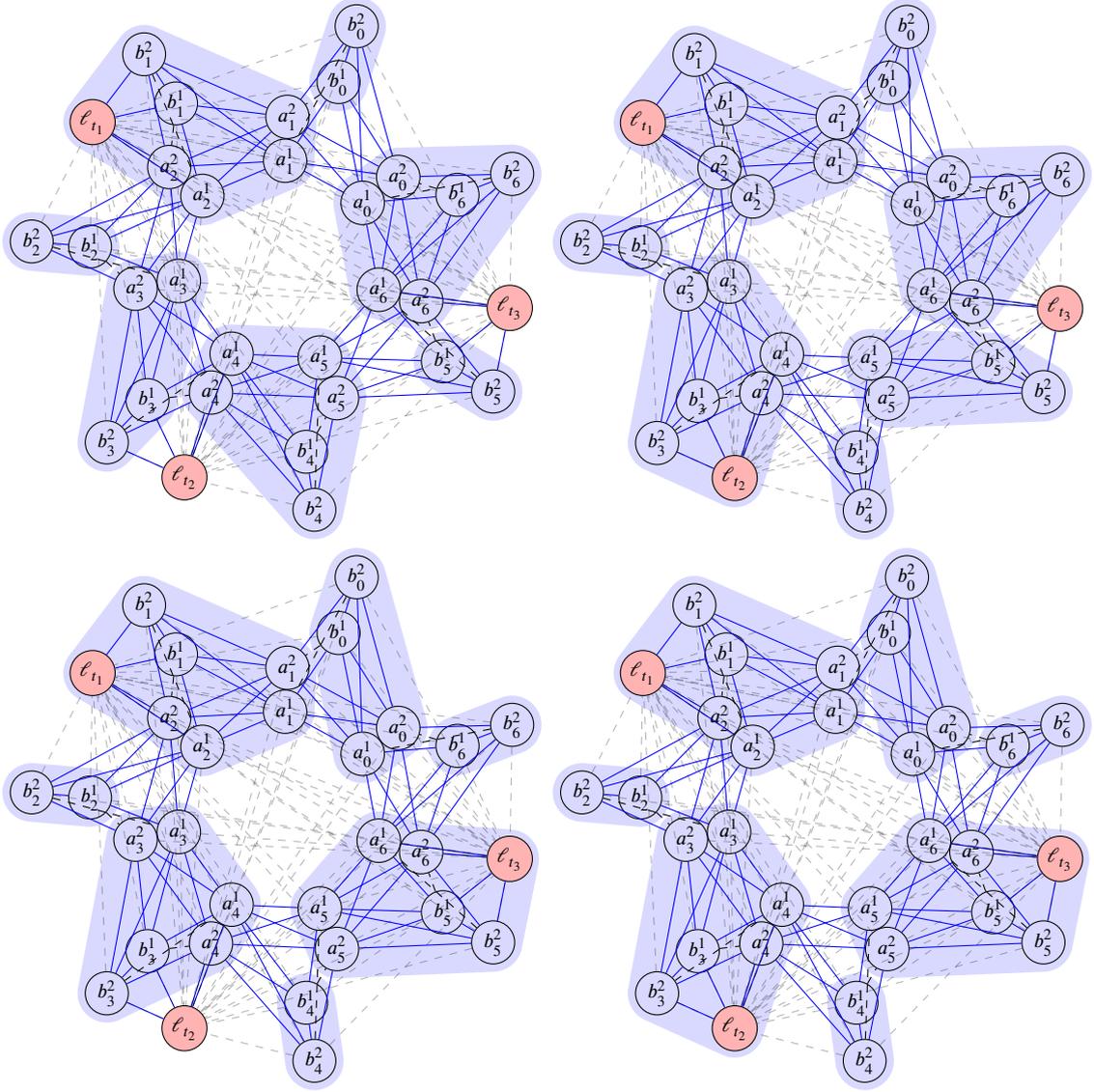

    \centering
    \begin{subfigure}[t]{0.45\textwidth}
    \begin{tikzpicture}[black, scale=0.4,vertex/.style = {draw, circle, inner sep=0.5mm}]
    \input{tikz-hypergraph}
            \sevenfig;

    \begin{pgfonlayer}{background}
    \draw[coal] \hedgeii{12}{12\labelTwo}{10mm};
    \draw[coal] \hedgeii{10}{10\labelTwo}{10mm};
    \draw[coal] \hedgeii{15}{15\labelTwo}{10mm};
    \draw[coal] \hedgem{3\labelTwo}{3}{13,13\labelTwo}{10mm};
   \draw[coal] \hedgem{4\labelTwo}{4}{5,5\labelTwo,14\labelTwo}{10mm};
    \draw[coal] \hedgem{6\labelTwo}{6}{0,0\labelTwo,16\labelTwo}{10mm};
    \draw[coal] \hedgem{1\labelTwo}{1}{2,2\labelTwo,\labell1,11\labelTwo}{10mm};
    \end{pgfonlayer}
    
\end{tikzpicture}
\end{subfigure}
\begin{subfigure}[t]{0.45\textwidth}
    \begin{tikzpicture}[black, scale=0.4,vertex/.style = {draw, circle,  inner sep=0.5mm}]
    \input{tikz-hypergraph}
    \sevenfig;
    
    \begin{pgfonlayer}{background}
    \draw[coal] \hedgeii{12}{12\labelTwo}{10mm};
    \draw[coal] \hedgeii{10}{10\labelTwo}{10mm};
    \draw[coal] \hedgeii{14}{14\labelTwo}{10mm};
    \draw[coal] \hedgem{5\labelTwo}{5}{15,15\labelTwo}{10mm};
   \draw[coal] \hedgem{6\labelTwo}{6}{0,0\labelTwo,16\labelTwo}{10mm};
    \draw[coal] \hedgem{3\labelTwo}{3}{4,4\labelTwo,\labell2,13\labelTwo}{10mm};
    \draw[coal] \hedgem{1\labelTwo}{1}{2,2\labelTwo,\labell1,11\labelTwo}{10mm};
    \end{pgfonlayer}
    
\end{tikzpicture}
\end{subfigure}
\begin{subfigure}[b]{0.45\textwidth}
    \begin{tikzpicture}[black, scale=0.4,vertex/.style = {draw, circle,  inner sep=0.5mm}]
    \input{tikz-hypergraph}
            \sevenfig;

    \begin{pgfonlayer}{background}
    \draw[coal] \hedgeii{12}{12\labelTwo}{10mm};
    \draw[coal] \hedgeii{16}{16\labelTwo}{10mm};
    \draw[coal] \hedgeii{14}{14\labelTwo}{10mm};
    \draw[coal] \hedgem{0\labelTwo}{0}{10,10\labelTwo}{10mm};
   \draw[coal] \hedgem{5\labelTwo}{5}{6,6\labelTwo,\labell3,15\labelTwo}{10mm};
    \draw[coal] \hedgem{3\labelTwo}{3}{4,4\labelTwo,13\labelTwo}{10mm};
    \draw[coal] \hedgem{1\labelTwo}{1}{2,2\labelTwo,\labell1,11\labelTwo}{10mm};
    \end{pgfonlayer}
    
\end{tikzpicture}
\end{subfigure}
\begin{subfigure}[b]{0.45\textwidth}
    \begin{tikzpicture}[black, scale=0.4,vertex/.style = {draw, circle,  inner sep=0.5mm}]
    \input{tikz-hypergraph}
            \sevenfig;

    \begin{pgfonlayer}{background}
    \draw[coal] \hedgeii{12}{12\labelTwo}{10mm};
    \draw[coal] \hedgeii{16}{16\labelTwo}{10mm};
    \draw[coal] \hedgeii{14}{14\labelTwo}{10mm};
    \draw[coal] \hedgem{0\labelTwo}{0}{10,10\labelTwo}{10mm};
   \draw[coal] \hedgem{5\labelTwo}{5}{6,6\labelTwo,\labell3,15\labelTwo}{10mm};
    \draw[coal] \hedgem{3\labelTwo}{3}{4,4\labelTwo,\labell2,13\labelTwo}{10mm};
    \draw[coal] \hedgem{1\labelTwo}{1}{2,2\labelTwo,\labell1,11\labelTwo}{10mm};
    \end{pgfonlayer}
\end{tikzpicture}
\end{subfigure}
    \caption{Partition from Step~\eqref{ceconstr:ICone} where $k = 1$ (top left), Step~\eqref{ceconstr:ICtwoa} where $k = 1$ (top right), Step~\eqref{ceconstr:ICtwob} (bottom left), and Step~\eqref{ceconstr:ICthree} (bottom right) from the proof of \cref{cen:nph}.
    Blue solid edges denote friendship relations and dashed black edges neutral relations. Lack of an edge denotes an enemy relation.}
    \label{fig:cenphardic}
\end{figure*}

(Only if) Suppose that $I$ admits a satisfying assignment \asg. We construct a core stable partition \partition as follows:
\begin{compactenum}
\item For every $x_\ii \in \vars$:
\begin{compactenum}
\item If $\asg[\ii] = \varT$, then:
\begin{compactenum}
\item Add the coalition $\{\fvarn \ii, \Ixx \ii {a^1_0}, \Ixx \ii {a^2_0}, \Ixx \ii {b^1_0},$ $\Ixx \ii {b^2_0},$ $\Ixx \ii {a^1_1},$ $\Ixx \ii {a^2_1} \} \cup \Rew \ii$ to \partition.
\label{ceconstr:varblock}
\item There are exactly two distinct clauses $C_{\cia_1}, C_{\cia_2} \in \clas$ and two indices $k_1, k_2 \in [3]$ such that $\bar{x}_i$ is the $k_1^{\text{th}}$ literal of~$C_{i_1}$ and $k_2^{\text{th}}$ literal of $C_{i_2}$.

Add the coalition $\{\tvarn \ii, \Icx {\cia_1} {a^1_{2k_1 - 1}},$ $\Icx {\cia_1} {a^2_{2k_1 - 1}},$ $\Icx {\cia_1} {b^1_{2k_1 - 1}},$ $\Icx {\cia_1} {b^2_{2k_1 - 1}},$ $\Icx{\cia_1} {a^1_{2k_1}},$ $\Icx {\cia_1} {a^2_{2k_1}}, \Icx {\cia_2} {a^1_{2k_2 - 1}},$ $\Icx {\cia_2} {a^2_{2k_2 - 1}},$ $\Icx {\cia_2} {b^1_{2k_2 - 1}},$ $\Icx {\cia_2} {b^2_{2k_2 - 1}},$ $\Icx{\cia_2} {a^1_{2k_2}},$ $\Icx {\cia_2} {a^2_{2k_2}}\}$ to \partition. \label{ceconstr:Cblock}
\end{compactenum}
\item If $\asg[\ii] = \varF$, then:
\begin{compactenum}
\item Add the coalition $\{\tvarn \ii, \Ixx \ii {a^1_0}, \Ixx \ii {a^2_0},$ $\Ixx \ii {b^1_0},$ $\Ixx \ii {b^2_0}, \Ixx \ii {a^1_1},$ $\Ixx \ii {a^2_1} \} \cup \Rew \ii$ to \partition.
\label{ceconstr:varblockF}
\item There are exactly two distinct clauses $C_{\cia_1}, C_{\cia_2} \in \clas$ and two indices $k_1, k_2 \in [3]$ such that $x_i$ is the $k_1^{\text{th}}$ literal of~$C_{i_1}$ and $k_2^{\text{th}}$ literal of $C_{i_2}$.

Add the coalition $\{\fvarn \ii, \Icx {\cia_1} {a^1_{2k_1 - 1}},$ $\Icx {\cia_1} {a^2_{2k_1 - 1}},$ $\Icx {\cia_1} {b^1_{2k_1 - 1}},$ $\Icx {\cia_1} {b^2_{2k_1 - 1}},$ $\Icx{\cia_1} {a^1_{2k_1}},$ $\Icx {\cia_1} {a^2_{2k_1}}, \Icx {\cia_2} {a^1_{2k_2 - 1}},$ $\Icx {\cia_2} {a^2_{2k_2 - 1}},$ $\Icx {\cia_2} {b^1_{2k_2 - 1}},$ $\Icx {\cia_2} {b^2_{2k_2 - 1}},$ $\Icx{\cia_2} {a^1_{2k_2}},$ $\Icx {\cia_2} {a^2_{2k_2}}\}$ to \partition. \label{ceconstr:CblockF}
\end{compactenum}
\item In either case, partition the remaining agents from \Ix \ii\ as follows: $\{b^1_1, b^2_1\}$, $\{a^2_1, a^2_2, b^1_2, b^2_2\}$, $\{a^1_3, a^2_3, b^1_3, b^2_3, a^1_4, a^2_4\}$, $\{b^1_4, b^2_4\}$.
This partition is depicted in \cref{fig:ceconstrix}.
\label{ceconstr:varrem}
\end{compactenum}
\item For every $C_\cia = \{\ell_{\cib_1}, \ell_{\cib_2},\ell_{\cib_3}\}\in \clas$, there is at least one literal which is true under \asg.
Parts of the partitions below are illustrated in \cref{fig:cenphardic}.
\begin{compactenum}
\item If there is only one $k \in [3]$ such that $\ell_{\cib_k}$ is true under \asg, then $\Icx \cia {a^1_{2k - 1}},$ $\Icx \cia {a^2_{2k - 1}}, \Icx \cia {b^1_{2k - 1}}, \Icx \cia {b^2_{2k - 1}},$ $\Icx \cia {a^1_{2k}},$ and $\Icx \cia {a^2_{2k}}$ are already assigned into a coalition. 
Let us partition the remaining agents from \Ic \cia\ as follows (the indices are computed modulo 7):
$\{b^1_{2k}, b^2_{2k}\}$,
$\{a^1_{2k + 1}, a^2_{2k + 1},$ $b^1_{2k + 1}, b^2_{2k + 1}\}$,
$\{a^1_{2k + 2}, a^2_{2k + 2}, b^1_{2k + 2}, b^2_{2k + 2}, a^1_{2k + 3},$ $ a^2_{2k + 3}\}$,  
$\{b^1_{2k + 3},$ $b^2_{2k + 3}\}$,
$\{a^1_{2k + 4}, a^2_{2k + 4}, b^1_{2k + 4}, b^2_{2k + 4}, a^1_{2k + 5}, a^1_{2k + 5}\}$, and
$\{b^1_{2k + 5},$ $b^2_{2k + 5}\}$.\label{ceconstr:ICone}

\item If there are exactly two distinct $k, k' \in [3], k < k'$ such that $\ell_{\cib_k}$ and $\ell_{\cib_{k'}}$ are true under \asg, then $\Icx \cia {a^1_{2\hat{k} - 1}},$ $\Icx \cia {a^2_{2\hat{k} - 1}},$ $\Icx \cia {b^1_{2\hat{k} - 1}}, \Icx \cia {b^2_{2\hat{k} - 1}},$ $\Icx \cia {a^1_{2\hat{k}}},$ and $\Icx \cia {a^2_{2\hat{k}}}$ are already assigned into coalitions for every $\hat{k} \in \{k, k'\}$.
Let us partition the remaining agents from \Ic \cia\ as follows (the indices are computed modulo 7):
\begin{compactenum}
\item If $k' = k + 1$:
$\{b^1_{2k}, b^2_{2k}\}$,
$\{b^1_{2k + 2}, b^2_{2k + 2}\}$,
$\{a^1_{2k + 3}, a^2_{2k + 3},$ $b^1_{2k + 3}, b^2_{2k + 3}\}$,
$\{a^1_{2k + 4}, a^2_{2k + 4}, b^1_{2k + 4}, b^2_{2k + 4}, a^1_{2k + 5},$ $ a^1_{2k + 5}\}$, and
$\{b^1_{2k + 5}, b^2_{2k + 5}\}$.\label{ceconstr:ICtwoa}
\item Else, i.e., $k = 1$ and $k' = 3$: 
$\{a^1_{0}, a^2_{0},$ $b^1_{0}, b^2_{0}\}$,
$\{b^1_{2}, b^2_{2}\}$,
$\{a^1_{3}, a^2_{3}, b^1_{3}, b^2_{3}, a^1_{4},$ $ a^2_{4}\}$, 
$\{b^1_{4}, b^2_{4}\}$, and
$\{b^1_{6}, b^2_{6}\}$.\label{ceconstr:ICtwob}
\end{compactenum}
\item If all literals of $C_\cia$ are true under \asg,  then $\Icx \cia {a^1_{2k - 1}},$ $\Icx \cia {a^2_{2k - 1}}, \Icx \cia {b^1_{2k - 1}}, \Icx \cia {b^2_{2k - 1}},$ $\Icx \cia {a^1_{2k}},$ and $\Icx \cia {a^2_{2k}}$ are already assigned into a coalition for every $k \in [3]$.
Let us partition the remaining agents from \Ic \cia\ as follows:
$\{a^1_{0}, a^2_{0},$ $b^1_{0}, b^2_{0}\}$,
$\{b^1_{2}, b^2_{2}\}$,
$\{b^1_{4}, b^2_{4}\}$, and
$\{b^1_{6}, b^2_{6}\}$.\label{ceconstr:ICthree}
\end{compactenum}
\end{compactenum}
It is straightforward to observe that the maximum coalition size is~$13$. 
The next observation also shows that we can modify this proof to work for $\maxNumberOfCoalitions = 8$.

\begin{obs}
We can create a modified coalition $\partition'$ with $|\partition'| = 8$ where every agent has no enemies and has the same friends in her coalition as under \partition.
\end{obs}
\begin{proof}\renewcommand{\qedsymbol}{$\diamond$}
\begin{compactitem}
\item[Coalition 1:]  Let this be the union of all the coalitions constructed in Step~\eqref{ceconstr:varblock}.
Clearly, every agent in this coalition gets the same friends as in \partition. Since there are no enemy relations between the agents in $\Rew \ii \cup \Ix \ii \cup \{\tvarn \ii, \fvarn \ii\}$ and the agents in $\Rew \ell \cup \Ix \ell \cup \{\tvarn \ell, \fvarn \ell\}$ when $\ii, \ell \in [n]$, $\ii \neq \ell$, no agent has enemies in this coalition.
\item[Coalition 2:]  Let this be the union of all the coalitions constructed in Step~\eqref{ceconstr:Cblock}.
Clearly, every agent in this coalition gets the same friends as in \partition. 
Observe that the only enemies the agents from~\Ic \cia, where $C_\cia \in \clas$, have outside of~$\Ic \cia$ are within $\bigcup_{x_\ii \in \vars} \Ix \ii$.
Also, for every $x_\ii \in \vars$, the only enemy $\tvarn \ii$ has is $\fvarn \ii$, and neither of them is in this coalition.
\item[Coalitions 3-8:] For every $x_\ii \in \vars$, the remaining agents from \Ix \ii\ are divided into four partitions.
For every $C_\cia \in \clas$, the remaining agents from $\Ic \cia$ are partitioned into at most six coalitions.
Let us merge these coalitions arbitrarily so that no two coalitions that originate from the same $\Ix \ii$ or $\Ic \cia$, $x_\ii \in \vars$, $C_\cia \in \clas$ are in the same final coalition.
Clearly, we can obtain six coalitions this way.
\end{compactitem}\end{proof}
It is clear that $\partition'$ is core stable if and only if $\partition$ is core stable.

Next, we show that \partition is core stable by showing that no agent can join a strictly blocking coalition \blockingCoalition.

No agent in the coalition constructed in Step~\eqref{ceconstr:varblock} can strictly prefer \blockingCoalition\ over her coalition under \partition:
The agents~$\Ixx \ii {b^1_0}$ and $\Ixx \ii {b^2_0}$ obtain all of their friends and no enemies.
The agents \Rew \ii\ have one friend -- \tvarn \ii\ -- outside of their current coalition, but this is an enemy of one agent in the current coalition -- \fvarn \ii. Thus the agents in \Rew \ii\ cannot prefer a coalition containing this agent over their current coalition.
The agents~$\Ixx \ii {a^1_0}$ and~$\Ixx \ii {a^2_0}$ each have three friends outside of their current coalition.
However, because these agents are all enemies with four friends in the current coalition, the agents~$\Ixx \ii {a^1_0}$ and~$\Ixx \ii {a^2_0}$ cannot prefer a coalition containing them over their coalition under \partition.
The agents~$\Ixx \ii {a^1_1}$ and~$\Ixx \ii {a^2_1}$ have each four friends outside of their current coalition.
However, because these agents are all enemies with four friends in the current coalition, the agents~$\Ixx \ii {a^1_0}$ and~$\Ixx \ii {a^2_0}$ cannot prefer a coalition containing them over their coalition under~\partition.
The agent~\fvarn \ii\ has eight friends in her current coalition.
She has eight friends outside of her current coalition: These are from sets~\Ic \cia\ such that~$x_\ii$ is in the clause~$C_\cia$ (recall that each literal appears exactly twice).
However, these agents are all enemies with eight agents in her current coalition, and thus ~\fvarn \ii\ cannot  prefer a coalition containing them over their coalition under~\partition.
The reasoning for the agents in the coalition constructed in Step~\eqref{ceconstr:varblockF} is analogous.

No agent in the coalition constructed in Step~\eqref{ceconstr:Cblock} can strictly prefer \blockingCoalition\ over her coalition under~\partition:
The agents~$\Icx \cia {b^1_{2k - 1}}$ and~$\Icx \cia {b^2_{2k - 1}}$ obtain all of their friends and now enemies.
The agents~$\Icx \cia {a^1_{2k - 1}}$ and~$\Icx \cia {a^2_{2k - 1}}$ have each three friends outside of their current coalition.
However, because these agents are all enemies with four friends in the current coalition, the agents~$\Icx \cia {a^1_{2k - 1}}$ and~$\Icx \cia {a^2_{2k - 1}}$ cannot prefer a coalition containing them over their coalition under \partition.
The agents~$\Icx \cia {a^1_{2k}}$ and~$\Icx \cia {a^2_{2k}}$ have each four friends outside of their current coalition.
However, because these agents are all enemies with four friends in the current coalition, the agents~$\Icx \cia {a^1_{2k}}$ and~$\Icx \cia {a^2_{2k}}$ cannot prefer a coalition containing them over their coalition under~\partition.
The agent~\tvarn \ii\ has eight friends outside of her current coalition: These are the agents from~\Rew \ii\ and \Ixx \ii {b^1_0}, \Ixx \ii {b^2_0}, \Ixx \ii {a^1_1}, and~\Ixx \ii {a^2_1}.
However, we have already shown that none of these agents can join a strictly blocking coalition.
The reasoning for the agents in the coalition constructed in Step~\eqref{ceconstr:CblockF} is analogous.

No agent in the coalitions constructed in Step~\eqref{ceconstr:varrem} can prefer \blockingCoalition\ over her coalition under \partition:
\begin{compactitem}
\item $\{a^1_3, a^2_3, b^1_3, b^2_3, a^1_4, a^2_4\}$: The agents $b^1_3$ and $b^2_3$ obtain all of their friends and no enemies, and thus cannot not prefer~\blockingCoalition\ over their coalition under \partition. 
The agents $a^1_3$ and $a^2_3$ have three friends outside of their current coalition, all of which are enemies of three currently obtained friends, so $a^1_3$ and $a^2_3$ cannot prefer a coalition containing them over their coalition under \partition.
The agents $a^1_4$ and $a^2_4$ obtain three friends, while they have four friends outside. However, we have shown that two of these friends, $a^1_0$ and $a^2_0$, cannot join a blocking coalition.
The two other friends outside, $b^1_4$ and~$b^2_4$ are both enemies with three currently obtained friends of $a^1_4$ and $a^2_4$, so $a^1_4$ and $a^2_4$ cannot prefer a coalition containing them over their coalition under~\partition.
\item $\{a^1_{2}, a^2_{2},$ $b^1_{2}, b^2_{2}\}$: The agents in this coalition have friends among $\{a^1_0, a^2_1, a^1_3, a^1_3\}$. However, we have shown that none of these agents can join a strictly blocking coalition.
The agent $a^1_2$ (resp.\ $a^2_2$) also has a friend $b^1_1$ (resp.\ $b^2_1$) outside of her current coalition, but since  $b^1_1$ (resp.~$b^2_1$) is enemies with two currently obtained friends of  $a^1_2$ (resp.~$a^2_2$), the agent $a^1_2$ (resp.~$a^2_2$) cannot prefer a coalition containing $b^1_1$ (resp.\ $b^2_1$) over her coalition under \partition.
\item $\{b^1_4, b^2_4\}$, $\{b^1_1, b^2_1\}$: In both cases, these agents have three friends outside of their current coalition, but we have shown that none of them can join a strictly blocking coalition.
\end{compactitem}

No agent in the coalitions constructed in Step~\eqref{ceconstr:ICone} can prefer \blockingCoalition\ over her coalition under \partition:
\begin{compactitem}
\item $\{a^1_{2k + 4}, a^2_{2k + 4}, b^1_{2k + 4}, b^2_{2k + 4}, a^1_{2k + 5}, a^1_{2k + 5}\}$: The reasoning for why this cannot block is analogous to why $\{a^1_3, a^2_3, b^1_3, b^2_3, a^1_4,$ $a^2_4\}$ cannot block in Step~\eqref{ceconstr:varrem} -- to see how, let $2k + 4 = 3$.
\item $\{a^1_{2k + 2}, a^2_{2k + 2}, b^1_{2k + 2}, b^2_{2k + 2}, a^1_{2k + 3},$ $ a^2_{2k + 3}\}$: The reasoning for why this cannot block is analogous to why  $\{a^1_3, a^2_3, b^1_3, b^2_3, a^1_4,$ $a^2_4\}$ cannot block in Step~\eqref{ceconstr:varrem} -- to see how, let $2k + 2 = 3$.  Recall that we have now shown that $\{a^1_{2k + 4}$ and $a^2_{2k + 4}$ cannot join a strictly blocking coalition.
\item $\{a^1_{2k + 1}, a^2_{2k + 1},$ $b^1_{2k + 1}, b^2_{2k + 1}\}$: The reasoning for this is case is analogous to why $\{a^1_{2}, a^2_{2},$ $b^1_{2}, b^2_{2}\}$ cannot block in Step~\eqref{ceconstr:varrem} -- to see how, let $2k + 1 = 2$. Recall that we have now shown that $a^1_{2k + 2}$ and $a^2_{2k + 2}$ cannot join a strictly blocking coalition.
\item $\{b^1_{2k}, b^2_{2k}\}$, $\{b^1_{2k + 3}, b^2_{2k + 3}\}$, $\{b^1_{2k + 5}, b^2_{2k + 5}\}$: We have shown that none of the friends these agents have outside of their coalition can join a strictly blocking coalition. Thus they cannot join a strictly blocking coalition.
\end{compactitem}

No agent in the coalitions constructed in Step~\eqref{ceconstr:ICtwoa} can prefer~\blockingCoalition\ over her coalition under \partition:
\begin{compactitem}
\item $\{a^1_{2k + 4}, a^2_{2k + 4}, b^1_{2k + 4}, b^2_{2k + 4}, a^1_{2k + 5}, a^1_{2k + 5}\}$: The reasoning for why this coalition cannot block is analogous to why it cannot block in Step~\eqref{ceconstr:ICone}.
\item $\{a^1_{2k + 3}, a^2_{2k + 3},$ $b^1_{2k + 3}, b^2_{2k + 3}\}$: The reasoning for this is case is analogous to why $\{a^1_{2}, a^2_{2},$ $b^1_{2}, b^2_{2}\}$ cannot block in Step~\eqref{ceconstr:varrem} -- to see how, let $2k + 3 = 2$. Recall that we have now shown that $a^1_{2k + 4}$ and $a^2_{2k + 4}$ cannot join a strictly blocking coalition.
\item $\{b^1_{2k}, b^2_{2k}\}$, $\{b^1_{2k + 2}, b^2_{2k + 2}\}$, $\{b^1_{2k + 5}, b^2_{2k + 5}\}$: We have shown that none of the friends these agents have outside of their coalition can join a strictly blocking coalition. Thus they cannot join a strictly blocking coalition.
\end{compactitem}

No agent in the coalitions constructed in Step~\eqref{ceconstr:ICtwob} can prefer~\blockingCoalition\ over her coalition under \partition:
\begin{compactitem}
\item $\{a^1_{3}, a^2_{3}, b^1_{3}, b^2_{3}, a^1_{4},$ $ a^2_{4}\}$: This is analogous to why $\{a^1_{3}, a^2_{3}, b^1_{3}, b^2_{3},$ $a^1_{4},$ $ a^2_{4}\}$ cannot block in Step~\eqref{ceconstr:varrem}.
\item $\{a^1_{0}, a^2_{0},$ $b^1_{0}, b^2_{0}\}$: The reasoning for this is case is analogous to why $\{a^1_{2}, a^2_{2},$ $b^1_{2}, b^2_{2}\}$ cannot block in Step~\eqref{ceconstr:varrem}. 
\item $\{b^1_{2}, b^2_{2}\}$, $\{b^1_{4}, b^2_{4}\}$, $\{b^1_{6}, b^2_{6}\}$:  We have shown that none of the friends these agents have outside of their coalition can join a strictly blocking coalition. Thus they cannot join a strictly blocking coalition.
\end{compactitem}

Finally, no agent in the coalitions constructed in Step~\eqref{ceconstr:ICthree} can prefer \blockingCoalition\ over her coalition under \partition:
\begin{compactitem}
\item $\{a^1_{0}, a^2_{0},$ $b^1_{0}, b^2_{0}\}$: The reasoning for this is case is analogous to why $\{a^1_{2}, a^2_{2},$ $b^1_{2}, b^2_{2}\}$ cannot block in Step~\eqref{ceconstr:varrem}. 
\item $\{b^1_{2}, b^2_{2}\}$, $\{b^1_{4}, b^2_{4}\}$, $\{b^1_{6}, b^2_{6}\}$:  We have shown that none of the friends these agents have outside of their coalition can join a strictly blocking coalition. Thus they cannot join a strictly blocking coalition.
\end{compactitem}

(If) Suppose that $I'$ admits a core stable partition \partition.
Recall that since $\partition$ is core stable, no agent has enemies in her coalition.

We start with the following observation that tells us that all the variable and clause blockers need variable agents to stop them from creating blocking coalitions.

\begin{obs}\label{clm:blockerreq}
For every $x_\ii \in \vars$  (resp.\ $C_\cia \in \clas$) there must be an agent $p$ in \Ix{\ii} (resp.\ \Ic{\cia}) such that $\partition(p)$ contains a friend of $p$ that is not from \Ix{\ii} (resp.\ \Ic{\cia}).
\end{obs}

\begin{proof}\renewcommand{\qedsymbol}{$\diamond$}
Assume, towards a contradiction, that no agent from \Ix{\ii} (resp.\ \Ic{\cia}) has a friend from outside of \Ix{\ii} (resp.\ \Ic{\cia}) in her coalition.
Then by a reasoning analogous to proof of \cref{thm:fensymno}, the partition~\partition cannot be stable; recall that neural agents do not affect agents' preferences over coalitions.
\end{proof}

This leads directly to the following observation:
\begin{obs}\label{obs:blockerreq}
\begin{compactenum}[(i)]
\item For every $x_\ii \in \vars$,  either $\{\Ixx \ii {b^1_0}$, $\Ixx \ii {b^2_0}$, $\Ixx \ii {a^1_1}$, $\Ixx \ii {a^2_1}\} \cap \partition(\fvarn \ii) \neq \emptyset$ or $\{\Ixx \ii {b^1_0}$, $\Ixx \ii {b^2_0}$, $\Ixx \ii {a^1_1}$, $\Ixx \ii {a^2_1}\} \cap \partition(\tvarn \ii) \neq \emptyset$.\label{obs:blockerreqX}
\item For every $C_\cia = \{\ell_{\cib_1}, \ell_{\cib_2},\ell_{\cib_3},\}\in \clas$, there is $k \in [3]$ such that $\{\Icx \cia {b^1_{2k - 1}}, \Icx \cia {b^2_{2k - 1}}, \Icx \cia {a^1_{2k}}, \Icx \cia {a^2_{2k}}\} \cap \partition(\lvarn{\cib_k}) \neq \emptyset$. \label{obs:blockerreqC} 
\end{compactenum}
\end{obs}
\begin{proof}\renewcommand{\qedsymbol}{$\diamond$}
Let us first show Statement~\eqref{obs:blockerreqX}.
By \cref{clm:blockerreq}, for every $x_\ii \in \vars$, there must be a coalition $\coalB$ such that an agent from~$\Ix \ii$ has a friend outside of $\Ix \ii$.
Since the only agents who have friends outside of $\Ix \ii$ are \Ixx \ii {b^1_0}, \Ixx \ii {b^2_0}, \Ixx \ii {a^1_1}, and \Ixx \ii {a^2_1}.
At least one of them must have a friend outside of $\Ix \ii$.
The agents \tvarn \ii\ and \fvarn \ii\ are the only agents who are friends with them and are outside of \Ix \ii. 
However, since they are enemies with each other, at most one of them can be in $\coalB$.

Next, we show Statement~\eqref{obs:blockerreqC}.
By \cref{clm:blockerreq}, for every $C_\cia = \{\ell_{\cib_1}, \ell_{\cib_2},\ell_{\cib_3}\} \in \clas$, there must be a coalition $\coalB$ such that an agent from~$\Ic \cia$ has a friend outside of $\Ic \cia$.
By construction, the only friendships between $\Ic \cia$ and agents outside of it are, for every $k \in [3]$, between~\lvarn{\cib_k} and the agents in $\{\Icx \cia {b^1_{2k - 1}}, \Icx \cia {b^2_{2k - 1}}, \Icx \cia {a^1_{2k}}, \Icx \cia {a^2_{2k}}\}$.
The statement follows directly.
\end{proof}

Let us now construct an assignment \asg\ as follows:
for every $x_\ii \in \vars$, if $\{\Ixx \ii {b^1_0}$, $\Ixx \ii {b^2_0}$, $\Ixx \ii {a^1_1}$, $\Ixx \ii {a^2_1}\} \cap \partition(\fvarn \ii) \neq \emptyset$, then 
$\asg[\ii] = \varT$.

Assume, towards a contradiction, that \asg\ is not a satisfying assignment.
Then there is a clause $C_\cia = \{\ell_{\cib_1}, \ell_{\cib_2},\ell_{\cib_3}\} \in \clas$ such that all its literals are set to false by \asg.
By \cref{obs:blockerreq}\eqref{obs:blockerreqC} there must be $k \in [3]$, such that $\{\Icx \cia {b^1_{2k - 1}},$ $\Icx \cia {b^2_{2k - 1}},$ $\Icx \cia {a^1_{2k}}, \Icx \cia {a^2_{2k}}\} \cap \partition(\lvarn{\cib_k}) \neq \emptyset$.

If $\ell_{\cib_k}$ is $\bar{x}_{\cib_k}$, i.e., $\ell_{\cib_k}$ is a false literal, then $\asg[\ii] = \varT$, as otherwise $\ell_{\cib_k}$ would be true under \asg.
By construction $\{\Ixx \ii {b^1_0}$, $\Ixx \ii {b^2_0}$, $\Ixx \ii {a^1_1}$, $\Ixx \ii {a^2_1}\} \cap \partition(\lvarn {\cib_k}) \neq \emptyset$.
However, since $\{\Ixx \ii {b^1_0}$, $\Ixx \ii {b^2_0}$, $\Ixx \ii {a^1_1}$, $\Ixx \ii {a^2_1}\}$ are all enemies with all the agents in $\{\Icx \cia {b^1_{2k - 1}}, \Icx \cia {b^2_{2k - 1}},$ $\Icx \cia {a^1_{2k}},$ $\Icx \cia {a^2_{2k}}\}$, they cannot be in the same coalition in \partition, a contradiction.

If $\ell_{\cib_k}$ is $x_{\cib_k}$, i.e., $\ell_{\cib_k}$ is a true literal, then $\asg[\ii] = \varF$, as otherwise $\ell_{\cib_k}$ would be true under \asg.
Thus $\{\Ixx \ii {b^1_0}$, $\Ixx \ii {b^2_0}$, $\Ixx \ii {a^1_1}$, $\Ixx \ii {a^2_1}\} \cap \partition(\nlvarn {\cib_k}) = \emptyset$.
Thus by \cref{obs:blockerreq}\eqref{obs:blockerreqX}, $\{\Ixx \ii {b^1_0}$, $\Ixx \ii {b^2_0}$, $\Ixx \ii {a^1_1}$, $\Ixx \ii {a^2_1}\} \cap \partition(\lvarn {\cib_k}) \neq \emptyset$. We reach a contradiction analogously to the previous case.
}

The following two results can be obtained through a straightforward modification of the proofs of \cref{prop:CE_Number_Coalitions_3+,prop:SCE_DegreeFriend_4}.
\begin{restatable}[\appsymb]{proposition}{propscenpartsize}
\neut{\SCE} is NP-hard even when $\maxDegreeFE =~12$. The result extends to \neut{\SCEBounded} even when $\maxNumberOfCoalitions = 3$.\label{prop:scenpartsize}
\end{restatable}

\appendixproofwithstatement{prop:scenpartsize}{\propscenpartsize*}{

The proof is a modification of the proof of \cref{prop:CE_Number_Coalitions_3+}.
We again reduce from \probname{3-Coloring} of a graph of degree 4, and construct two dummies for each vertex. Let the original graph be $G = (V, E)$. The enemy relations are constructed as previously; it is clear that every agent has at most 6 enemies.

We construct the friendship relations as follows: For every $v_i \in V$, the agents from $\{i, d_{i}, d'_i\}$ are all friends with all the agents from $\{i + 1, d_{i+1}, d'_{i + 1}\}$ (we take $3n + 1 = 1$), except $i$ and $i + 1$ if $\{v_i, v_{i+1}\} \in E$. Observe that because the relations are symmetric, this implies that the agents from $\{i, d_{i}, d'_i\}$ are all also friends with all the agents from $\{i - 1, d_{i-1}, d'_{i - 1}\}$ (we take $1- 1 = 3n$), except $i$ and $i - 1$ if $\{v_i, v_{i-1}\} \in E$.
All the unmentioned relations are neutral.

Thus every agent has at most 6 friends, which means that the maximum degree is 12. 

We again claim that there is a strictly core stable partition if and only if $G$ admits a 3-coloring.

(If) We construct a partition from the coloring as previously. Observe that since the coalition contains one agent from each  $\{i, d_{i}, d'_i\}$, where $v_i \in V$, and no pair of agents in the coalition are enemies, every agent obtains two friends. 
Observe that no agent may obtain more than two friends in a coalition where no pair of agents is enemies.
This is because $\{i, d_{i}, d'_i\}$ are mutual enemies for every $v_i \in V$.
Thus this partition is strictly core stable.

(Only if) We assume there is a strictly core stable partition \partition.
Assume, towards a contradiction, that there is $v_i \in V$, such that for some $a \in \{i, d_i, d'_i\}$, the agent $a$ obtains  strictly fewer than two friends under \partition.
Observe that every agent obtains two friends in the coalition $\{a\} \cup \{d_j \mid v_j \in V \setminus \{v_i\}\}$.
By earlier reasoning, no agent may obtain more than two friends in any strictly core stable partition.
Thus $\{a\} \cup \{d_j \mid v_j \in V \setminus \{v_i\}\}$ blocks \partition, a contradiction.

Therefore every agent must obtain two friends under \partition. 
Let $\coalB$ be an arbitrary coalition in \partition.
Since it is nonempty, it must contain $a_i \in \{i, d_i, d'_i\}$ for some $v_i \in V$.
Since $a_i$ must obtain two friends, there must be $a_{i + 1} \in \{i + 1, d_{i + 1}, d'_{i }\}$ such that $a_{i + 1} \in \coalB$.
By iterating this argument we obtain that $\coalB$ must contain precisely $n$ agents.
From here the proof is analogous to the proof of \cref{prop:CE_Number_Coalitions_3+}.
}

\begin{restatable}[\appsymb]{proposition}{propscencoalsize}
\neut{\SCE} is NP-hard even when $\maxDegreeFE =~12$.
The result extends to \neut{\SCEBoundedCoal} even when $\maxCoalitionSize = 3$.\label{prop:scencoalsize}
\end{restatable}

\appendixproofwithstatement{prop:scencoalsize}{\propscencoalsize*}{
The proof is similar to the proof of \cref{prop:SCE_DegreeFriend_4}.
We again reduce from \probname{Triangle Packing}, where each vertex of the input graph has one of two specific neighborhoods, described in \cref{fig:2b_3a}.
Let $G = (V, E)$ be the original graph.
The friendship relations are as described in the proof of \cref{prop:SCE_DegreeFriend_4}.
The enemy relations are as follows: For every $v_i, v_j, v_k \in V$, if $\{v_i, v_j\}, \{v_j, v_k\} \in E$ but $\{v_i, v_k\} \notin E$, then the agents $i$ and $k$ are enemies. In other words, the agents are enemies with the friends of their friends, who are not already their friends. 
We observe that for every $v_i \in V$, the vertex $v_i$ has four neighbors, and each of these neighbors has at most two friends who are not friends of~$v_i$.
Thus every agent has at most eight enemies, and thus a total degree of at most 12.

Observe that if $\coalB$ is a coalition where no pair of agents are enemies with each other, then for every pair $j, \ell$ such that $j, \ell \in \coalB$ and $j, \ell$ are friends of $i$, the agents $j$ and $\ell$ must also be friends with each other.
By construction, they would otherwise be enemies.
Also, no agent $i$ may obtain three friends in $\coalB$. If this were the case, then by previous argumentation these friends would have to also be mutual friends, and they would correspond to $K_4$ in $G$, a contradiction.

Thus every connected component of the subgraph induced on~$G$ by the agents corresponding to $\coalB$ must be a triangle, pair, or singleton.
Since the agents in different connected components must be neutral towards each other, we can, without loss of generality, divide $\coalB$ into smaller coalitions based on these connected components.
Thus the best possible coalition for an agent corresponds to a triangle containing her vertex.

From here on, the proof proceeds similarly to the proof of \cref{prop:SCE_DegreeFriend_4}.
}
The following problem is used in the remaining three reductions:
\begin{definition}[Exact Cover]
Given a set $\elements$ and a collection $\ecov = \{\sset 1, \dots, \sset k\}$ of subsets of $\elements$, we say that \ecov\ is an \myemph{exact cover} of \elements\ if for every $i \in \elements$, there is exactly one set $\sset j$ in $\ecov$ such that $i \in \sset j$.
\end{definition}
\decprob{Exact Cover by 3-Sets}{An element set~$\elements$ and a collection~$\sets = \{\sset{1}, \dots, \sset{m}\}$ of size-3 subsets of $\elements$.}{Is there a subset $\ecov \subseteq \sets$ such that $\ecov$ is an exact cover of \elements?}

\probname{Exact Cover by 3-Sets} is NP-complete even when every element appears in at most three sets~\cite{DF86}.
\begin{restatable}[\appsymb]{theorem}{thmscenbipartite}
\neut{SCE} is NP-hard even when $\friendshipGraph \cup \enemyGraph$ is bipartite.\label{thm:scenbipartite}
\end{restatable}

\newcommand{\dumcountc}{\ensuremath{3(m -n) + 4}}
\newcommand{\dumcount}{\ensuremath{[\dumcountc]}}

\appendixproofwithstatementsketch{thm:scenbipartite}{\thmscenbipartite*}{
We reduce from \probname{Exact Cover by 3-Sets}.
We construct agents for both the sets and elements.
The idea of the proof is that all the element agents need to be with a set agent that contains her element for the coalition to be stable.
We also construct $m - n$ set holders so that the set agents who do not get their elements get the same number of friends as the sets who get their elements.
We additionally construct some dummies to enforce that any coalition may contain at most one set and at most one set holder.}{Let us reduce from \probname{Exact Cover by 3-Sets}.
Let $I = (\elements, \sets = \{S_1, \dots, S_{m}\})$ be an instance of \probname{Exact Cover by 3-Sets}.

\begin{figure}
    \centering
\begin{tikzpicture}[black]
             \input{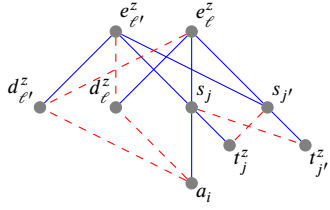}
         \foreach  \x / \y / \n / \nn / \d in
         {3/2/ai/{a_i}/3,
         4/2/sj/{s_j}/1,
         3.5/2.5/tj/{t^z_j}/3,
         4/3/sk/{s_{j'}}/1,
         3.5/3.5/tk/{t^z_{j'}}/3,
         5/2/el/{e^z_{\ell}}/1,
         5/1/elp/{e^z_{\ell'}}/1,
         4/1/dl/{d^z_{\ell}}/0,
         4/0/dlp/{d^z_{\ell'}}/0}
         {
           \node[pn] at (\y, \x) (\n) {};
           \myVertexName{$\nn$}{\n}{\d};
         }
         
		\foreach  \x / \y / \n in
         {}
         {
           \node[] at (\x, \y) (\n) {};
         }         
         
        \foreach \x / \y / \r / \n in {}
        {
			\node[rotate=\r, inner sep=0pt] at (\x, \y) (\n) {{$\ldots$}};    
        }

     	\foreach \a / \b / \r in {
            ai/sj/0,
            sj/tj/0,sk/tk/0,
            sj/el/0,sj/elp/0,sk/el/0,sk/elp/0,
            el/dl/0,elp/dlp/0
        } { 
             \draw[-, friend] (\a) edge[bend right=\r] (\b);
           }
           
      \foreach \a / \b / \r in {sj/tk/0,sk/tj/0,
            el/dlp/0,elp/dl/0,
            ai/dl/0,ai/dlp/0} { 
             \draw[-, enemy] (\a) edge[bend right=\r] (\b);
           }

\end{tikzpicture}
    \caption{A figure for the proof of \cref{thm:scenbipartite}. We assume $i \in S_j$. The black solid edges signify friendships and the red dashed edges enemy-relations.}
    \label{fig:scenbipartite}
\end{figure}

Let us construct an instance $I'$ of \neut{\SCE}. The construction is also illustrated in \cref{fig:scenbipartite}.
We construct the agents as follows:
\begin{compactitem}
\item For every $i \in \elements$, construct an element agent $a_i$.
\item For every $\sset j \in \sets$, construct a set agent $s_j$, and for every $z \in \dumcount$, construct a set dummy $t^z_j$.
\item For every $\ell \in [m - n]$, construct for every $z \in [3]$ a set holder $e^z_\ell$ and for every $w \in [m + 1]$ a set holder dummy~$d^w_\ell$.
\end{compactitem}
The idea of the proof is that the element agents need to be with the set agents for the coalition to be stable.
We also construct $m - n$ set holders so that the set agents who do not get their elements get the same number of friends as the sets who get their elements.
The set dummies are used to make sure that different set agents cannot be in the same coalition.
Similarly, the set holder dummies enforce that two different set holders cannot be in the same coalition.

The friendship relations are as follows:
\begin{compactitem}
\item For every $\sset j \in \sets$:
\begin{compactitem}
\item For every $i \in \sset j$, the agent $a_i$ is friends with the agent~$s_j$, i.e., the elements are friends with the sets that contain them.
\item For every $z \in \dumcount$, the agent $s_j$ is friends with the agent $t^z_j$, i.e., the set agents are friends with their respective dummies.
\item For every $\ell \in [m - n], z \in [3]$, the agent $s_j$ is friends with the agent $e^z_\ell$, i.e., the set agents are friends with the set holders.
\end{compactitem}
\item For every $\ell \in [m - n], z \in [3], w \in [m + 1]$, the agents $e^z_\ell$ and $d^w_\ell$ are friends, i.e., the set holders are friends with their respective dummies.
\end{compactitem}

The enemy relations are as follows:
\begin{compactitem}
\item For every $\sset j, \sset k \in \sets$ such that $k \neq j$, for every $z \in \dumcount$, the agent $s_j$ is enemies with $t^z_j$. In other words, the set agents are enemies with the dummies of the other sets.
\item For every $\ell, j \in [n  - m]$ such that $\ell \neq j$, for every $z \in [3], w \in [m + 1]$, the agents $e^z_\ell$ and $d^w_j$ are enemies.
In other words, the set holders are enemies with the dummies of the other set holders.
\item For every $\ell \in [n - m], i \in \elements, z \in [m + 1]$, the agents $a_i$ and $d^z_\ell$ are enemies. In other words, the element agents are enemies with the set holder dummies.
\end{compactitem}
The graph is clearly bipartite by a partition whose first set is $\{a_i \mid i \in \elements\} \cup \{t^z_j \mid \sset j \in \sets, z \in \dumcount\} \cup \{e^z_\ell \mid \ell \in [m - n], z \in [3]\}$ and the second set is $\{s_j \mid j \in [m]\} \cup \{d^z_\ell \mid \ell \in [m - n], z \in [m + 1]\}$.

Next, we show that $I$ admits an exact cover if and only if $I'$ admits a strictly core stable partition.

(Only if) Assume $I$ admits an exact cover \ecov. We construct a partition~\partition as follows:
\begin{compactitem}
\item For every $\sset j \in \ecov$, add the set $\{a_i \mid i \in \sset j\} \cup \{s_j\} \cup \{t^z_j \mid z \in \dumcount\}$ to \partition.
\item Let $\sset {j_1}, \dots, \sset {j_{m - n}}$ be an arbitrary but fixed order of $\sets \setminus \ecov$.
For every $\sset {j_i} \in \sets \setminus \ecov$, add the set $\{s_{j_i}\} \cup \{t^z_{j_i} \mid z \in \dumcount\} \cup \{e^z_i \mid z \in [3]\} \cup \{d^z_i \mid z \in [m + 1]\}$ to \partition.
\end{compactitem}

We now show that there cannot be a coalition \blockingCoalition\ that weakly blocks \partition.
We first observe that all the dummy set holders $d^z_i, i \in [m -n ], z \in [3]$ obtain all of their friends and no enemies in \partition, so they cannot prefer \blockingCoalition\ over their coalition under \partition. 
Similarly, all the set dummies $\{t^z_j \mid \sset j \in \sets, z \in \dumcount\}$ obtain all of their friends and no enemies in \partition, and also cannot prefer \blockingCoalition\ over their coalition under \partition.

Next, we show that no agent $a_i, i \in \elements$ can prefer \blockingCoalition\ over her coalition under \partition.
Assume, towards a coalition, that~$a_i$ prefers \blockingCoalition\ over her coalition under \partition. Because~$a_i$ obtains one friend under~$\partition$, she must obtain two friends $s_j, s_k$, where $\sset j, \sset k \in \sets$ and $i \in \sset j \cap \sset k$ in~\blockingCoalition.
Since $s_j$ and $s_k$ have $\dumcountc + 3$ friends under~\partition, they must obtain at least this many in~\blockingCoalition.
Observe that $s_j$ has $3(m - n) + 3$ friends in $I'$ without the agents $t^z_j, z \in \dumcount$.
Thus there must be $w \in \dumcount$ such that $t^w_j \in \blockingCoalition$.
However, by construction~$t^w_j$ is enemies with $s_k$, a contradiction.
Thus $a_i$ cannot prefer \blockingCoalition\ over~$\partition(a_i)$.

Next, we show that no agent $s_j$, $\sset j \in \sets$ can prefer \blockingCoalition\ over her coalition under \partition.
Observe that regardless of whether $\sset j \in \ecov$ or not, $s_j$ has $\dumcountc +3$ friends.
To obtain more, there needs to be $i \in \sset j, \ell \in [n -m]$, and $z \in [3]$ such that $a_i$ and $e^z_\ell$ are in \blockingCoalition, as otherwise $s_j$ would have at most $\dumcountc + 3$ friends.
Since $e^z_\ell$ obtains $m + 2$ friends under \partition, she must obtain at least this many in~\blockingCoalition.
Because she has only $m$ friends in $I'$ without the set holder dummies, there must be at least one $w \in [m + 1]$ such that $d^w_\ell \in \blockingCoalition$. However, $d^w_\ell$ and $a_i$ are enemies, a contradiction.

Finally, we show that no set holder $e^z_\ell, \ell \in [m -n], z \in [3]$ can prefer \blockingCoalition\ over her coalition under \partition.
The agent $e^z_\ell$ has $m + 2$ friends under \partition.
As she has all the possible dummy set holder friends, she must have two set friends $s_j, s_k$, where $\sset j, \sset k \in \sets$.
However, as reasoned earlier, there must be $w \in \dumcount$ such that $t^w_j \in \blockingCoalition$.
However, $t^w_j$ is enemies with $s_k$, a contradiction.

As no agent can prefer \blockingCoalition\ over her coalition under \partition, the partition~\partition must be stable.

(If) Assume that $I$ admits a core stable partition \partition.

We start with the following observation:
\begin{obs}\label{obs:dummyjob}
For every $\sset j \in \sets$, there exist a coalition $\coalB \in \partition$ such that $\{s_j\} \cup \{t^z_j \mid z \in \dumcount\} \subseteq B$.
Moreover, for every $\sset k \in \sets \setminus \{\sset j\}$, $s_k \notin \coalB$.

Similarly, for every $\ell \in [m -n]$, there exist a coalition $\coalB \in \partition$ such that $\{e^z_\ell \mid z \in [3]\} \cup \{d^z_\ell \mid z \in [m + 1]\} \subseteq B$.
Moreover, for every $k \in [m - n] \setminus \{\ell\}, w \in [3]$, $e^w_k \notin \coalB$.
\end{obs}
\begin{proof}\renewcommand{\qedsymbol}{$\diamond$}
Observe that $s_j$ has $3(m - n) + 3$ friends without the set dummies, which is strictly fewer than the number of set dummies \dumcountc.
Also, the set dummies have no friends other than~$s_j$.
Thus, if $s_j$ is not with any set dummies, then $\{s_j\} \cup \{t^z_j \mid z \in \dumcount\}$ forms a blocking coalition.
If $s_j$ is with some of her set dummies but not all of them, then adding the remaining set dummies creates a weakly blocking coalition: All the set dummies of $s_j$ have the same enemies, so no agent will obtain new enemies.
The added set dummies and $s_j$ will obtain more friends and thus prefer this coalition.
Also, since for every $\sset k \in \sets \setminus \{\sset j\}$ we have that $s_k$ is enemies with all the agents in $\{t^z_j \mid z \in \dumcount\}$, the agent $s_k$ cannot be in this coalition.

The proof for the second statement is analogous.
\end{proof}

Let us construct an exact cover \ecov\ as follows: For every $\sset j \in \sets$, if $\partition(s_j) \nsubseteq \partition(e^z_\ell)$ for some $z \in [3], \ell \in [m - n]$, then $\sset j \in \ecov$.

Let us show the following observation:
\begin{claim}\label{obs:ecovdefalt}
For every $\sset j \in \ecov$, $i \in \sset j$, we have that $a_i \in \partition(s_j)$.
For every $\sset k \in \sets \setminus \ecov$, we have that $a_{i'} \notin \partition(s_k)$ for any $i' \in \elements$.
\end{claim}
\begin{proof}\renewcommand{\qedsymbol}{$\diamond$}
By \cref{obs:dummyjob}, there can be no two set agents who are in the same coalition.
Thus for every $\ell \in [m - n], z \in [3]$, the set holder agent $e^z_\ell$ can have at most $m + 1 + 1$ friends -- her dummies and one set agent.
Thus if the agent $s_j$ obtains at most $\dumcountc + 2$ friends, the coalition $\{s_j\} \cup \{t^z_j \mid z \in [3]\} \cup \{e^z_\ell \mid z \in [3]\} \cup \{d^z_\ell \mid z \in [m + 1]\}$ weakly blocks \partition: The agent $s_j$ obtains $\dumcountc + 3$ friends, the agents $e^z_\ell$, $z \in [3]$ obtain $m + 1 + 1$ friends, and the dummies obtain all of their friends. No agent obtains enemies.
Therefore $s_j$ must obtain $\dumcountc + 3$ friends.
As she does not have set holder friends and she can obtain \dumcountc\ friends from her dummies, $a_i \in \partition(s_j)$ for every $i \in \sset j$.

Next assume, towards a contradiction, that there is $i' \in \elements$ such that $a_{i'} \in \partition(s_k)$. By construction of $\ecov$ we have that $s_k \subseteq \partition(e^z_\ell)$ for some $\ell \in [m -n], z \in [3]$.
Therefore $a_{i'} \in \partition(e^z_\ell)$.
But $d^w_\ell \in \partition(e^z_\ell)$ for every $w \in [m + 1]$ by \cref{obs:dummyjob}.
The agents $a_{i'}$ and $d^w_\ell$ are enemies, a contradiction.
\end{proof}

There can be two reasons \ecov\ is not an exact cover: (1) There is an element $i \in \elements$ such that no set in \ecov\ contains it, or (2) there is an element $i \in \elements$ such that two sets in \ecov\ contain it.

We first show that Case (2) leads to a contradiction.
Assume that there is some element $i \in \elements$ such that there are two distinct sets $\sset j, \sset k \in \ecov$ such that $i \in \sset j \cap \sset k$.
But then we have by \cref{obs:ecovdefalt} that $a_i \in \partition(s_j)$ and $a_i \in \partition(s_k)$ and thus $\partition(s_j) = \partition(s_k)$, which contradicts \cref{obs:dummyjob}.

Next, we show that Case (1) leads to a contradiction.
Assume that there is an element $i \in \elements$ who is not in any set in \ecov.
By \cref{obs:ecovdefalt}, the agent $a_i$ can have no friends in her partition.
Consider an arbitrary set $\sset j \in \sets$ such that $i \in \sset j$. By assumption that $i$ is not in a set in $\ecov$, we have that $\sset j \notin \ecov$.
By \cref{obs:ecovdefalt}, the agent $s_j$ does not have any element agent friends.
Since by \cref{obs:dummyjob} there can be at most one $\ell \in [m - n]$ such that $s_j$ is in a coalition with the agents $e^z_\ell, z \in [3]$, we have that the agent $s_j$ has at most $\dumcountc + 3$ friends.
Thus the coalition $\coalB \coloneqq \{a_i \mid i \in \sset j\} \cup \{s_j\} \cup \{t^z_j \mid z \in \dumcount\}$ blocks \partition; we can verify that $a_i$ strictly prefers \coalB\ and every other agent in \coalB\ weakly prefers \coalB.}

\begin{restatable}[\appsymb]{theorem}{thmcvneutrhard}
\neut{\CV} is NP-complete even when $\friendshipGraph \cup \enemyGraph$ is bipartite, $\maxDegreeFE$ is a constant, and either $\maxNumberOfCoalitions = 2$ or $\maxCoalitionSize = 3$.\label{thm:cvneutrhard}
\end{restatable}

\appendixproofwithstatement{thm:cvneutrhard}{\thmcvneutrhard*}{
We reduce from \probname{Exact Cover by 3-Sets},  where every element appears in at most three sets.

\begin{figure}
    \centering
\begin{tikzpicture}[black]
             \input{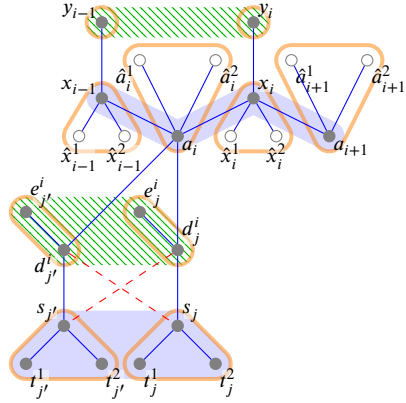}
         \foreach  \x / \y / \n / \nn / \d in
         {6/6.5/ai/{a_i}/3,
         6/4/sj/{s_j}/1,
         5.5/3.5/tj1/{t^1_j}/3,
         6.5/3.5/tj2/{t^2_j}/3,
         4.5/4/sk/{s_{j'}}/0,
         4/3.5/tk1/{t^1_{j'}}/3,
         5/3.5/tk2/{t^2_{j'}}/3,
         5.5/5.5/eij/{e^i_{j}}/1,
         6/5/dij/{d^i_{j}}/1,
         4/5.5/eik/{e^i_{j'}}/1,
         4.5/5/dik/{d^i_{j'}}/2,
         7/7/xi/{x_i}/1,
         7/8/yi/{y_i}/1,
         8/6.5/aip/{a_{i+1}}/3,
         5/7/xim/{x_{i-1}}/0,
         5/8/yim/{y_{i-1}}/0}
         {
           \node[pn] at (\x, \y) (\n) {};
           \myVertexName{$\nn$}{\n}{\d};
         }

                  \foreach  \x / \y / \n / \nn / \d in
         {7.5/7.5/hai1p/{\hat{a}^1_{i+1}}/3,
         8.5/7.5/hai2p/{\hat{a}^2_{i+1}}/3,
         4.7/6.5/hxi1m/{\hat{x}^1_{i-1}}/4,
         5.3/6.5/hxi2m/{\hat{x}^2_{i-1}}/4,
         6.7/6.5/hxi1/{\hat{x}^1_i}/4,
         7.3/6.5/hxi2/{\hat{x}^2_i}/4,
         5.5/7.5/hai1/{\hat{a}^1_i}/2,
         6.5/7.5/hai2/{\hat{a}^2_i}/3}
         {
           \node[pndummy] at (\x, \y) (\n) {};
           \myVertexName{$\nn$}{\n}{\d};
         }
         
		\foreach  \x / \y / \n in
         {}
         {
           \node[] at (\x, \y) (\n) {};
         }         
         
        \foreach \x / \y / \r / \n in {}
        {
			\node[rotate=\r, inner sep=0pt] at (\x, \y) (\n) {{$\ldots$}};    
        }

     	\foreach \a / \b / \r in {
        ai/dij/0,ai/dik/0,
        dij/eij/0,dik/eik/0,
        dij/sj/0,dik/sk/0,
        sj/tj1/0,sj/tj2/0,sk/tk1/0,sk/tk2/0,
        ai/xi/0,xi/aip/0,xim/ai/0,
        xim/yim/0,ai/hai1/0,ai/hai2/0,xi/yi/0,aip/hai1p/0,aip/hai2p/0,
        xi/hxi1/0,xi/hxi2/0,xim/hxi1m/0,xim/hxi2m/0
        } { 
             \draw[-, friend] (\a) edge[bend right=\r] (\b);
           }
           
      \foreach \a / \b / \r in {dij/sk/0,dik/sj/0} { 
             \draw[-, enemy] (\a) edge[bend right=\r] (\b);
           }
    \begin{pgfonlayer}{background}

      \draw[coal] \hedgem{tk1}{sk}{sj,tj2}{2mm};
      \draw[coal] \hedgeii{xim}{ai}{2mm};
      \draw[coal] \hedgeii{xi}{ai}{2mm};
      \draw[coal] \hedgeii{aip}{xi}{2mm};
      \draw[coalgreen] \hedgem{dik}{eik}{eij,dij}{2mm};
      \draw[coalgreen] \hedgeii{yim}{yi}{2mm};

      \draw[coalline]  \hedgeiii{sj}{tj2}{tj1}{2mm};
      \draw[coalline]  \hedgeiii{sk}{tk2}{tk1}{2mm};
      \draw[coalline]  \hedgeii{eij}{dij}{2mm};
      \draw[coalline]  \hedgeii{eik}{dik}{2mm};
      \draw[coalline] \hedgeiii{ai}{hai1}{hai2}{2mm};
      \draw[coalline] \hedgeiii{aip}{hai1p}{hai2p}{2mm};
      \draw[coalline] \hedgeiii{xi}{hxi2}{hxi1}{2mm};
      \draw[coalline] \hedgeiii{xim}{hxi2m}{hxi1m}{2mm};
      \draw[coalline] \hedgei{yim}{2mm};
      \draw[coalline] \hedgei{yi}{2mm};
    \end{pgfonlayer}

\end{tikzpicture}
    \caption{A figure for the proof of \cref{thm:cvneutrhard}. We assume $i \in S_j \cap S_{j'}$. The black solid edges signify friendships and the red dashed edges enemy-relations.
    The blue solid and green dashed regions signify the initial partition with two coalitions.
    The orange lines signify the coalitions in the initial partition with coalitions of size at most $3$.
    The nodes with white inside only appear in the construction for the case where coalitions are of size at most $3$.}
    \label{fig:cvneutrhard}
\end{figure}

Let $I = (\elements, \sets)$ be an instance of \probname{Exact Cover by 3-Sets}.
Let us construct the agents as follows:
The construction is also illustrated in \cref{fig:cvneutrhard}.
\begin{compactitem}
\item For every $i \in \elements$, construct the agents $a_i, x_i, y_i$.
\item For every set $\sset j \in \sets$, construct the agents $s_j, t^1_j, t^2_j,$ and for every $i \in \sset j$ the agents  $d^{i}_j, e^{i}_j$.
\end{compactitem}

Let us construct the friendship relations as follows, where $3n + 1 =~1$:
\begin{compactitem}
\item For every set $\sset j \in \sets$,  $s_j$ is friends with  $t^1_j, t^2_j,$ and $d^{i}_j$ for every $i \in \sset j$. For every $i \in \sset j$, the agent $d^{i}_j$ is friends with $a_{i}$ and~$e^{i}_j$.
\item For every $i \in \elements$, the agent $x_i$ is friends with $y_i, a_i,$ and~$a_{i + 1}$.
\end{compactitem}
The enemy relations are as follows:
\begin{compactitem}
\item For every $i \in \elements$, for every $\sset j, \sset k \in \sets$ such that $i \in \sset j \cap \sset k$ and $\sset j \neq \sset k$, the agent $d^i_j$ is enemies with the agent $s_k$.
\end{compactitem}

The union of the \friendshipGraph and \enemyGraph is clearly bipartite by a partition where the first set is $\{\{a_i, y_i \mid i \in \elements\} \cup \{s_j \mid \sset j \in \sets\}  \cup \{e^i_j \mid \sset j \in \sets, i \in \sset j\} \}$ and the second set is $\{\{x_i \mid i \in \elements\} \cup \{d^i_j \mid \sset j \in \sets, i \in \sset j\} \cup \{t^1_j, t^2_j \mid \sset j \in \sets\}\}$.

Let us observe that $\maxDegreeFE$ is a constant:
\begin{compactitem}
\item For every $i \in \elements$:
\begin{compactitem}
\item The agent $a_i$ has the friends $x_{i - 1}$ (where $1 -1 = 3n$), $x_i$, and $d^i_j$ for every set $\sset j \in \sets$ such that $i \in \sset j$, in total five agents. She has no enemies.
\item The agent $x_i$ is friends with $a_i$, $a_{i + 1}$ (where $3n + 1 = 1$), and $y_i$, in total three agents. She has no enemies.
\item The agent $y_i$ is friends only with $x_i$, and has no enemies.
\end{compactitem}
\item For every $\sset j \in \sets$:
\begin{compactitem}
\item The agent $s_j$ is friends with $d^i_j$ for every $i \in \sset j$ and the agents $t^1_j$ and $t^2_j$, in total five friends. She is enemies with~$d^{i'}_\ell$ for every $\sset \ell \in \sets, i' \in \sset \ell \cap \sset k$. Since every element $i \in \sset j$ is in at most two other sets, and there are three elements in \sset j, this is in total at most six enemies.
Thus~$s_j$ has degree at most 11.
\item The agents $t^1_j, t^2_j$ have one friend and no enemies.
\item For every $i \in \sset j$:
\begin{compactitem}
\item The agent $d^i_j$ has three friends and is enemies with $s_\ell$ for every $\sset \ell \in \sets$ such that $i \in \sset j \cap \sset \ell$. There are at most two such sets, as $i$ appears in at most three sets. 
Thus $d^i_j$ has degree at most four.
\item The agent $e^i_j$ has one friend and no enemies.
\end{compactitem}
\end{compactitem}
\end{compactitem}
Thus the maximum degree in $I'$ is 11.

We first show the proof for two initial coalitions.
Let the initial partition be $\partition = \{\{a_i, x_i \mid i \in \elements\} \cup \{s_j, t^1_j, t^2_j \mid \sset j \in \sets\}, \{y_i \mid i \in \elements\} \cup \{d^i_j, e^i_j \mid \sset j \in \sets, i \in \sset j\}\}$.

We make the following observation about $\partition$:
\begin{obs}\label{obs:cvfriends}
In \partition:
\begin{compactenum}[(i)]
\item For every $i \in \elements$, the agents~$a_i$ and $x_i$ obtain two friends in their initial coalition, and $y_i$ obtains no friends in her initial coalition.\label{obs:cvfriendsi}
\item For every $\sset j \in \sets, i \in \sset j$, the agent $s_j$ obtains two friends in her initial coalition, the agents $t^1_j$ and $t^2_j$ each obtain one friend, and~$d^i_j$ and $e^i_j$ obtain one friend each.\label{obs:cvfriendss}
\item No agent obtains enemies in her coalition.
\end{compactenum}
\end{obs}

We show that $I$ admits an exact cover if and only if $\partition$ admits a strictly blocking coalition.

(Only if) $I$ admits an exact cover \ecov.
Let us construct a blocking coalition $\blockingCoalition$ as follows:
\begin{compactitem}
\item For every $i \in \elements$, add the agents $a_i, x_i, y_i$ to $\blockingCoalition$.
\item For every $\sset j \in \ecov, i \in \sset j$, add the agents $d^i_j, s_j$ to $\blockingCoalition$.
\end{compactitem}

We proceed to show that every agent obtains strictly more friends in $\blockingCoalition$ than in her partition under \partition.
\begin{compactitem}
\item For every $i \in \elements$, the agent $a_i$ obtains three friends: $x_i, x_{i - 1}$ (where $1 - 1 = 3n$), and $d^i_j$ for some $\sset j \in \sets$ such that $i \in \sset j$. Since \ecov\ is an exact cover, such a set must exist. By \cref{obs:cvfriends}\eqref{obs:cvfriendsi}, she only obtains two friends under~\partition. As she has no enemies, she may not obtain any.
\item For every $i \in \elements$, the agent $x_i$ obtains three friends: $a_i, a_{i + 1}$ (where $3n + 1 = 1)$, and $y_i$. 
By \cref{obs:cvfriends}\eqref{obs:cvfriendsi}, she only obtains two under \partition. As she has no enemies, she may not obtain any.
\item For every $i \in \elements$, the agent $y_i$ obtains one friend $x_i$.
By \cref{obs:cvfriends}\eqref{obs:cvfriendsi}, she obtains none under \partition. 
As she has no enemies, she may not obtain any.
\item For every $\sset j \in \ecov, i \in \sset j$, the agent $d^i_j$ obtains two friends:~$s_j$ and~$a_i$. 
By \cref{obs:cvfriends}\eqref{obs:cvfriendss}, she obtains one under \partition. 
Since \ecov\ is an exact cover, there may be no set $\sset k \in \sets$ such that $k \neq j$ and $\sset j \cap \sset k \neq \emptyset$.
Thus $d^i_j$ obtains no enemies.
\item For every $\sset j \in \ecov$, the agent $s_j$ obtains three friends: $d^i_j$ for every $i \in \sset j$. 
By \cref{obs:cvfriends}\eqref{obs:cvfriendss}, she obtains two under \partition. 
Since \ecov\ is an exact cover, there may be no set $\sset k \in \sets$ such that $k \neq j$ and $\sset j \cap \sset k \neq \emptyset$.
Thus $s_j$ obtains no enemies.
\end{compactitem}
Thus every agent prefers \blockingCoalition\ over her coalition under \partition.

(If): Assume $\partition$ admits a strictly blocking coalition \blockingCoalition. 
We first make the following observation of agents which may not be contained in it:
\begin{obs}\label{obs:cvcantimpr}
For every $\sset j \in \sets, i \in \sset j$, the agents $e^i_j, t^1_j$, and $t^2_j$ may not join a blocking coalition.
\end{obs}
\begin{proof}\renewcommand{\qedsymbol}{$\diamond$}
All of these agents obtain all their friends and no enemies under \partition. Thus they may not prefer \blockingCoalition\ over their coalitions under~\partition.
\end{proof}

Let us construct an exact cover $\ecov \coloneqq \{\sset j \mid \sset j \in \sets, s_j \in \blockingCoalition\}$.
Assume, towards a contradiction, that \ecov\ is not an exact cover.
Then there are two possible cases: (1) There is an element that is not contained in any of the sets in \ecov, or (2) there is an element that is contained in two sets.

We first show that Case (2) leads to a contradiction.
Assume that there are two sets $\sset j, \sset k \in \ecov$ such that there is $i' \in \sset j \cap \sset k \neq \emptyset$ and $j \neq k$.
Since~$s_j$ obtains two friends in $\partition$ (\cref{obs:cvfriends}\eqref{obs:cvfriendss}), she must obtain three in \blockingCoalition. 
Since~$t^1_j$ and~$t^2_j$ cannot be in \blockingCoalition\ by \cref{obs:cvcantimpr}, the agent $s_j$ must obtain the friend~$d^{i}_j$ for every $i \in \sset j$. This includes $d^{i'}_j$, who is an enemy of the agent~$s_k$, a contradiction.

Now we show that Case (1) leads to a contradiction.
Assume that there is an element $i \in \elements$ that is contained in no set in \ecov. 

Thus no agent $s_j, \sset j \in \sets$ such that $i \in \sset j$ may be in \blockingCoalition.
By \cref{obs:cvfriends}\eqref{obs:cvfriendss}, if $d^i_j \in \blockingCoalition$, where $\sset j \in \sets$ such that $i \in \sset j$, she must obtain two friends, and by \cref{obs:cvcantimpr}, the agent~$e^i_j$ cannot be one of then. But then~$s_j$ must be in~\blockingCoalition, a contradiction.
Therefore~$d^i_j$ cannot be in \blockingCoalition\ for any $\sset j \in \sets$ such that $i \in \sset j$.

Thus $a_i$ cannot be in \blockingCoalition, because she needs to obtain three friends by \cref{obs:cvfriends}\eqref{obs:cvfriendsi}, but she can obtain only two: $x_i$ and $x_{i - 1}$.
Similarly, $x_i$ cannot join a blocking coalition, because she needs three friends by \cref{obs:cvfriends}\eqref{obs:cvfriendsi}, but can obtain only two: $y_i$ and $a_{i + 1}$.
The agent $y_i$ cannot join \blockingCoalition since $x_i$ does not join~\blockingCoalition, as she is her only friend.

The agent $a_{i + 1}$ needs three friends by \cref{obs:cvfriends}\eqref{obs:cvfriendsi}.
We have shown that $x_i$ cannot join \blockingCoalition, so the remaining friends are $s_j$ for every $\sset j \in \sets$ such that $i + 1 \in \sset j$ and the agent $x_{i + 1}$.
Thus for $a_{i + 1}$ to be in \blockingCoalition, there must be two sets $s_j, s_k \in \sets$ such that $i \in \sset j \cap \sset k$ and $d^{i + 1}_j, d^{i + 1}_k \in \blockingCoalition$.
By \cref{obs:cvfriends}\eqref{obs:cvfriendss} and \cref{obs:cvcantimpr}, we must have that $s_j \in \blockingCoalition$, but she is enemies with $d^{i + 1}_k$.
Thus $a_{i + 1}$ cannot be in \blockingCoalition.

By iterating this argument we obtain that no agent in $\{a_i, x_i, y_i \mid i \in \elements\}$ may be in \blockingCoalition.
By \cref{obs:cvcantimpr}, no agent in  $\{t^1_j, t^2_j \mid \sset j \in \sets\} \cup \{e^i_j \mid \sset j \in \sets, i \in \sset j\}$ may be in \blockingCoalition.

Thus there must be some $\sset j \in \sets$ such that $s_j$ or $d^i_j$ for some $i \in \sset j$ is in~\blockingCoalition, as otherwise \blockingCoalition\ is empty.
If $s_j$ is in~\blockingCoalition, to obtain a friend there must be some  $i \in \sset j$ such that~$d^i_j$ is in~\blockingCoalition.
To join \blockingCoalition, the agent~$d^i_j$ needs two friends by \cref{obs:cvfriends}\eqref{obs:cvfriendss}, and one of them cannot be~$e^i_j$. Thus~$a_i$ must be in~\blockingCoalition, a contradiction.

This concludes the (If)-direction.
To show that the problem remains NP-hard even when each initial coalition is of a constant size, let us construct for each $i \in \elements$ four extra agents: $\hat{a}^1_i, \hat{a}^2_i, \hat{x}^1_i,$ and $\hat{x}^2_i$.
The agents $\hat{a}^1_i$ and $\hat{a}^2_i$ are friends only with the agent $a_i$, and the agents $\hat{x}^1_i$ and $\hat{x}^2_i$ are friends only with the agent $x_i$. They are neutral towards all the other agents.
We can observe this does not change the maximum degree of the instance.

The initial partition is as follows: $\bigcup_{i \in \elements}\{\{a_i, \hat{a}^1_i, \hat{a}^2_i\}\} \cup$ $\bigcup_{i \in \elements}$ $\{\{x_i,$ $ \hat{x}^1_i, \hat{x}^2_i\}\}$ $\cup \bigcup_{i \in \elements} \{y_i\}$ $\cup$ $\bigcup_{\sset j \in \sets, i \in \sset j} \{\{d^i_j, e^i_j\}\} \cup$ $\bigcup_{\sset j \in \sets} \{\{s_j,$ $t^1_j, t^2_j\}\}$.

Observe that every agent has the same number of friends and enemies in her initial coalition as previously. The newly created agents obtain all of their friends under $\partition$ and can thus not join a strictly blocking coalition. The rest of the proof proceeds as previously.}

\begin{restatable}[\appsymb]{theorem}{thmscvneutrhard}
\neut{\SCV} is NP-complete even when $\friendshipGraph \cup \enemyGraph$ is bipartite,~$\maxDegreeFE$ is a constant, and either $\maxNumberOfCoalitions = 2$ or $\maxCoalitionSize$ is a constant.\label{thm:scvneutrhard}
\end{restatable}
\appendixproofwithstatement{thm:scvneutrhard}{\thmscvneutrhard*}{
We reduce from \probname{Exact Cover by 3-Sets}, where every element is in at most three sets.

\begin{figure}
    \centering
\begin{tikzpicture}[black]
             \input{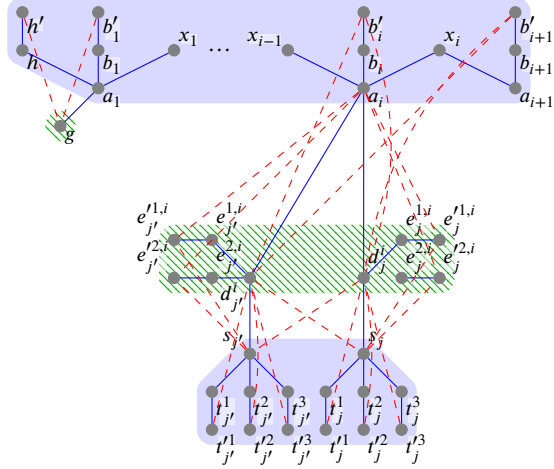}
         \foreach  \x / \y / \n / \nn / \d in
         {2.5/7.5/a1/{a_1}/3,
         3.5/8/x1/{x_1}/1,
         2.5/8/b1/{b_1}/3,
         2.5/8.5/bp1/{b'_1}/3,
         1.5/8/h/{h}/3,
         1.5/8.5/hp/{h'}/3,
         2/7/g/{g}/3,
         6/7.5/ai/{a_i}/3,
         6/8/bi/{b_i}/3,
         6/8.5/bpi/{b'_i}/3,
         6/4/sj/{s_j}/1,
         5.5/3.5/tj1/{t^1_j}/3,
         6/3.5/tj2/{t^2_j}/3,
         6.5/3.5/tj3/{t^3_j}/3,
         5.5/3/tpj1/{t'^1_j}/3,
         6/3/tpj2/{t'^2_j}/3,
         6.5/3/tpj3/{t'^3_j}/3,
         4.5/4/sk/{s_{j'}}/0,
         4/3.5/tk1/{t^1_{j'}}/3,
         4.5/3.5/tk2/{t^2_{j'}}/3,
         5/3.5/tk3/{t^3_{j'}}/3,
         4/3/tpk1/{t'^1_{j'}}/3,
         4.5/3/tpk2/{t'^2_{j'}}/3,
         5/3/tpk3/{t'^3_{j'}}/3,
         6.5/5.5/eij/{e^{1,i}_{j}}/1,
         6.5/5/eij2/{e^{2,i}_{j}}/1,
         7/5.5/epij/{e'^{1,i}_{j}}/1,
         7/5/epij2/{e'^{2,i}_{j}}/1,
         6/5/dij/{d^i_{j}}/1,
         4/5.5/eik/{e^{1,i}_{j'}}/1,
         3.5/5.5/epik/{e'^{1,i}_{j'}}/0,
         4/5/e2ik/{e^{2,i}_{j'}}/1,
         3.5/5/ep2ik/{e'^{2,i}_{j'}}/0,
         4.5/5/dik/{d^i_{j'}}/2,
         7/8/xi/{x_i}/1,
         8/7.5/aip/{a_{i+1}}/3,
         8/8/bip/{b_{i+1}}/3,
         8/8.5/bpip/{b'_{i+1}}/3,
         5/8/xim/{x_{i-1}}/0}
         {
           \node[pn] at (\x, \y) (\n) {};
           \myVertexName{$\nn$}{\n}{\d};
         }

                  \foreach  \x / \y / \n / \nn / \d in
         {3/6.5/ha11p/{\hat{b'}^1_{1}}/3,
         3/7/ha11/{\hat{b}^1_1}/2,
         7.5/7/hai1p/{\hat{b}^1_{i+1}}/3,
         8.5/7/hai2p/{\hat{b}^2_{i+1}}/3,
         5.5/7/hai1/{\hat{b}^1_i}/2,
         6.5/7/hai2/{\hat{b}^2_i}/3,
         7.5/6.5/hpai1p/{\hat{b'}^1_{i+1}}/3,
         8.5/6.5/hpai2p/{\hat{b'}^2_{i+1}}/3,
         5.5/6.5/hpai1/{\hat{b'}^1_i}/2,
         6.5/6.5/hpai2/{\hat{b'}^2_i}/3}
         {
                               }
         
		\foreach  \x / \y / \n in
         {}
         {
           \node[] at (\x, \y) (\n) {};
         }         
         
        \foreach \x / \y / \r / \n in {4.1/8/0/adots}
        {
			\node[rotate=\r, inner sep=0pt] at (\x, \y) (\n) {{$\ldots$}};    
        }
     
     	\foreach \a / \b / \r in {
        a1/x1/0,a1/b1/0,b1/bp1/0,a1/g/0,a1/h/0,h/hp/0,
                        ai/dij/0,ai/dik/0,
        ai/bi/0,bi/bpi/0,aip/bip/0,bip/bpip/0,
                dij/eij/0,dik/eik/0,dik/e2ik/0,eij/epij/0,eij2/epij2/0,eik/epik/0,e2ik/ep2ik/0,
        dij/sj/0,dik/sk/0,
        sj/tj1/0,sj/tj2/0,sj/tj3/0,sk/tk1/0,sk/tk2/0,sk/tk3/0,
        tj1/tpj1/0,tj2/tpj2/0,tj3/tpj3/0,tk1/tpk1/0,tk2/tpk2/0,tk3/tpk3/0,
        ai/xi/0,xi/aip/0,xim/ai/0
        } { 
             \draw[-, friend] (\a) edge[bend right=\r] (\b);
           }
           
      \foreach \a / \b / \r in {dij/sk/0,dik/sj/0,
      hp/g/0,bp1/g/0,
      bpi/dij/-20,bpi/dik/0,bpip/dij/20,bpip/dik/0,
      tpk1/dik/0,tpk2/dik/10,tpk3/dik/0,tpj1/dij/0,tpj2/dij/10,tpj3/dij/0,
      sj/epij2/0,sj/epij/0,sk/epik/0,sk/ep2ik/0,
      ai/epij2/0,ai/epij/0,ai/ep2ik/0,ai/epik/0} { 
             \draw[-, enemy] (\a) edge[bend right=\r] (\b);
           }
    \begin{pgfonlayer}{background}

      \draw[coal] \hedgem{tpk1}{tk1}{sk,sj,tj3,tpj3}{2mm};
      \draw[coalgreen] \hedgem{ep2ik}{epik}{epij,epij2}{2mm};
      \draw[coal] \hedgem{h}{hp}{bpip,aip,a1}{2mm};
      \draw[coalgreen] \hedgei{g}{2mm};

                    \end{pgfonlayer}

\end{tikzpicture}
    \caption{A figure for the proof of \cref{thm:scvneutrhard}. We assume $i \in S_j \cap S_{j'}$. The black solid edges signify friendships and the red dashed edges enemy-relations.
    The blue solid and green dashed regions signify the initial partition with two coalitions.}
    \label{fig:scvneutrhard}
\end{figure}

Let $I = (\elements, \sets = \{S_1, \dots, S_m\})$ be an instance of \probname{Exact Cover by 3-Sets}.
Let us construct the agents as follows: This is also illustrated in \cref{fig:scvneutrhard}.
\begin{compactitem}
\item For every $i \in \elements$, construct the agents $a_i, b_i, b'_i$, and if $i \neq 3n$, then $x_i$.
\item For every set $\sset j \in \sets$, construct the agent $s_j,$ the agents $t^z_j, t'^z_j$ for every $z \in [3]$, and for every $i \in \sset j$, \begin{compactitem}
\item the agent $d^i_j$, and
\item for every $z \in [2]$, the agents $e^{z,i}_j, e'^{z,i}_j$.
\end{compactitem}
\item Finally construct the agents $g, h, h'$.
\end{compactitem}

Let us construct the friendship relations as follows:
\begin{compactitem}
\item For every set $\sset j \in \sets, i \in \sset j$,  the agent~$s_j$ is friends with the agent~$d^i_j$ and the agent $d^i_j$ is friends with $a_i, e^{1,i}_j$, and $e^{2,i}_j$.
For every $w \in [3]$, $s_j$ is friends with $t^w_j$, and $t^w_j$ is also friends with $t'^w_j$.
For every $z \in [2]$, $e^{z,i}_j$ is friends with $e'^{z,i}_j$.
\item For every $i \in [3n - 1]$, the agent $x_i$ is friends with $a_i$ and~$a_{i + 1}$.
\item For every $i \in [3n]$, the agent $a_i$ is friends with $b_i$, and the agent $b_i$ is also friends with $b'_i$.
\item The agent $a_1$ is friends with $h$ and $g$, and $h$ is friends with $h'$.
 \end{compactitem}
The enemy relations are as follows:
\begin{compactitem}
\item For every $i \in \elements$, for every $\sset j, \sset k \in \sets$ such that $i \in \sset j \cap \sset k$ and $\sset j \neq \sset k$, the agent $d^i_j$ is enemies with the agent $s_k$.
\item For every $i \in \elements$, the agent $b'_i$ is enemies with~$d^i_j$ for every $\sset j \in \sets$ such that $i \in \sset j$ and if $i \neq 1$, then with~$d^{i - 1}_j$ for every $\sset j \in \sets$ such that $i - 1 \in \sset j$.
\item For every $\sset j \in \sets, i \in \sset j, z \in [2]$, the agent $e'^{z,i}_j$ is enemies with $a_i$ and $s_j$.
\item For every $\sset j \in \sets, i \in \sset j, z \in [3]$, the agent $t'^{z}_j$ is enemies with $d^i_j$.
\item The agent $g$ is enemies with $h'$ and $b'_1$.
\end{compactitem}

The union of the \friendshipGraph and \enemyGraph is clearly bipartite by a partition whose first set is $\{\{a_i, b'_i \mid i \in \elements\} \cup \{s_j, t'^1_j, t'^2_j, t'^3_j \mid \sset j \in \sets\}  \cup \{e^i_j \mid \sset j \in \sets, i \in \sset j\} \cup \{h'\}\}$ and the second set is $\{ \{x_i \mid i \in [3n - 1]\} \cup \{b_i \mid i \in \elements\} \cup \{d^i_j, e'^{1,i}_j, e'^{2,i}_j, e'^{3,i}_j \mid \sset j \in \sets, i \in \sset j\} \cup \{t^1_j, t^2_j, t^3_j \mid \sset j \in \sets\} \cup \{h,g\}\}$.

Let us observe that the agents have the following numbers of friends and enemies:
\begin{compactitem}
\item For every $i \in \elements$:
\begin{compactitem}
\item The agent $a_i$ has the friends $x_{i - 1},$ (or $g$ and $h$ if $i = 1$), $x_i, b_i$, and $d^i_j$ for every set $\sset j \in \sets$ such that $i \in \sset j$, in total at most six agents. She is enemies with $e'^{z,i}_j$ for every $\sset j \in \sets$ such that $i \in \sset j$ (at most three such sets), for every $z \in [2]$, in total at most six agents.
Thus her degree is at most 12.
\item If $i \neq 3n$, the agent $x_i$ is friends with $a_i$ and $a_{i + 1}$, in total two agents. 
She has no enemies.
\item The agent $b_i$ has two friends and no enemies.
\item The agent $b'_i$ has one friend, and she is enemies with $g$ (if $i = 1$), the agents $d^i_j$ for every $\sset j \in \sets$ such that $i \in \sset j$, if $i \neq 1$ the agents $d^{i - 1}_j$ for every $\sset j \in \sets$ such that $i - 1\in \sset j$, in total at most six enemies. Thus her degree is at most seven.
\end{compactitem}
\item For every $\sset j \in \sets$:
\begin{compactitem}
\item The agent $s_j$ is friends with $d^i_j$ for every $i \in \sset j$ and the agents $t^z_j$ for every $z \in [3]$, in total six friends. She is enemies with $d^{i'}_\ell$ for every $\sset \ell \in \sets, i' \in \sset \ell \cap \sset k$ and $e^{z,i}_j$ for every $i \in \sset j, z \in [2]$. Since every element $i \in \sset j$ is in at most two other sets, and there are three elements in $\sset j$, this is in total at most 12 enemies.
Thus $s_j$ has degree at most 18.
\item For every $z \in [3]$, the agent $t^z_j$ has two friends, and no enemies. The agent $t'^z_j$ has one friend and is enemies with~$d^i_j$ for every $i \in \sset j$, giving in total three enemies. Thus her degree is four.
\item For every $i \in \sset j$:
\begin{compactitem}
\item The agent $d^{z,i}_j$ has four friends and is enemies for $s_\ell$ for every $\sset \ell \in \sets$ such that $i \in \sset j \cap \sset \ell$. There are at most two such sets, as $i$ appears in at most three sets. 
She is also enemies with $b_i$ and $b_{i + 1}$ (if $i \neq 3n$), and $t'^z_j$ for every $z \in [3]$, giving in total at most seven enemies.
Thus her degree is at most 11.
\item For every $z \in [2]$, the agent $e^{z,i}_j$ has two friends, and no enemies. The agent $e'^{z,i}_j$ has one friend and is enemies with $a_i$ and $s_j$, giving in total two enemies. 
Thus her degree is four.
\end{compactitem}
\end{compactitem}
\item The agent $g$ has one friend and two enemies, thus degree three. The agent $h$ has two friends and no enemies. The  agent~$h'$ has one friend and one enemy, thus degree two.
\end{compactitem}
Thus the maximum degree in $I'$ is 18.

We first show the proof for two initial coalitions.
Let the initial partition be $\partition = \{\{a_i, b_i, b'_i \mid i \in \elements\} \cup \{x_i \mid i \in [3n - 1]\}  \cup \{s_j \mid \sset j \in \sets\} \cup \{t^z_j, t'^z_j \mid \sset j \in \sets, z \in [3]\} \cup \{h,h'\},  \{d^i_j \mid \sset j \in \sets, i \in \sset j\} \cup \{e^{z,i}_j, e'^{z,i}_j \mid \sset j \in \sets, i \in \sset j, z \in [3]\}\} \cup \{g\} \}$.

We make the following observation about $\partition$:
\begin{obs}\label{obs:csvfriends}
In \partition:
\begin{compactenum}[(i)]
\item For every $i \in [3n - 1]$, the agent~$a_i$ obtains three friends in her initial coalition, $x_i$ two friends, $b_i$ two friends, and $b'_i$ one friend. The agent $a_{3n}$ obtains two friends, $b_{3n}$ two friends, and $b'_{3n}$ one friend.\label{obs:csvfriendsi}
\item For every $\sset j \in \sets$, the agent $s_j$ obtains three friends in her initial coalition, and for every $z \in [3]$, the agent $t^z_j$ gets two friends, and the agent $t'^z_j$ one friend. 

For every $i \in \sset j$, the agent $d^i_j$ obtains two friends. For every $w \in [2]$, the agent $e^{w, i}_j$ obtains two friends and $e'^{w, i}_j$ one.\label{obs:csvfriendss}
\item The agent $h$ obtains two friends in her initial coalition, $h'$ one, and $g$ zero.\label{obs:scvfriendsg}
\item No agent obtains enemies in her coalition.
\end{compactenum}
\end{obs}

We show that $I$ admits an exact cover if and only if $\partition$ admits a weakly blocking coalition.

(Only if) $I$ admits an exact cover \ecov.
Let us construct a blocking coalition $\blockingCoalition$ as follows:
\begin{compactitem}
\item For every $i \in \elements$, add the agents $a_i, x_i$ (if $i = 3n$, add only~$a_i$) to $\blockingCoalition$.
\item For every $\sset j \in \ecov, i \in \sset j$, add the agents $d^i_j, s_j$ to $\blockingCoalition$.
\item Add the agent $g$ in \blockingCoalition.
\end{compactitem}

We proceed to show that every agent weakly prefers $\blockingCoalition$ over her coalition under \partition.
\begin{compactitem}
\item For every $i \in \elements$, the agent $a_i$ obtains three friends in \blockingCoalition: $x_i, x_{i - 1}$ (unless $i = 3n$), and $s_j$ for some $\sset j \in \sets$ such that $i \in \sset j$. Since~\ecov\ is an exact cover, such a set must exist. By \cref{obs:csvfriends}\eqref{obs:csvfriendsi}, she obtains three (two if $i = 3n$) under \partition. Her only enemies are agents $e'^{z,i}_j$ for some $\sset j \in \sets, z \in [2]$, and these are not in \blockingCoalition.
Thus $a_i$ weakly prefers \blockingCoalition\ over $\partition(a_i)$.
\item For every $i \in [3n - 1]$, the agent $x_i$ obtains two friends in~\blockingCoalition: the agents~$a_i$ and $a_{i + 1}$. 
By \cref{obs:csvfriends}\eqref{obs:csvfriendsi}, she obtains two under~\partition. As she has no enemies, she may not obtain any.
Thus she weakly prefers \blockingCoalition\ over $\partition(x_i)$.
\item For every $\sset j \in \ecov, i \in \sset j$, the agent $d^i_j$ obtains two friends: the agents~$s_j$ and $a_i$. 
By \cref{obs:csvfriends}\eqref{obs:csvfriendss}, she obtains two under~\partition. 
Since~\ecov\ is an exact cover, there is no set $\sset k \in \sets$ such that $k \neq j$ and $\sset j \cap \sset k \neq \emptyset$.
She is additionally enemies with~$b'_i$, who is not in \blockingCoalition.
Thus $d^i_j$ obtains no enemies.
Therefore~$d^i_j$ weakly prefers \blockingCoalition\ over $\partition(d^i_j)$.
\item For every $\sset j \in \ecov$, the agent $s_j$ obtains three friends: $d^i_j$ for every $i \in \sset j$. 
By \cref{obs:csvfriends}\eqref{obs:csvfriendss}, she obtains three under~\partition. 
Since \ecov\ is an exact cover, there may be no set $\sset k \in \sets$ such that $k \neq j$ and $\sset j \cap \sset k \neq \emptyset$.
She is additionally enemies with~$t'^{z,i}_j$ for every $i \in \sset j, z \in [3]$, but these are not in \blockingCoalition.
Thus $s_j$ obtains no enemies.
She weakly prefers \blockingCoalition\ over~$\partition(s_j)$.
\item The agent $g$ obtains one friend, whereas she initially obtains none. She is only enemies with $h'$ and $b'_1$, who are not in \blockingCoalition.
Thus she prefers \blockingCoalition\ over $\partition(g)$.
\end{compactitem}
Thus every agent in \blockingCoalition\ weakly prefers \blockingCoalition\ over her coalition under \partition, and $g$ prefers \blockingCoalition\ over $\partition(g)$.
Therefore \blockingCoalition\ indeed weakly blocks \partition.

(If): Assume $\partition$ admits a weakly blocking coalition \blockingCoalition. 

\begin{claim}\label{clm:scv_exactcov}
For every $i \in \elements$, if $a_i \in \blockingCoalition$, there can be at most one set $\sset j \in \sets$ such that $d^i_j \in \blockingCoalition$.
\end{claim}
\begin{proof}\renewcommand{\qedsymbol}{$\diamond$}
Assume, towards a contradiction, that there are two distinct sets $\sset j, \sset k \in \sets$ such that $i \in \sset j \cap \sset k$ and $a_i, d^i_j, d^i_k \in \blockingCoalition$.
Since~$t'^{1,i}_j$ and~$t'^{2,i}_j$ are enemies with $a_i$, they will not join \blockingCoalition. Thus neither will their friends $t^{1,i}_j$, $t^{2,i}_j$, as they would obtain fewer friends in \blockingCoalition.
Therefore $d^i_j$ loses one existing friend and must thus obtain one new friend. The only option is $s_j$. However, $s_j$ is enemies with~$d^i_k$, a contradiction.
\end{proof}

\begin{claim}\label{obs:csvcantimpr}
The only agent that can prefer $\blockingCoalition$ over her coalition under~\partition\ is $g$.
\end{claim}

\begin{proof}\renewcommand{\qedsymbol}{$\diamond$}
The agents $h, h'$, $b_i, b'_i$, where $i \in \elements$, $x_i$, where $i \in [3n - 1]$, $e^{z,i}_j, e'^{z,i}_j$, where $z \in [2], \sset j \in \sets, i \in \sset j$, $t^w_j, t'^w_j$, where $w \in [3], \sset j \in \sets$ already have all their friends and no enemies.

The agent $a_i$, where $i \in \elements$ can prefer \blockingCoalition\ over $\partition(a_i)$ if she obtains friends~$d^i_j$ for a set $\sset j \in \sets$ such that $i \in \sset j$.
However, every such agent is an enemy of $b'_i$, who will then not join \blockingCoalition. Her friend~$b_i$ needs all of her friends to join a blocking coalition, so $b_i$ will not join either.
Thus $a_i$ loses a friend, and must obtain another friend~$d^k_j$, where $\sset k \in \sets, i \in \sset k$. By \cref{clm:scv_exactcov} this cannot happen.

The agent $d^i_j$, where $\sset j \in \sets, i \in \sset j$, needs three friends to prefer \blockingCoalition\ over $\partition(d^i_j)$ by \cref{obs:csvfriends}\eqref{obs:csvfriendss}. Since she is friends with $e^{1,i}_j, e^{2,i}_j, a_i$, and~$s_j$, the agent $e^{1,i}_j$ or $e^{2,i}_j$ must be included, as must~$a_i$ or~$s_j$. However, for every $z \in [2]$, the agents $a_i$ and~$s_j$ are both enemies with~$e'^{z,i}_j$, and~$e^{z,i}_j$ will not join a blocking coalition without~$e'^{z,i}_j$, as she obtains all of her friends in \partition. Thus $d^i_j$ cannot prefer~\blockingCoalition\ over~$\partition(d^i_j)$.

The agent $s_j$, where $\sset j \in \sets$, needs four friends to prefer~\blockingCoalition\ over~$\partition(s_j)$ by \cref{obs:csvfriends}\eqref{obs:csvfriendss}.
However, then she will need at least one friend~$d^i_j$, where $i \in \sset j$, and at least one friend $t^z_j$, where $z \in [3]$. To join~\blockingCoalition\ the agent~$t^z_j$ needs to get all of her friends as she has them under \partition, and thus $t'^z_j \in \blockingCoalition$. However, the agent~$t'^z_j$ is enemies with~$d^i_j$, a contradiction.
\end{proof}

Let us construct an exact cover $\ecov \coloneqq \{\sset j \mid \sset j \in \sets, s_j \in \blockingCoalition\}$.
We proceed to show that \ecov\ is an exact cover.

Since $g$ is the only agent that can prefer \blockingCoalition\ over her coalition under \partition, she must be in \blockingCoalition\ and prefer it over $\partition(g)$.
Thus $a_1 \in \blockingCoalition$, since $g$ needs a friend in \blockingCoalition.
Because $g$ is enemies with~$h$ and $b'_1$, neither they nor their respective friends~$h$ and~$b_1$ are in \blockingCoalition.
Thus by \cref{obs:csvfriends}\eqref{obs:scvfriendsg}, the agent $a_1$ must get two friends among the agents $x_1$ and $d^1_j$, where $\sset j \in \sets, 1 \in \sset j$.
By \cref{clm:scv_exactcov}, there can be at most one agent~$d^1_j$, such that $\sset j \in \sets, 1 \in \sset j$.
Thus~$x_1$ and exactly one~$d^1_j$, such that $\sset j \in \sets, 1 \in \sset j$ are in \blockingCoalition.

The agent $x_1$ must get all of her friends in \blockingCoalition. Thus $a_2$ must be in \blockingCoalition.
Because there is $d^1_j \in \blockingCoalition$ such that $1 \in \sset j$, the agent $b'_2$ has an enemy in \blockingCoalition\ and will not join. Neither will her friend $b_2$.
Thus by \cref{obs:csvfriends}\eqref{obs:scvfriendsg}, the agent $a_2$ must get two friends among $x_2$, and $d^2_j$ such that $\sset j \in \sets, 2 \in \sset j$.
By \cref{clm:scv_exactcov}, there can be at most one agent $d^2_j \in \blockingCoalition$, such that $\sset j \in \sets, 2 \in \sset j$.
Thus $x_2$ and exactly one~$d^2_j$, such that $\sset j \in \sets, 2 \in \sset j$ are in \blockingCoalition.

By repeating this argument we obtain that for every $i \in  [3n - 1]$, $a_i, x_i \in \blockingCoalition$ and there is exactly one $\sset j \in \sets$ such that $d^i_j \in \blockingCoalition$.
For $3n$, we have by \cref{obs:csvfriends}\eqref{obs:scvfriendsg} and \cref{clm:scv_exactcov} that the agent $a_{3n}$ must get exactly one friend among~$d^{3n}_j$, where $\sset j \in \sets, 3n \in \sset j$.

Since $a_i \in \blockingCoalition$ for every $i \in \elements$, no element  $e'^{z,i}_j$, $z \in [2], \sset j \in \sets$ can be in \blockingCoalition, and neither can her respective friend~$e^{z,i}_j$.

For every $i \in \elements$, let us consider $\sset j \in \sets$ for which $d^i_j$.
The agent $d^i_j$ must obtain two friends.
Thus $s_j \in \blockingCoalition$.
As this holds for every element, the set \ecov\ must cover all the elements.

Moreover, for every $\sset j \in \ecov$, the agent $s_j$ must obtain three friends, which must be precisely $d^i_j$ for every $i \in \sset j$. 
By \cref{clm:scv_exactcov}, for every element $\ell$ there can be at most one $d^\ell_k \in \blockingCoalition$ such that $\ell \in \sset k$.
Thus \ecov\ must be an exact cover.

This concludes the (Only if)-direction.
To show that the problem remains NP-hard even when each initial coalition is of size at most three, construct the additional agents $\hat{b}^z_i, \hat{b}'^z_i$ for every $i \in \elements, z \in [2]$ (unless $i \in \{1, 3n\}$, in which case $z \in [1]$) and $y^z_i, y'^z_i$ for every $i \in [3n - 1], z \in [2]$.

For every $i \in \elements, z \in [2]$ (unless $i \in \{1, 3n\}$, in which case $z \in [1]$), let $\hat{b}^z_i$ be friends with $a_i$ and $\hat{b}'^z_i$, and let $\hat{b}'^z_i$ be enemies with all the enemies of $b'_i$. Additionally, let $b'_i$, $\hat{b}'^1_i$, and $\hat{b}'^2_i$ be enemies with $x_i$ if $i \neq 3n$ and $x_{i - 1}$ if $i \neq 1$.

For every $i \in [3n - 1], z \in [2]$, let the agent $y^z_i$ be friends with~$x_i$ and~$y'^z_i$.
Let~$y'^z_i$ be enemies with $a_i$ and $a_{i + 1}$.

This increases the degree of the agents $a_i, i \in \elements$ by four. Thus the maximum degree of the new instance is still~18.

The initial partition is as follows: $\{\{h, h', a_1, b_1, b'_1, \hat{b}^1_1, \hat{b}'^1_1\}\} \cup \bigcup_{i \in [3n - 1]} \} \{\{a_i, b_i, b'_i, \hat{b}^1_i,  \hat{b}^2_i,  \hat{b}'^1_i, \hat{b}'^2_i\}\} \cup \{ \{a_{3n}, b_{3n},$ $b'_{3n}, \hat{b}^1_{3n},   \hat{b}'^1_{3n}\} \}\cup \bigcup_{i \in \elements \setminus \{1\}} \{\{x_i,$ $ y^1_i,$ $y^2_i, y'^1_i, y'^2_i\}\}$ $\cup$ $\bigcup_{\sset j \in \sets, i \in \sset j} \{\{d^i_j, e^{1,i}_j, e^{2,i}_j, e'^{1,i}_j,$ $e'^{2,i}_j\}\} \cup$ $\bigcup_{\sset j \in \sets} \{\{s_j,$ $t^1_j,$ $t^2_j, t^3_j, t'^1_j,$ $t'^2_j, t'^3_j\}\} \cup \{\{g\}\}$.

Observe that every old agent has the same number of friends and enemies in her initial coalition as previously.
Thus the (If)-direction proceeds analogously.

If the modified instance admits a blocking coalition \blockingCoalition, the proof that only $g$ can prefer \blockingCoalition\ over her coalition under \partition\ proceeds analogously.
Since $g \in \blockingCoalition$, the agents $h, h', b_1, b'_1, \hat{b}^1_1,$ and $ \hat{b}'^1_1$ cannot join~\blockingCoalition\ through a reasoning similar to the original case.
We again obtain that $a_1$ must get two further friends: $d^1_j$ for some $\sset j \in \sets$ and $x_1$.
Since $y'^1_i$ and $y'^2_i$ are enemies with $a_1$, the agents $y^1_i$ and~$y^2_i$ cannot be in \blockingCoalition. Thus $x_1$ needs to get an additional friend in \blockingCoalition, and the only such friend is $a_{2}$.
By repeating this argument, we get that $a_i$ must be in \blockingCoalition\ for every $i \in \elements$, and she needs to obtain one friend $d^i_j$, where $\sset j \in \sets, i \in \sset j$.
From here the proof proceeds as previously.}

\section{Conclusion}
In this paper, we have studied the parameterized complexity of hedonic games under \enemyav, both with and without neutrals.
In the case with neutrals, a few open questions remain.
We know that all the studied problems remain NP-hard when $\friendshipGraph \cup \enemyGraph$ is an interval graph, since clique is an interval graph and the problems are NP-hard without neutrals.
How about the case where only $\friendshipGraph$ or $\enemyGraph$ is an interval graph? 
We also do not know the complexity of \neut{CE} when $\friendshipGraph \cup \enemyGraph$ is bipartite.
The parameterized complexity regarding treewidth is also unknown, although Peters~\cite{peters2016graphical} shows that the problems are in FPT with respect to $\maxDegreeFE + \treewidth$.
\label{sec:conclusion}

\paragraph*{Acknowledgements.}
Martin Durand and Sofia Simola have been funded by the Vienna Science and Technology Fund (WWTF) grant [10.47379/VRG18012].
We would like to thank Jiehua Chen for helpful conversations and insights.

\bibliographystyle{elsarticle-num} 
\bibliography{bib}

\begin{appendix}
\newpage
\appendixtext
\end{appendix}

\end{document}